\documentclass[twocolumn,showpacs,nofootinbib,preprintnumbers,amsmath,amssymb,showkeys,superscriptaddress]{revtex4}
\usepackage{epsfig}
\usepackage{amssymb}
\usepackage{amsmath}
\usepackage{graphicx}
\usepackage{bm}
\usepackage{color}
\usepackage{ulem}
\usepackage{dsfont}

\newcommand{\beq}{\begin{equation}}
\newcommand{\eeq}{\end{equation}}
\newcommand{\bea}{\begin{eqnarray}}
\newcommand{\eea}{\end{eqnarray}}
\newcommand{\nn}{\nonumber \\}

\newcommand\eqn[1]{(\ref{#1})}      
\newcommand\Eqn[1]{Eq.~(\ref{#1})}  
\newcommand\Fig[1]{Fig.~\ref{#1}}  

\newcommand{\tr}{\hbox{tr}}

\begin{document}

\title{Yang-Mills correlators across the deconfinement phase transition }
\author{U. Reinosa}%
\affiliation{%
Centre de Physique Th\'eorique, \'Ecole Polytechnique, CNRS,  Universit\'e Paris-Saclay, F91128 Palaiseau Cedex, France.\vspace{.1cm}}%
\author{J. Serreau}
\affiliation{Astro-Particule et Cosmologie (APC), CNRS UMR 7164, Universit\'e Paris Diderot,\\ 10, rue Alice Domon et L\'eonie Duquet, 75205 Paris Cedex 13, France.\vspace{.1cm}}
\author{M. Tissier}
\affiliation{{ Laboratoire de Physique Th\'eorique de la Mati\`ere Condens\'ee, UPMC, CNRS UMR 7600, Sorbonne
Universit\'es, 4 Place Jussieu,75252 Paris Cedex 05, France.}\vspace{.1cm}}
\affiliation{ Instituto de F\'\i sica, Facultad de Ingenier\'\i a, Universidad de la Rep\' ublica,\\
Julio Herreira y Reissig 565, 11000 Montevideo, Uruguay.\vspace{.1cm}}
\author{A. Tresmontant}
\affiliation{Astro-Particule et Cosmologie (APC), CNRS UMR 7164, Universit\'e Paris Diderot,\\ 10, rue Alice Domon et L\'eonie Duquet, 75205 Paris Cedex 13, France.\vspace{.1cm}}
\affiliation{{ Laboratoire de Physique Th\'eorique de la Mati\`ere Condens\'ee, UPMC, CNRS UMR 7600, Sorbonne
Universit\'es, 4 Place Jussieu,75252 Paris Cedex 05, France.}\vspace{.1cm}}

\date{\today}

\begin{abstract}
We compute the  finite temperature ghost and gluon propagators of Yang-Mills theory in the Landau-DeWitt gauge. The background field that enters the definition of the latter is intimately related with the (gauge-invariant) Polyakov loop and serves as an equivalent order parameter for the deconfinement transition. We use an effective gauge-fixed description where the nonperturbative infrared dynamics of the theory is parametrized by a gluon mass which, as argued elsewhere, may originate from the Gribov ambiguity. In this scheme, one can perform consistent perturbative calculations down to infrared momenta, which have been shown to correctly describe the phase diagram of Yang-Mills theories in four dimensions as well as the zero-temperature correlators computed in lattice simulations. In this article, we provide the one-loop expressions of the finite temperature Landau-DeWitt ghost and gluon propagators for a large class of gauge groups and present explicit results for the SU($2$) case. These are substantially different from those previously obtained in the Landau gauge, which corresponds to a vanishing background field. The nonanalyticity of the order parameter across the transition is directly imprinted onto the propagators in the various color modes. In the SU($2$) case, this leads, for instance, to a cusp in the electric and magnetic gluon susceptibilities as well as similar signatures in the ghost sector.  We mention the possibility that such distinctive features of the transition could be measured in lattice simulations in the background field gauge studied here.
\end{abstract}


\maketitle

\section{Introduction}\label{sec:introduction}

The phase diagram of strong interactions at finite temperature, chemical potential, magnetic field, etc. is the subject of intense theoretical and experimental studies \cite{QM2015}.  Many features have been firmly assessed by numerical lattice simulations, such as the existence and order of a confinement-deconfinement phase transition at finite temperature in Yang-Mills theories which turns into a crossover in QCD with realistic quark masses \cite{Bazavov:2014pvz,Borsanyi:2013bia}. However, various questions remain unanswered, the most prominent one being related to the physics at finite chemical potential (where Monte Carlo techniques are plagued by the sign problem) and, in particular, the possible existence of a critical point. 

The correlation functions of the elementary QCD fields are the basic ingredients of continuum approaches \cite{Alkofer00,Pawlowski:2005xe}. Their calculation necessarily involves approximations, but it has the advantage that it does not suffer from the severe sign problem of lattice techniques.\footnote{We mention, though, that there is a milder sign problem in continuum approaches, related to the nonconvexity of the effective action at nonzero real chemical potential \cite{Fukushima:2006uv,Nishimura:2014kla,Reinosa:2015oua}.} It is then of first importance to have a detailed understanding of these basic correlators in order to assess the validity of the approximations employed in such approaches. Moreover, gauge-invariant observables are often pretty difficult to access through continuum approaches and it is thus of great interest to check whether the physics under scrutiny, e.g., the value of a critical temperature, can be read off directly at the level of the basic correlation functions of the theory. Finally, the determination of gauge-invariant observables in continuum approaches requires a good understanding of the properties of correlation functions.

For these reasons, a great deal of activity has been devoted to compute Euclidean Yang-Mills correlation functions in the vacuum both with (gauge-fixed) lattice calculations \cite{Cucchieri_08b,Bornyakov2008,Bogolubsky09,Cucchieri09,Bornyakov09,Dudal10,Boucaud:2011ug,Cucchieri:2012ii,Maas:2011se} and with continuum approaches \cite{Ellwanger96,vonSmekal97,Alkofer00,Boucaud06,Aguilar07,Aguilar08,Boucaud08,Fischer08,RodriguezQuintero10,Tissier:2010ts,Pelaez:2013cpa,Huber:2012kd,Quandt:2013wna,Siringo:2014lva,Machado:2016cij,Cyrol:2016tym} in the Landau gauge.\footnote{ Continuum studies in the Hamiltonian approach have also been performed in the Coulomb gauge \cite{Feuchter:2004mk,Reinhardt:2004mm}; see also \cite{Reinhardt:2013iia} for an extension to finite temperatures.} The main conclusion is that the ghost propagator behaves as that of a massless field whereas the gluon propagator saturates to a finite value at vanishing momentum, corresponding to a nonzero screening mass. It also violates spectral positivity, which indicates that the associated excitation is not an asymptotic state, as expected from confinement. This line of investigation has been naturally extended at finite temperature; see, e.g. Refs.~\cite{Heller95,Cucchieri00,Cucchieri07,Fischer:2010fx,Cucchieri11,Aouane:2011fv,Maas:2011ez,Maas:2011se,Silva:2013maa,Mendes:2014gva} for lattice calculations in the Landau gauge. One of the questions studied in these works concerns the possibility that the nonanalytic behavior of the gauge-invariant order parameter of the transition, the Polyakov loop, is imprinted in the correlators of the gluon and ghost degrees of freedom. This is certainly a nontrivial question since such correlators are gauge-dependent quantities. For instance, one would expect such a scenario when the gluon field is directly related to the Polyakov loop, such as in the Polyakov gauge \cite{Marhauser:2008fz}. It is, however, far from clear that the same is true in a generic gauge.

Lattice calculations in the Landau gauge found no sign of the phase transition, neither in the ghost propagator (which is, in fact, essentially independent of the temperature) nor in the so-called magnetic gluon propagator, which is roughly speaking associated with the correlation function for the spatial components of the gluon field. The situation is less clear for the so-called electric sector, which involves the time component of the gluon field, more directly connected to the Polyakov loop. Despite early indications that the electric susceptibility---the electric propagator at vanishing frequency and momentum---might be a sensitive probe of the transition~\cite{Fischer:2010fx,Maas:2011ez},  simulations with larger volumes showed no clear signature \cite{Cucchieri11}. In fact, existing calculations of the electric susceptibility in the Landau gauge show an extreme sensitivity to both the lattice size and the lattice spacing for temperatures slightly below the transition temperature \cite{Mendes:2014gva}, for reasons that are not fully understood; see, however, Ref.~\cite{Silva:2016onh}.

Continuum calculations of the finite-temperature ghost and gluon propagators have been performed in the Landau gauge using a wide variety of approaches  \cite{Fister11,Fischer:2012vc,Fukushima:2013xsa,Huber:2013yqa,Reinosa:2013twa,Quandt:2015aaa}. One typically finds a slight nonmonotonous behavior of the electric susceptibility below the transition, but no clear sign of the transition, in qualitative agreement with the lattice results on larger volumes \cite{Cucchieri11,Mendes:2014gva}.

Finally, we mention that an important drawback of the Landau gauge is that it explicitly breaks the center symmetry of the finite temperature problem: the set of configurations compatible with the gauge condition is not invariant under the corresponding transformations. This makes it difficult to monitor the transition---which is controlled by the spontaneous breaking of the center symmetry---within a given approximation scheme. Incidentally, this could also be related to the convergence issues of lattice calculations. Moreover, the fact that no sign of the transition is seen in the lowest order correlation functions may be understood from the fact that the influence of the order parameter on the correlation functions is very indirect. For instance, in continuum approaches, the Polyakov loop does not enter at any level in the definition or calculation of the correlators.

In the present work, we undertake the study of the gluon and the ghost propagators at finite temperature in the Landau-DeWitt (LDW) gauge, which is a straightforward generalization of the Landau gauge in presence of a background field \cite{DeWitt:1967ub,Abbott:1980hw,Weinberg:1996kr}. More precisely, we consider the massive extension of the LDW gauge put forward in Ref.~\cite{Reinosa:2014ooa}, which allows for a perturbative description of the phase transition, see also \cite{Reinosa:2014zta,Reinosa:2015gxn}. { The motivations for such a massive extension have been discussed in these articles and basically originate from the decoupling behavior of the vacuum ghost and gluon propagators observed in Landau gauge lattice calculations. The massive extension of the Landau gauge---a particular case of the Curci-Ferrari model \cite{Curci:1976bt}---has been shown to give an accurate description of the vacuum \cite{Tissier:2010ts,Pelaez:2013cpa} and, to some extent, of the finite temperature \cite{Reinosa:2013twa} Yang-Mills correlators at one-loop order. The present work is a direct background field generalization of the calculation of Ref.~\cite{Reinosa:2013twa}. }The major interest of this approach as compared to the Landau gauge is that the center symmetry is explicit \cite{Braun:2007bx}.  In fact, one can show that certain background fields---obtained by minimizing an appropriate potential to be defined below---provide alternative order parameters for the center symmetry, equivalent to the (gauge-invariant but more difficult to access) Polyakov loop \cite{Herbst:2015ona,Reinosa:2015gxn}. The relevant background field potential has been evaluated both with nonperturbative continuum approaches \cite{Braun:2007bx,Braun:2010cy,Fukushima:2012qa,Fister:2013bh,Herbst:2015ona,Quandt:2016ykm} and from perturbative calculations, either in the Gribov-Zwanziger approach at one-loop order \cite{Canfora:2015yia}, or in the massive extension of the LDW gauge considered here at one- and two-loop orders \cite{Reinosa:2014ooa,Reinosa:2014zta,Reinosa:2015gxn}. Such calculations correctly reproduce the phase structure of Yang-Mills theories with values of the transition temperatures in good agreement with lattice results. 

Interestingly, these calculations also show that the background field takes nonzero values below and, to some extent, above the transition temperature,\footnote{Two-loop perturbative calculations in \cite{Reinosa:2014zta,Reinosa:2015gxn} show that the background only vanishes at asymptotically large temperatures.} and that the vanishing background field, which corresponds to the Landau gauge, is never a minimum of the background potential in the relevant range of temperatures.
This suggests that the correlators in the LDW gauge might be more directly sensitive to the phase transition than the ones in the Landau gauge. It would be of interest to study the possible implementation of the LDW gauge in lattice calculations, e.g., along the lines of Ref.~\cite{Cucchieri:2012ii}.

We compute the basic Yang-Mills two-point correlators  at one-loop order in the (massive) LDW gauge. The whole calculation is essentially analytical, which allows us to consistently keep track of the background field at all steps. We give the general expressions of the ghost and gluon self-energies for a large class of gauge groups and we present explicit results for the SU($2$) theory. We find striking differences as compared to the case with vanishing background which was already treated in \cite{Reinosa:2013twa}. In particular, all propagators show a clear nonanalytic behavior across the transition.

The paper is organized as follows. In Sec. \ref{sec_model}, we present the model, recall how the center symmetry can be controlled in presence of a background field and describe the Feynman rules of the theory. In Sec.~\ref{sec_propag}, we present the one-loop calculation of the self-energies. Section~\ref{sec_perturbativ_results} is devoted to the description of our results. We finally conclude. Several appendixes describe some technical aspects of the calculations.


\section{The (massive) Landau-DeWitt gauge}
\label{sec_model}
\subsection{Gauge-fixing}
\label{sec_gaugefixing}
We consider the finite temperature Euclidean Yang-Mills action for a finite dimensional compact Lie group $G$ with a simple Lie algebra $\mathcal{G}$ in $\smash{d=4-2\epsilon}$ dimensions, with $\epsilon$ an ultraviolet regulator. The classical action reads
\beq 
\label{eq:SYM}
 S_{\rm YM}[A]=\frac{1}{2}\int_x \tr \left\{ F_{\mu\nu}F_{\mu\nu}\right\}, 
\eeq 
where $\smash{\int_x\equiv \int_0^\beta  d \tau \int d^{d-1}x}$, with $\smash{\beta=1/T}$ the inverse temperature. The field strength tensor $ F_{\mu\nu}$ writes
\beq
\label{eq:fst}
 F_{\mu\nu}=\partial_\mu A_\nu-\partial_\nu A_\mu -ig_0 [A_\mu,A_\nu]\,,
\eeq
where $g_0$ is the (bare) coupling constant. The (matrix) gauge field $\smash{iA_\mu=iA_\mu^a t^a}$ belongs to the algebra ${\cal G}$, with the group generators $t^a$ normalized as $\tr\{t^a t^b \}=\delta^{ab}/2$. 

We quantize the theory using background field methods \cite{DeWitt:1967ub,Abbott:1980hw,Weinberg:1996kr}. Specifically, we write $A_\mu=\bar A_\mu+a_\mu$, with a given background field $\bar A_\mu$ and we choose the Landau-DeWitt (LDW) gauge
\beq
\label{eq_gaugecondition}
 \bar{D}_\mu a_\mu=0\,,
\eeq
where $\bar D_\mu\varphi=\partial_\mu\varphi -ig_0 [\bar A_\mu,\varphi]$ for any field $i\varphi$ in the algebra ${\cal G}$. Here, we study the following gauge-fixed action
\beq
\label{eq_gf}
 S=\!\int_x\tr\left\{{1\over2}F_{\mu\nu}F_{\mu\nu}\!+\!{m_0^2}a_\mu a_\mu\!+\!2\bar D_\mu\bar cD_\mu c\!+\!2ih\bar D_\mu a_\mu\right\}\!,
\eeq
with $ih$ a Nakanishi-Lautrup (Lagrange multiplier) field and $(ic,i\bar c)$ a pair of Faddeev-Popov (FP) ghost/antighost fields, all in the Lie algebra of the group. Apart from the bare mass term $\propto m_0^2$, the action \eqn{eq_gf} is nothing but the standard FP gauge-fixed action corresponding to the condition \eqn{eq_gaugecondition}. However, the latter presents Gribov ambiguities, i.e., it fixes the gauge only up to a discrete set of configurations, an issue that the FP procedure simply disregards. As discussed at length elsewhere \cite{Reinosa:2014ooa,Reinosa:2014zta,Reinosa:2015gxn}, the bare mass $m_0^2$ for the gluon field $a_\mu$ is an effective way to account for the Gribov problem in the present gauge.\footnote{See Refs.~\cite{Serreau:2012cg} for an explicit realization of this model in relation with a Gribov-consistent gauge fixing procedure and Ref.~\cite{Serreau:2013ila} for a generalization to a broader class of (nonlinear) covariant gauges.} It explicitly breaks the usual nilpotent BRST symmetry of the FP gauge-fixed action, however without hampering the renormalizability of the theory. 

In terms of the field $a_\mu$, we have
\begin{align}
\label{eq:Fieldstrength}
 F_{\mu\nu}&=\bar F_{\mu\nu}+\bar D_\mu a_\nu-\bar D_\nu a_\mu-ig_0 [a_\mu, a_\nu]\,,
\end{align}
with $\bar F_{\mu\nu}$ the field strength tensor \eqn{eq:fst} evaluated at $A=\bar A$, and 
\beq
 D_\mu \varphi=\partial_\mu \varphi-ig_0 [A_\mu,\varphi]=\bar D_\mu\varphi-ig_0 [a_\mu,\varphi]\,.
\eeq
The action \eqn{eq_gf} has the obvious property
\beq
\label{eq_ginvbare}
 S[\bar A, \varphi]=S[\bar A^U, U\varphi U^{-1}]\,,
\eeq
where $U$ is an element of the gauge group, $\varphi=(a,c,\bar c,h)$, and 
\begin{equation}
\label{eq:transfofo}
  \bar A_\mu^U=U\bar A_\mu U^{-1}+\frac i {g_0}U\partial_\mu U^{-1}.
\end{equation}
At the level of the (quantum) effective action $\Gamma$ this implies \cite{Weinberg:1996kr}
\beq
\label{eq_ginv}
 \Gamma[\bar A, \varphi]=\Gamma[\bar A^U, U\varphi U^{-1}]\,,
\eeq
provided that $U$ preserves the periodicity of the fields along the Euclidean time direction.

For a given background $\bar A$, the vertex functions of the theory are obtained as
\beq
\label{eq:vertexfunc}
 \Gamma^{(n)}[\bar A]=\left.\frac{\delta^{(n)}\Gamma[\bar A, \varphi]}{\delta\varphi\cdots\delta\varphi}\right|_{\varphi_{\rm min}[\bar A]},
\eeq
where $\varphi_{\rm min}[\bar A]=(a_{\rm min}(\bar A),0,0,0)$, with $a_{\rm min}(\bar A)$ the absolute minimum of $\Gamma[\bar A,a,0,0,0]$ for fixed $\bar A$.\footnote{ More precisely, the extremization with respect to $h$ forces the gauge condition $\bar D_\mu a_\mu=0$. Under this contraint, the effective action is independent of $h$.} They depend, of course, on the chosen background $
\bar A$. In contrast, (gauge-invariant) observables do not depend, in principle, on the chosen background field $\bar A$. In practice however, it is convenient to choose so-called self-consistent backgrounds $\bar A^s$, defined as $a_{\rm min}[\bar A^s]=0$ or, equivalently, $A_{\rm min}[\bar A^s]=\bar A^s$. As discussed, for instance, in \cite{Reinosa:2015gxn}, these correspond to the absolute minima of the following background field functional 
 \beq
 \label{eq:bckfunc}
  \tilde \Gamma[\bar A]=\Gamma[\bar A,0].
\eeq 
Self-consistent background fields thus allow one to study the possible states of the system through the simpler functional \eqn{eq:bckfunc}. The latter is invariant under background field gauge transformations \eqn{eq:transfofo} that preserve the time-periodicity of the fields,
\beq
\label{eq_ginvtilde}
 \tilde \Gamma[\bar A]=\tilde \Gamma[\bar A^U],
\eeq
as follows from \Eqn{eq_ginv}. This encodes, in particular, the center symmetry of the theory. As we recall below, see also \cite{Herbst:2015ona,Reinosa:2015gxn}, one deduces from \Eqn{eq_ginvtilde} that the minima of the functional \eqn{eq:bckfunc} are order parameters of this symmetry---and thus of the confinement-deconfinement transition for static color charges---. 
The relation with the usual gauge-invariant order parameter, the Polyakov loop $\ell$, is as follows
\beq
\label{eq:decadix}
\ell=\frac{1}{N}{\rm tr}\,\left\langle P \exp \int_0^\beta \!\!d\tau \, ig_0\left(\bar A_0+a_0\right)\right\rangle\,,
\eeq
where, on the right-hand side, it is understood that $\bar A_0$ is a minimum of $\tilde \Gamma[\bar A]$ and, consequently, that\footnote{One virtue of self-consistent background field configurations is that tadpolelike diagrammatic insertions are automatically cancelled, which greatly simplifies perturbative calculations.} $\langle a_0\rangle=0$. This relation between the background field and the Polyakov loop has been computed to next-to-leading order in \cite{Reinosa:2015gxn}.

\subsection{Center symmetry}\label{sec:sym}
We now specialize to self-consistent background fields that explicitly preserve the symmetries of the finite temperature problem at hand, that is, homogeneous and in the temporal direction:\footnote{More generally, one could consider background fields configurations which preserve the Euclidean space symmetries up to a gauge transformation.}
\beq
\label{eq:form}
 \bar{A}_\mu(x)=\bar{A} \delta_{\mu 0}\,.
\eeq
Using the symmetry under global transformations of $G$, we can choose the constant matrix $\bar A$ in the Cartan subalgebra of ${\cal G}$ without loss of generality. We write $g_0\beta\bar A=r^j t^j$, where $t^j$ are the generators in the Cartan subalgebra, and we introduce the background field effective potential \cite{Braun:2007bx}
\beq
\label{eq:potential}
V(r)\equiv\frac{1}{\beta\Omega}\tilde\Gamma[\bar A]\,,
\eeq
where $\Omega$ is the spatial volume. It follows from \Eqn{eq_ginvtilde} that the function $V(r)$ is invariant under gauge transformations \eqn{eq:transfofo} that  preserve the periodicity of the fields as well as the form \eqn{eq:form} of the background field. { These are given by particular global color rotations,} called Weyl transformations, together with specific time-dependent gauge transformation which are $\beta$-periodic up to an element of the center of the group. In particular, these include standard, periodic gauge transformations. Exploiting the fact that the latter, together with the Weyl transformations, do not change the physical observables, one can restrict the study of $V(r)$ to a finite domain called the Weyl chamber \cite{Dumitru:2012fw,vanBaal:2000zc,Herbst:2015ona,Reinosa:2015gxn}. The latter is typically invariant under a discrete group whose elements correspond to center transformations and possibly other discrete symmetries of the theory such as charge conjugation. Depending on their location in the Weyl chamber, the minima of the potential $V(r)$ are left invariant or not by such transformations, which demonstrates that they are order parameters { for the corresponding discrete symmetries \cite{Reinosa:2015gxn}.}\footnote{ That the background field is an order parameter of the center symmetry in the LDW gauge has been widely assumed in the literature; see, e.g., Refs.~\cite{Braun:2007bx,Braun:2010cy,Fister11,Quandt:2016ykm}. A rigorous proof has been given for the large class of gauge groups considered here, together with a discussion of charge conjugation symmetry, using the concept of Weyl chambers, in Ref.~\cite{Reinosa:2015gxn}; see also Ref.~\cite{Herbst:2015ona} for a discussion of center symmetry in the SU($2$) and SU($3$) groups.}

As an illustration, consider the the SU($2$) case. The Cartan subalgebra is one-dimensional and the Weyl chamber can be taken as the segment $r\in[0,2\pi]$. Its reflexion symmetry about $r=\pi$ corresponds to the center $\mathds{Z}_2$. A minimum at $r\neq\pi$ corresponds to a state of spontaneously broken center symmetry, that is, a deconfined phase. 
Clearly, the center-symmetric value $r=\pi$ leads to a vanishing Polyakov loop in \Eqn{eq:decadix} at all loop orders \cite{Reinosa:2014zta}. Note that the reverse is not necessarily true, although it has been explicitly checked at next-to-leading order in the SU($2$) and in the SU($3$) theories \cite{Reinosa:2014zta,Reinosa:2015gxn}. In the general case, demanding the vanishing of the Polyakov loop \eqn{eq:decadix} is not enough to uniquely select the center-symmetric value of the background field in a given Weyl chamber. One must, at least, demand that a collection of Polyakov loops, one for each group representation of nonzero $N$-ality, vanish; see Ref.~\cite{Reinosa:2015gxn} for a detailed discussion.

The potential \eqn{eq:potential} has been computed in perturbation theory in the present massive model at one- and two-loop orders for any compact gauge group with a simple Lie algebra in Refs.~\cite{Reinosa:2014ooa,Reinosa:2014zta,Reinosa:2015gxn}. In the SU($2$) and SU($3$) cases, the obtained phase structure and the values of the transition temperatures agree well with known lattice results and with nonperturbative continuum approaches \cite{Braun:2007bx,Maas:2011ez,Fister:2013bh}. In the present work, we shall use the two-loop results of Ref.~\cite{Reinosa:2014zta} for the SU($2$) theory. An extension to SU($3$) is straightforward, following \cite{Reinosa:2015gxn}.  Let us comment that, unlike their one-loop counterparts, these two-loop potentials have been shown to yield a positive entropy at all temperatures \cite{Reinosa:2014zta,Reinosa:2015gxn}.

\subsection{Feynman rules}\label{Feynman_rules}
{
The background field introduces preferred directions in color space, those of the Cartan subalgebra. Different color modes couple differently to the background which lifts their degeneracy and induces a nontrivial color structure of the various correlators. In particular, the standard cartesian basis $t^a$ is not the most appropriate one to discuss the Feynman rules. Instead, it is  preferable to work in so-called canonical, or Cartan-Weyl bases \cite{Dumitru:2012fw,Reinosa:2015gxn} where the background covariant derivative $\bar D_\mu$ and, thus, the free propagators are diagonal. In the case of the SU($2$) group, where the Cartan subalgebra has a single element, say $t^3=\sigma^3/2$, a canonical basis is given by the generators}
\beq
\label{eq:basis}
 t^0=t^3\,, \quad t^\pm=\frac{t^1\pm i t^2}{\sqrt{2}},
\eeq 
which satisfy
\beq\label{eq:basis2}
 [t^\kappa,t^\lambda]=\varepsilon^{\kappa\lambda\tau}t^{-\tau}\quad{\rm and} \quad \tr\{t^\kappa t^{-\lambda}\}=\frac{\delta^{\kappa\lambda}}{2}\,,
\eeq
{ with} $\kappa,\lambda,\tau\in\{0,+,-\}$ and where $\varepsilon^{\kappa\lambda\tau}$ is the completely antisymmetric tensor, with $\varepsilon^{0+-}=1$. A given matrix field in the Lie algebra decomposes as $i\varphi=i\varphi^\kappa t^{\kappa}$, with $(\varphi^\kappa)^*=\varphi^{-\kappa}$ for appropriately choosen $t^\kappa$'s. For instance, the background field writes $r=g_0\beta\bar A=r^0t^0$. In the following, we shall write $r^0=r$ for simplicity since there is only one component and thus no ambiguity.

In Fourier space, with the convention $\partial_\mu\to-iK_\mu$, the action of the background covariant derivative reduces to
\beq
\label{eq:barderivative2}
(\bar D_\mu \varphi)^\kappa \to -iK_\mu^{\kappa}\varphi^\kappa(K),
\eeq
where $K_\mu^{\kappa}=K_\mu+\kappa rT\delta_{\mu0} $ defines a generalized momentum.\footnote{We reserve the set of greek letters $(\mu, \nu, \rho, \sigma)$ to denote Euclidean space indices and the other set $(\kappa, \lambda, \eta, \xi,\tau)$ to denote color states. With this convention, $K_\mu$ denotes the $\mu$ component of the four-vector $K$, whereas $K^\kappa$ refers to the shifted momentum $K+\kappa g_0\bar An$ with $n=(1,{\bf 0})$.}  { The latter is conserved} by virtue of the  invariance under translation in Euclidean space and under the residual global SO($2$) symmetry corresponding to those color rotations that leave the background invariant. The index $\kappa=0,\pm$ labels the corresponding Noether charges and we see that the covariant derivative simply shifts the Matsubara frequencies of the charged color modes by  $\pm rT$. In this sense, $rT$ plays the role of an imaginary chemical potential for the color charge measured by $t^0$ in the adjoint representation.

{ The above considerations extend to any compact Lie group $G$ with a simple Lie algebra}. In general, there are as many neutral modes, denoted by $0^{(j)}$, as there are generators  $t^j\equiv t^{0^{(j)}}$ in the Cartan subalgebra. The charged modes correspond to certain combinations $t^\alpha$ of the remaining generators and it is convenient to label them in terms of the roots $\alpha$ of the algebra. Such roots are real vectors with as many components $\alpha^j$ as there are dimensions in the Cartan subalgebra. The set of generators $t^\kappa$, with $\kappa=0^{(j)}$ or $\kappa=\alpha$, forms a canonical basis in which the action of the covariant derivative in Fourier space amounts to the multiplication of the corresponding color mode by the shifted, or generalized, momentum $K_\mu^\kappa\equiv K_\mu+(\kappa^jr^j)T\delta_{\mu0}$. For each neutral mode, $\kappa=0^{(j)}$ is the null vector, $\kappa^j=0$, whereas for charged modes $\kappa^j=\alpha^j$. The generators in the canonical basis satisfy the relations \eqn{eq:basis2}  with generic structure constant $\varepsilon^{\kappa\lambda\tau}\to f^{\kappa\lambda\tau}$. The conservation of the color charge is encoded in the fact that  $f^{\kappa\lambda\tau}$ is zero if $\kappa +\lambda +\tau\neq 0$. It can also be shown that
{ 
\beq
\label{eq:structure}
  (f^{\kappa\lambda\tau})^*=-f^{(-\kappa)(-\lambda)(-\tau)}.
\eeq 
Finally, we shall consider theories with real structure constants, $(f^{\kappa\lambda\tau})^*=f^{\kappa\lambda\tau}$, whose Lagrangian is invariant under the combined charge-conjugation transformation of the background, $\bar A^j\to-\bar A^j$, and of the fluctuating fields, $\varphi^\kappa\to-\varphi^{-\kappa}$.}

\begin{figure}[t!]
\begin{center}
\epsfig{file=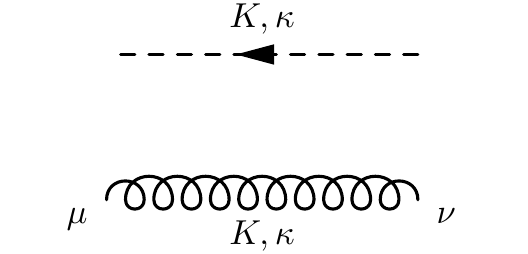,width=4.2cm}
 \caption{Diagrammatic representation of the ghost (dashed) and gluon (curly) propagators for momentum $K$ and color charge $\kappa$. The common orientation of the flow of momentum and color charge is arbitrary.}\label{fig:propags2}
\end{center}
\end{figure}

Let us recall the Feynman rules in the canonical basis for a general gauge group $G$ \cite{Reinosa:2015gxn}. The tree-level propagators, see { \Fig{fig:propags2},} read
\begin{eqnarray}
  \langle c^{-\kappa}(-K)\bar{c}^{\kappa}(K)\rangle & = & G_0(K^\kappa)\,,\\\label{eq_propagcc}
  \langle a_\mu^{-\kappa}(-K)a_\nu^{\kappa}(K)\rangle & = & P^{\perp}_{\mu \nu}(K^\kappa) G_{m_0}(K^\kappa)\,,  \label{eq_propagAA}
\end{eqnarray}
where $P^{\perp}_{\mu \nu}(K) = \delta_{\mu \nu}-K_\mu K_\nu/K^2$ and
\beq
G_{m_0}(K)\equiv\frac{1}{K^2+m^2_0}\,.
\eeq
The gluon propagator in each color mode is transverse with respect to the corresponding generalized momentum. Notice also that, thanks to the identity 
\beq
 \label{eq:idmom}
 (-K_\mu)^{-\kappa}=-K_\mu^\kappa\,,
\eeq
the orientation of the generalized momentum in the diagrams of \Fig{fig:propags2} is arbitrary.

\begin{figure}[t!]  
\begin{center}
\epsfig{file=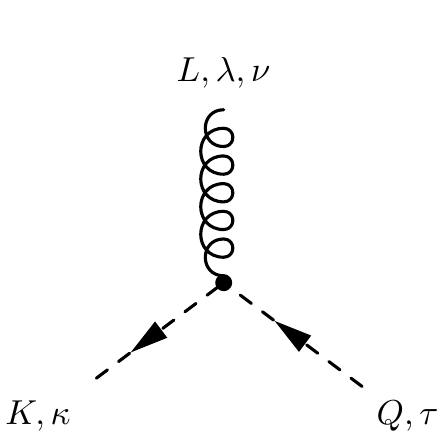,width=3.4cm}\qquad\epsfig{file=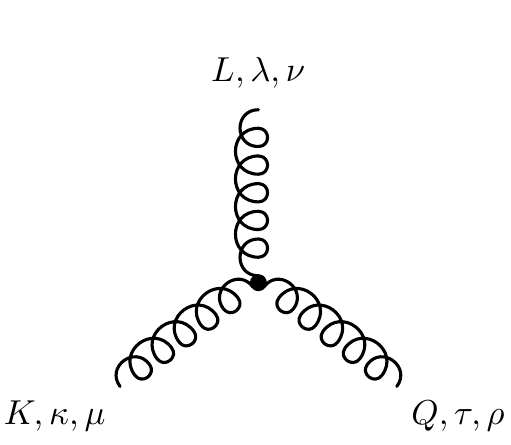,width=3.7cm}\\
 \caption{Diagrammatic representation of the cubic vertices. }\label{fig:vertex3}
\end{center}
\end{figure}

\begin{figure}[t!]  
\begin{center}
\epsfig{file=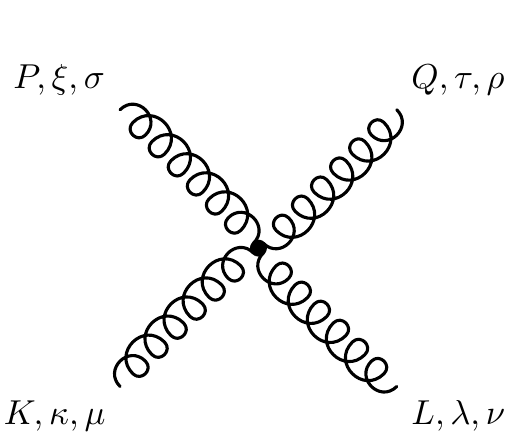,width=4cm}\\
 \caption{Diagrammatic representation of the four-gluon vertex.}\label{fig:vertex4}
\end{center}
\end{figure}

The interaction vertices are displayed in \Fig{fig:vertex3}-\ref{fig:vertex4} and are the standard YM ones, that is, the ghost-antighost-gluon vertex and the three- and four-gluon vertices.
The expression of the ghost-antighost-gluon vertex is given by
\beq\label{eq:v31}
g_0 f_{\kappa \lambda \tau} K_\nu^\kappa\,,
\eeq
where we use as convention that all momenta and color charges are outgoing. With similar conventions, the three gluon vertex reads
\beq\label{eq:v32}
\frac{g_0}{6} f_{\kappa \lambda \tau}\!\Big[ \delta_{\mu \rho}\!\left(K_\nu^\kappa - Q^\tau_\nu \right)\!+\! \delta_{\mu \nu}\!\left(L_\rho^\lambda - K^\kappa_\rho \right)\!+\!\delta_{\rho \nu}\!\left(Q_\mu^\tau - L^\lambda_\mu \right) \Big].
\eeq
 The structure constants ensure that the color charges are conserved at both cubic vertices, which, together with the usual conservation of momenta, leads to the announced conservation rule for the generalized momenta: $K^\kappa+Q^\tau+ L^\lambda=0$.
Finally, the four gluon vertex, represented in \Fig{fig:vertex4}  with all charges outgoing, is given by 
\begin{eqnarray}\label{eq:v4}
\frac{g_0^2}{24} \sum_{\eta} & & \Big[ f_{\kappa \lambda \eta} f_{\tau \xi (-\eta)}\left(\delta_{\mu \rho}\delta_{\nu \sigma}-\delta_{\mu \sigma} \delta_{\nu \rho} \right)\nonumber\\
 & &+ f_{\kappa \tau \eta} f_{\lambda \xi (-\eta)}\left(\delta_{\mu \nu}\delta_{\rho \sigma}-\delta_{\mu \sigma} \delta_{\nu \rho} \right)\nonumber\\
 & & +f_{\kappa \xi \eta} f_{\tau \lambda (-\eta)}\left(\delta_{\mu \rho}\delta_{\nu \sigma}-\delta_{\mu \nu} \delta_{\sigma \rho}\right) \Big].
\end{eqnarray}
Again, color charge and momentum conservation lead to $K^\kappa+Q^\tau+ L^\lambda+ P^\xi=0$.

The Feynman rules for the vertices are given in Eqs.~(\ref{eq:v31})--(\ref{eq:v4}) with the convention that the momenta and color charges are all outgoing. In
general, if they are all taken {ingoing} instead, one needs to replace the structure constants by their complex conjugates. However, this is of no effet for the groups where the structure constants $f^{\kappa\lambda\tau}$ are real, as considered here. We now apply these Feynman rules to the computation of the two-point correlators of the theory at one-loop order.

\section{Propagators at one-loop order}
\label{sec_propag}

{The ghost and gluon two-point correlators are matrices in color space. As emphasized before, the nontrivial background yields preferred directions in color space and the correlators are not simply proportional to the unit matrix. However, symmetries allow us to constrain the form of that matrix. In particular, global color transformations of the form $\exp\{i\theta^jt^j\}$, with $t^j$ the Cartan generators clearly leave the background invariant and thus remain symmetries of the theory. The fields transform as \cite{Reinosa:2015gxn}
\beq
 \varphi^\kappa\to e^{i\kappa\cdot \theta}\varphi^\kappa.
\eeq
Of course, neutral modes $\kappa=0^{(j)}$ are left invariant. One concludes that two-point correlators are block diagonal, with blocks in the neutral and charged sectors, but no mixing between the neutral and charged sectors. Moreover the block in the charged sector is diagonal. In contrast, the block in the neutral sector does not need to be diagonal and could involve off-diagonal elements, coupling different neutral modes, for groups with a Cartan subalgebra of dimension larger than one. However, in certain cases, like the SU($3$) theory, one can show that the neutral block is diagonal if the state of the system (as defined by the background field) is charge-conjugation invariant.\footnote{The proof goes as follows. The SU($3$) theory has two neutral modes corresponding to the generators $t^3$ and $t^8$ of the Cartan subalgebra in the Gell-Mann basis. Weyl chambers are equilateral triangles in the plane $(r_3,r_8)$, for instance, the one with vertices $(0,0)$ and $(2\pi,\pm 2\pi/\sqrt{3})$. As discussed in Ref.~\cite{Reinosa:2015gxn} the charge-conjugation transformation of both the background and the fluctuating fields reads $\varphi^\kappa\to-\varphi^{-\kappa}$, while there exists a particular Weyl transformation (which, we recall, is nothing but a particular global color rotation) given by $(r_3,r_8)\to(-r_3,r_8)$ and  similarly for the fluctuating fields in the neutral sector. Combining both transformations and making the background dependence of the correlator explicit, as ${\cal G}^{\kappa\lambda}_{(r_3,r_8)}$, we conclude, first, that the charge-conjugation invariant backgrounds in the above-mentioned Weyl chamber are located at $r^8=0$ and, second, that ${\cal G}^{0^{(3)}0^{(8)}}_{(r^3,r^8)}=-{\cal G}^{0^{(3)}0^{(8)}}_{(r^3,-r^8)}$. It follows that, if charge-conjugation invariance is not broken, ${\cal G}^{0^{(3)}0^{(8)}}_{(r^3,0)}=0$, as announced.} This is the actual situation in the pure Yang-Mills theory or in the presence of quarks at vanishing chemical potential but it is not true anymore at finite chemical potential, in which case we expect a mixing between the two SU($3$) neutral components, $0^{(3)}$ and $0^{(8)}$. The formulae to be derived below for a general group assume that the background is such that there is no mixing between the various color components.

We define the color components of the ghost and gluon propagators as 
\beq
\label{eq:propertyiy}
 {\cal G}^{\kappa\lambda}(K)=\delta^{-\kappa,\lambda}{\cal G}^\lambda(K)\,\,,\quad{\cal G}_{\mu\nu}^{\kappa\lambda}(K)=\delta^{-\kappa,\lambda}{\cal G}_{\mu\nu}^\lambda(K)
\eeq
 and those of the corresponding self-energies as,
\beq
 \Sigma^{\kappa\lambda}(K)=\delta^{-\kappa,\lambda}\Sigma^\kappa(K)\,\,,\quad\Pi_{\mu\nu}^{\kappa\lambda}(K)=\delta^{-\kappa,\lambda}\Pi_{\mu\nu}^\kappa(K).
\eeq
With these conventions, we have, for the ghost propagator}
\beq
\label{eq:lefantome}
{ {\cal G}^\lambda(K)=\frac{1}{\left(K^\lambda\right)^2+g^2_0\Sigma^\lambda(K)}\,,}
\eeq
where we have extracted a factor $g^2_0$ for later convenience. As for the gluon propagator, the LDW gauge condition~\eqn{eq_gaugecondition} implies that ${\cal G}^\lambda_{\mu\nu}(K)$ is transverse with respect to the generalized momentum: $\smash{K_\mu^\lambda {\cal G}^\lambda_{\mu \nu}(K)= {\cal G}^\lambda_{\mu \nu}(K)K_\nu^\lambda=0}$. 
It thus admits the following tensorial decomposition
\beq\label{eq:decomp}
{\cal G}^\lambda_{\mu \nu}(K)= {\cal G}^\lambda_T(K) P_{\mu \nu}^T(K^\lambda)+{\cal G}^\lambda_L(K) P_{\mu \nu}^L(K^\lambda)\,,
\eeq
where $P_{\mu \nu}^T(K)$ and $P_{\mu \nu}^L(K)$ are the transverse and longitudinal projectors with respect to the frame of the thermal bath, defined as [we write $K=(\omega,{\bf k})$ and $k=|{\bf k}|$]
\beq\label{eq_projT}
P^T_{\mu \nu}(K)=\left(1-\delta_{\mu 0} \right)\left(1-\delta_{\nu 0} \right) \left(\delta_{\mu \nu}-\frac{K_\mu K_\nu}{k^2}\right)
\eeq
and
\beq\label{eq_projL}
P^L_{\mu \nu}(K)+P_{\mu \nu}^T(K)=P_{\mu \nu}^\perp(K)= \delta_{\mu \nu}-\frac{K_\mu K_\nu}{K^2}.
\eeq
{ It follows in particular that ${\cal G}^\lambda_{\mu\nu}(K)={\cal G}^\lambda_{\nu\mu}(K)$. In terms of the projected self-energies
\bea
\label{eq_decompo_propag-2}
\Pi^\lambda_{T}(K) & = & \frac{P^T_{\mu \nu}(K^\lambda)\Pi_{\mu\nu}^\lambda(K)}{d-2}\,,\\
\label{eq_decompo_propag-3}
\Pi^\lambda_{L}(K) & = & P^L_{\mu \nu}(K^\lambda)\Pi_{\mu\nu}^\lambda(K)\,,
\eea
the scalar components of the gluon propagator read 
\beq\label{eq_decompo_propag}
{ {\cal G}^\lambda_{T/L}(K)=\frac{1}{\left(K^\lambda\right)^2+m^2_0+g^2_0\Pi^\lambda_{T/L}(K)}.}
\eeq}
The longitudinal and transverse sectors are referred to as electric and magnetic respectively. For states satisfying \Eqn{eq:propertyiy}, the propagators are real and have the property
 \begin{align}
\label{eq:prop1}
 {\cal G}^\lambda(K)&={\cal G}^{-\lambda}(-K)\,,\\ 
\label{eq:LT}
{\cal G}^\lambda_{L,T}(K)&={\cal G}^{-\lambda}_{L,T}(-K)\,,
\end{align}
and similarly for the self-energies.\footnote{One can always chose the group generators such that $\smash{t^\dagger_\kappa=t_{-\kappa}}$. We thus have $\varphi_\kappa^*=\varphi_{-\kappa}$  for the Hermitian matrix fields $\varphi=(a_\mu,c,\bar c,h)$ or, in momentum space, $\varphi_\kappa^*(Q)=\varphi_{-\kappa}(-Q)$. Using the ghost conjugaison symmetry $(c,\bar c)\to(\bar c,-c)$, one concludes that ${\cal G}^{\kappa\lambda}(K)={\cal G}^{\lambda\kappa}(-K)=\left[{\cal G}^{(-\lambda)(-\kappa)}(K)\right]^*$ and ${\cal G}_{\mu\nu}^{\kappa\lambda}(K)={\cal G}_{\nu\mu}^{\lambda\kappa}(-K)=\left[{\cal G}_{\nu\mu}^{(-\lambda)(-\kappa)}(K)\right]^*$. Eqs.~\eqn{eq:prop1} and \eqn{eq:LT} then follow from the properties \eqn{eq:idmom} and \eqn{eq:propertyiy} and the decomposition \eqn{eq:decomp}.}

\subsection{Ghost self-energy}\label{sec:twopt}
\begin{figure}[t]
  \centering
  \includegraphics[width=.5\linewidth]{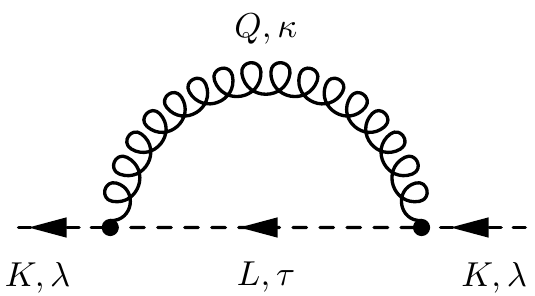}\quad
   \caption{One-loop contribution to the ghost self-energy. Momenta and color charges are all either incoming or outgoing at the vertices.}
  \label{fig_ccb}
\end{figure}

We compute the one-loop contribution to the ghost self-energy represented { in \Fig{fig_ccb}.} 
We note $K=(\omega,{\bf k})$ and $\lambda$ the external momentum and color charge, and $K^\lambda=(\omega^\lambda,{\bf k})$ the corresponding shifted/generalized momentum. The internal loop momentum is denoted $Q\equiv (\omega_n,{\bf q})$ and the shifted/generalized one $Q^\kappa=(\omega_n^\kappa,{\bf q})$, with the Matsubara frequency $\omega_n\equiv 2\pi n T$, $n\in\mathbb{Z}$. Finally we use the notation
\beq
\int_Q f(Q)\equiv \mu^{2\epsilon}\,T\sum_{n\in\mathds{Z}}\int \frac{d^{d-1}q}{(2\pi)^{d-1}}f(\omega_n,{\bf q})\,,
\eeq
where $\mu$ is the arbitrary scale associated with dimensional regularization (recall that $d=4-2\epsilon$).

A direct application of the Feynman rules of Sec.~\ref{Feynman_rules} to the diagram of  \Fig{fig_ccb} yields
\begin{align}
 \Sigma^\lambda(K)&=-\sum_{\kappa,\tau}{ f^{\lambda\kappa\tau}f^{(-\tau)(-\kappa)(-\lambda)}}\nonumber\\
&\times \int_Q  P^\perp_{\mu\nu}(Q^\kappa) (-L)_\mu^{-\tau} K_\nu^\lambda G_m(Q^\kappa)G_0(L^\tau)\,,
\end{align}
where $K+Q+L=0$ and where the sum runs over all color states. Using the anti-symmetry of the structure constant tensor as well as the identities \eqn{eq:structure} and \eqn{eq:idmom}, we arrive at
\beq
\label{eq:houhou}
 \Sigma^\lambda(K)=-\sum_{\kappa,\tau}{\cal C}_{\kappa\lambda\tau}\!\int_Q  \!P^\perp_{\mu\nu}(Q^\kappa) L_\mu^\tau K_\nu^\lambda G_m(Q^\kappa)G_0(L^\tau)\,,
\eeq
with the totally symmetric tensor ${\cal C}_{\kappa\lambda\tau}=|f_{\kappa\lambda\tau}|^2$. As discussed previously, $f^{\kappa\lambda\tau}$ and, consequently, { ${\cal C}_{\kappa\lambda\tau}$ vanish if $\kappa+\lambda+\tau\neq0$}, which implies the conservation of the generalized momentum at the vertices: $Q^\kappa+K^\lambda+L^\tau=0$. We note the close resemblance of the above expression with the corresponding one-loop expression in the Landau gauge; see Eq.~(15) of Ref.~\cite{Reinosa:2013twa}. For instance, one checks that \Eqn{eq:houhou} reduces to the Landau gauge expression in the case of a vanishing background field. 

More generally, the relation between loop calculations in the Landau and in the LDW gauges has been discussed in Ref.~\cite{Reinosa:2014zta} and { is useful to shortcut} intermediate calculations of the loop integrals. In particular, the conservation of the generalized momentum allows us to closely follow those manipulations performed in Ref.~\cite{Reinosa:2013twa} which only relied on momentum conservation and which aimed at expressing the one-loop self-energy in terms of simple scalar tadpolelike and bubblelike integrals. We shall thus not repeat these manipulations here but simply point out the differences with the results of Ref.~\cite{Reinosa:2013twa}. In the present case, the only difference is that the integral $\int_Q (K\cdot Q)[G_0(Q)-G_m(Q)]=0$, which appeared in the Landau gauge calculation, generalizes to $\int_Q (K^\lambda\cdot Q^\kappa)[G_0(Q^\kappa)-G_m(Q^\kappa)]\neq0$ in the LDW gauge. We obtain
\begin{align}
 \label{eq:ghost_self}
\Sigma^\lambda(K)  &= \sum_{\kappa,\tau}{\cal C}_{\kappa \lambda \tau} \Bigg[\frac{K_\lambda^2-m^2}{4m^2} \left(J_m^\kappa-J_0^\kappa \right)+\frac{K_\lambda^4}{4m^2}I_{00}^{\kappa\tau}(K)\nn&-\frac{\left(K_\lambda^2+m^2\right)^2}{4m^2} I_{m0}^{\kappa\tau}(K)-\frac{\omega^\lambda}{2m^2}(\tilde J_m^\kappa - \tilde J_0^\kappa)\Bigg],
\end{align}
to be compared with Eq.~(22) of Ref.~\cite{Reinosa:2013twa}. The new contribution mentioned above is the last term between brackets, proportional to the shifted frequency $\omega^\lambda$. Here, for simplicity,  we noted $K_\lambda^2=(K^\lambda)^2=K_\mu^\lambda K_\mu^\lambda$ (no sum over $\lambda$) and $K_\lambda^4=(K_\lambda^2)^2$ and we introduced the following scalar tadpole sum-integrals
\begin{align}\label{eq_integrals}
J^{\kappa}_m & = \int_Q G_m(Q^\kappa)\,,\\
\tilde J_{m}^\kappa&=\int_Q { \omega^\kappa} G_m(Q^\kappa)\,,
\end{align}
{ as well as} the scalar bubble sum-integral ($K+Q+L=0$)
\beq
\label{eq_integralsss}
 I_{m_1m_2}^{\kappa\tau}(K)=\int_Q G_{m_1}(Q^\kappa)G_{m_2}(L^\tau)\,.
\eeq
Clearly, $I_{m_1m_2}^{\kappa\tau}(K)=I_{m_2m_1}^{\tau\kappa}(K)$. Moreover, it follows from the identity \eqn{eq:idmom} and from $G_m(Q)=G_m(-Q)$ that $I_{m_1m_2}^{\kappa\tau}(-K)=I_{m_1m_2}^{(-\kappa)(-\tau)}(K)$. Similarly, one shows that $J_m^\kappa=J_m^{-\kappa}$ and $\tilde J_m^\kappa=-\tilde J_m^{-\kappa}$.

\begin{center}
\begin{figure}[t]
    \includegraphics[width=1\linewidth]{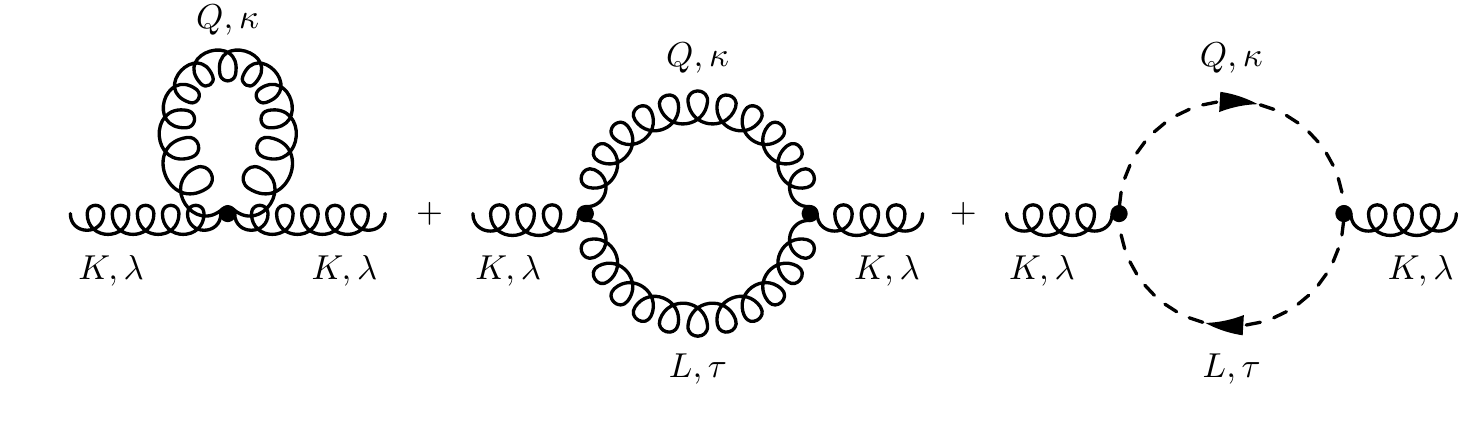}
  \caption{One-loop diagrams for the gluon self-energy.}
  \label{fig_AA}
\end{figure}
\end{center}
\subsection{Gluon self-energy}
The one-loop contributions to the gluon self-energy are shown in Fig.~\ref{fig_AA}. 
The calculation of each diagram proceeds along the same lines as for the ghost self-energy. As before, the obtained expressions are similar to the corresponding ones in the Landau gauge, with the various momenta in the internal lines replaced by appropriate generalized momenta. Again, one can follow closely those manipulations of Ref.~\cite{Reinosa:2013twa} which only relied on momentum conservation. The reduction to simple tadpole and bubble loop integrals is detailed in { Appendix~\ref{app:gl_self}.} Although the complete tensorial expression of the self-energy is shown there, it follows from { Eqs.~\eqn{eq_decompo_propag-2}--\eqn{eq_decompo_propag}} that it is sufficient to retain only those contributions which are transverse with respect to the external generalized momentum $K^\lambda$. These read, using the same notational conventions as before,
\begin{widetext}
\begin{align}
\label{eq:PLT}
\Pi_{T/L}^\lambda(K) &= \sum_{\kappa,\tau}{\cal C}_{\lambda\kappa\tau}\Bigg\{\left(1-\frac{K^4_\lambda}{2m^4}\right)\!\{I_{T/L}^\lambda\}_{00}^{\kappa\tau}(K)+\left(1+\frac{K^2_\lambda}{m^2}\right)^{\!\!2}\!\{I_{T/L}^\lambda\}_{m0}^{\kappa\tau}(K)-2\!\left[d-2+\left(1+\frac{K^2_\lambda}{2m^2}\right)^{\!\!2}\right]\!\{I_{T/L}^\lambda\}_{mm}^{\kappa\tau}(K)\nonumber\\
&+ (d-2)J_m^\kappa+\frac{K_\lambda^2+m^2}{m^2}\left(J_m^\kappa-J_0^\kappa\right)-\frac{2\omega^\lambda}{m^2}(\tilde J^\kappa_m-\tilde J_0^\kappa)+\frac{(K^2_\lambda+m^2)^2}{m^2}I_{m0}^{\kappa\tau}(K)-K^2_\lambda\left(4+\frac{K^2_\lambda}{m^2}\right)I_{mm}^{ \kappa\tau}(K)\Bigg\}.
\end{align}
\end{widetext}
We have introduced the following integrals
\begin{align} \label{eq:integrals2}
\{I_{T}^\lambda\}_{m_1m_2}^{\kappa\tau}(K) & =  \frac{P^T_{\mu \nu}(K^\lambda)\{I_{\mu\nu}\}_{m_1m_2}^{\kappa\tau}(K)}{d-2}\,,\\
 \{I_L^\lambda\}_{m_1m_2}^{\kappa\tau}(K) & =  P^L_{\mu \nu}(K^\lambda)\{I_{\mu\nu}\}_{m_1m_2}^{\kappa\tau}(K)\,,  \label{eq:integrals3}
\end{align}
where
\beq
\label{eq:hehehe}
\{I_{\mu\nu}\}_{m_1m_2}^{\kappa\tau}(K)\equiv\int_Q { Q_\mu^\kappa Q_\nu^\kappa} G_{m_1}(Q^\kappa)G_{m_2}(L^\tau)\,,
\eeq
with $\smash{K+Q+L=0}$. Using similar arguments as before, one easily shows that $\{I^\lambda_{T/L}\}_{m_1m_2}^{\kappa\tau}(K)=\{I^\lambda_{T/L}\}_{m_2m_1}^{\tau\kappa}(K)=\{I^\lambda_{T/L}\}_{m_1m_2}^{(-\kappa)(-\tau)}(-K)$. Equation \eqn{eq:PLT} is to be compared to  Eq.~(28) of Ref.~\cite{Reinosa:2013twa} in the Landau gauge.\footnote{To this purpose, notice the identity $m^2 I_{m0}^{\kappa(-\kappa)}(0)=J_0^\kappa-J_m^\kappa$.} As for the ghost self-energy discussed above, only the third contribution on the second line, proportional to the external shifted frequency $\omega^\lambda$, is structurally new in the LDW gauge. The sum-integrals \eqn{eq_integrals}--\eqn{eq_integralsss} and \eqn{eq:integrals2}--\eqn{eq:hehehe} { are evaluated in Appendix~\ref{app:sum_int}.}

\subsection{Renormalization}
\label{sec_renormalization}

We introduce renormalized parameters and fields, related to the corresponding bare quantities in the usual way: 
\beq
{ m_0^2=Z_{m^2}m^2\,,\quad g_0=Z_g g\,,}
\eeq
and
\beq
\label{eq:renormfactors}
\begin{tabular}{ccccccc}
 ${\bar A}$&$=$& $\sqrt{Z_{\bar A}}\,\bar A_R$&$,$&$\quad a$&$=$& $\sqrt{Z_a}\,a_R\,,$\\
 \\
 $c$&$=$& $\sqrt{Z_c}\,c_R$&$,$&$\quad \bar c$&$=$&$ \sqrt{Z_c}\,\bar c_R\,.$
 \end{tabular}
\eeq
Notice that the background field $\bar A$ and the fluctuating field $a$ receive independent renormalizations \cite{Binosi:2013cea}. The background field gauge symmetry \eqn{eq_ginv} implies that the product $g_0\bar A$ is finite \cite{Weinberg:1996kr}. Imposing the renormalization condition 
\beq
\label{eq:condgA}
 Z_g\sqrt{Z_{\bar A}}=1
\eeq
for the finite parts as well, we have { $g_0\bar A= g\bar A_R$.}

We define the renormalized self-energies $ \Sigma^{\lambda}_R (K)$ and $\Pi^\lambda_{R,T/L}(K)$ from the renormalized propagators 
\begin{align}
 {\cal G}^\lambda_{R}(K)&=Z_c^{-1}{\cal G}^\lambda(K)\,,\\
 {\cal G}^\lambda_{R,T/L}(K)&=Z_A^{-1}{\cal G}^\lambda_{T/L}(K)\,,
\end{align} as in Eqs.~\eqn{eq:lefantome} and \eqn{eq_decompo_propag} by simply replacing { $m^2_0\to m^2$ and $g_0\to g$ (we already anticipated this replacement in the loop expressions evaluated above).}
On very general grounds, one can separate a thermal and a vacuum contribution as 
\beq
\label{eq:notation}
  \Sigma^{\lambda}_R (K)= \Sigma^{\lambda,{\rm th}}_R (K)+ \Sigma^{{\rm vac}}_R (K^2_\lambda),
 \eeq
 and similarly for $\Pi^\lambda_{R,T/L}(K)$, with $\Pi^{\rm vac}_{R,T/L}(K^2)=\Pi^{\rm vac}_{R}(K^2)$ by Lorentz symmetry. The vacuum parts are defined as the $T=0$ contributions at fixed background field $\bar A_R$. The background-field  gauge symmetry { (\ref{eq_ginv})} guarantees that these only depend on $\bar A_R$ through the shifted momentum, as emphasized in the notation of \Eqn{eq:notation}. The functions $\Sigma^{{\rm vac}}_R (K^2)$ and $ \Pi^{\rm vac}_{R}(K^2)$ are thus nothing but the renormalized self-energies in the (massive) Landau gauge and one can write 
\begin{align}
\label{eq:renghost}
 {\cal G}^\lambda_{R}(K)&=\frac{1}{K_\lambda^2F^{-1}_{R,{\rm vac}}(K^2_\lambda)+{ g^2\Sigma^{\lambda,{\rm th}}_{R}(K)}}\,,\\
\label{eq:rengluon}
 {\cal G}^\lambda_{R,T/L}(K)&=\frac{1}{{\cal G}^{-1}_{R,{\rm vac}}(K^2_\lambda)+{ g^2\Pi^{\lambda,{\rm th}}_{R,T/L}(K)}}\,,
\end{align}
where $F_{R,{\rm vac}}(K^2)$ and ${\cal G}_{R,{\rm vac}}(K^2)$ are, respectively,  the zero-temperature renormalized ghost dressing function and gluon propagator in the Landau gauge. These have been computed at one-loop order in the Curci Ferrari model (i.e., the Landau gauge limit of the present model) for the groups SU($N$) in Ref.~\cite{Tissier:2010ts} using the set of renormalization conditions 
\beq\label{eq:ren_cond}
 \Sigma_R^{\rm vac}(K^2=\mu^2)=\Pi_R^{\rm vac}(K^2=0)=\Pi_R^{\rm vac}(K^2=\mu^2)=0,
\eeq
where $\mu$ is the renormalization scale. In principle, one could implement temperature and/or background-field dependent renormalization conditions. For simplicity, we use the set of renormalization conditions \eqn{eq:ren_cond}. The one-loop vacuum ghost dressing function and gluon propagator in Eqs.~\eqn{eq:renghost} and \eqn{eq:rengluon} can thus be taken from Eqs.~($17$) of Ref.~\cite{Tissier:2010ts}, which we recall here for completeness,
\begin{align}
\label{eq:vacpieces1}
 F_{R,{\rm vac}}^{-1}(K^2)&=1+\frac{g^2 N}{64 \pi^2} \Big[f(K^2/m^2)-f(\mu^2/m^2)\Big]
\end{align}
and
\begin{align}
\label{eq:vacpieces2}
 G_{R,{\rm vac}}^{-1}(K^2)&=K^2+m^2\nn
 &+\frac{g^2 NK^2}{384 \pi^2} \Big[g(K^2/m^2)-g(\mu^2/m^2)\Big]\!,
\end{align}
where 
\beq
f(s)=-s \log s +(s+1)^3 s^{-2} \log(s+1)- s^{-1}
\eeq
and 
\begin{align}
g(s)&=111s^{-1}-2 s^{-2}+(2-s^2)\log s\nn
&+2(s^{-1}+1)^3\!\left(s^2-10 s+1\right) \log(1+s)\nonumber\\
&+(4 s^{-1} +1)^{3/2}\!\left(s^2-20 s+12\right)\log \!\left(\!\frac{\sqrt{4+s}-\sqrt{s}}{\sqrt{4
   +s}+\sqrt{s}}\right)\!.
\end{align}
The vacuum expressions for more general groups are obtained after replacing $N$ in Eqs.~(\ref{eq:vacpieces1}) and \eqn{eq:vacpieces2} by the Casimir of the corresponding adjoint representation. In general, a fifth prescription is needed for the coupling renormalization factor $Z_g$. One could, for instance, use a background field generalization of the Taylor scheme \cite{Taylor:1971ff} often used in the Landau gauge and its massive extension \cite{Tissier:2010ts}. However, this is not needed at the order of approximation considered here. In the following, we simply set $Z_g\to1$ in the one-loop expressions.

\section{Results for the SU($2$) theory}
\label{sec_perturbativ_results}

In this section, we apply the above results to the SU($2$) group. The background field has only one component, denoted $r$, and  the nonzero structure constants are given by the permutations of the Levi-Civita tensor $\epsilon^{0+-}=1$. At the order of approximation considered here, we minimize the two-loop potential of Ref.~\cite{Reinosa:2014zta}. The parameters used in that reference, namely $g=7.5$, $m=0.68$~GeV, and $\mu=1$~GeV, were taken from fits of the lattice ghost and gluon propagators in the Landau gauge at zero temperature \cite{Tissier:2010ts}. This choice was motivated by the fact that, since the background $\bar A\propto T$, the LDW gauge reduces to the Landau gauge at $T=0$; see also \cite{Reinosa:2013twa}. However, in the present case, this set of parameters leads to unphysical features for temperatures around $T_c$: The susceptibility of neutral color modes, defined below, turns negative, which yields a pole at nonvanishing momentum in the Euclidean propagator at zero Matsubara frequency. As discussed below, this is a consequence of the too large value of the coupling.

Here, it is worth emphasizing that, although it makes sense to fit the zero temperature propagators of the LDW gauge against the lattice data in the Landau gauge, there is {\it a priori} no reason to expect the parameters not to vary with temperature. This could arise, for instance, from renormalization group effects or from our assumption that the present massive model effectively accounts for Gribov ambiguities, which do depend on the Euclidean spacetime volume and thus on the temperature. As a matter of fact, fitting the one-loop propagators against lattice data in the Landau gauge at finite temperature \cite{Reinosa:2013twa} indeed reveals that the best value for the coupling decreases from $g\approx7$ at $T=0$ to about $g\approx5$ close to $T_c$. At present, we have no way to predict the possible temperature dependence of our parameters and there exists no lattice data in the present LDW gauge. As a rough guide, we shall use the value $g(\mu)=5$ at $\mu=1$ GeV obtained for temperatures in the vicinity of $T_c$ in Ref.~\cite{Reinosa:2013twa} in the Landau gauge. We adjust the mass parameter accordingly to the value $m(\mu)=0.75$ GeV such that the transition temperature remains fixed to $T_c^{\rm 2loop}=0.285$ GeV, obtained from the minimization of the background effective potential at two-loop order \cite{Reinosa:2014zta}. 

 \begin{figure}[ht]
  \centering
  \includegraphics[width=0.9\linewidth]{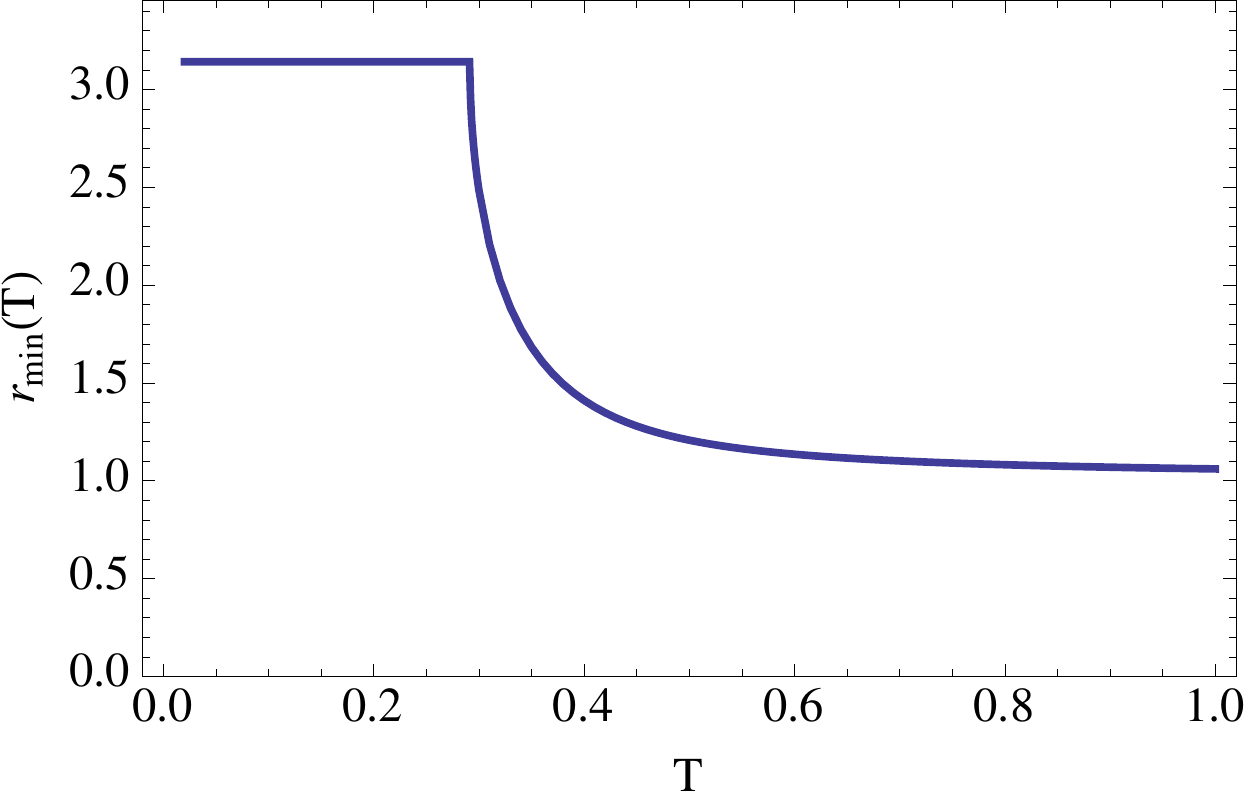}\\
  \vglue2mm
  \includegraphics[width=0.9\linewidth]{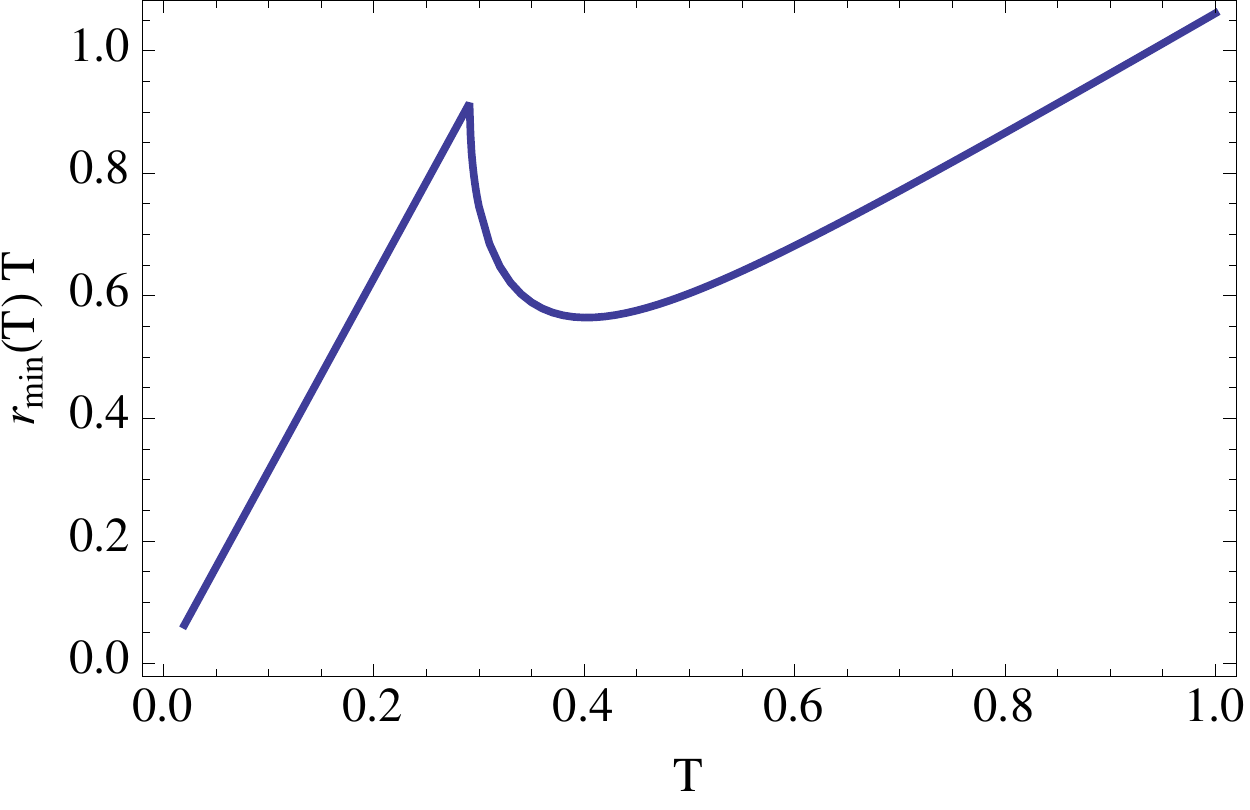}
   \caption{The physical background $r_{\rm min}(T)$ for the SU($2$) theory as a function of the temperature, obtained from the minimization of the potential \eqn{eq:potential} at two-loop order (top panel). For $\smash{T< T_c}$, the minimum sits at the $\mathds{Z}_2$-symmetric point $r=\pi$. The symmetry is spontaneously broken for $T>T_c$ and the transition is continuous. The bottom plot shows the  dimensionful background $r_{\rm min}(T)T\propto\bar A_{\rm min}(T)$.}
  \label{fig_rT}
\end{figure} 

\begin{figure}[ht]
  \centering
  \includegraphics[width=0.9\linewidth]{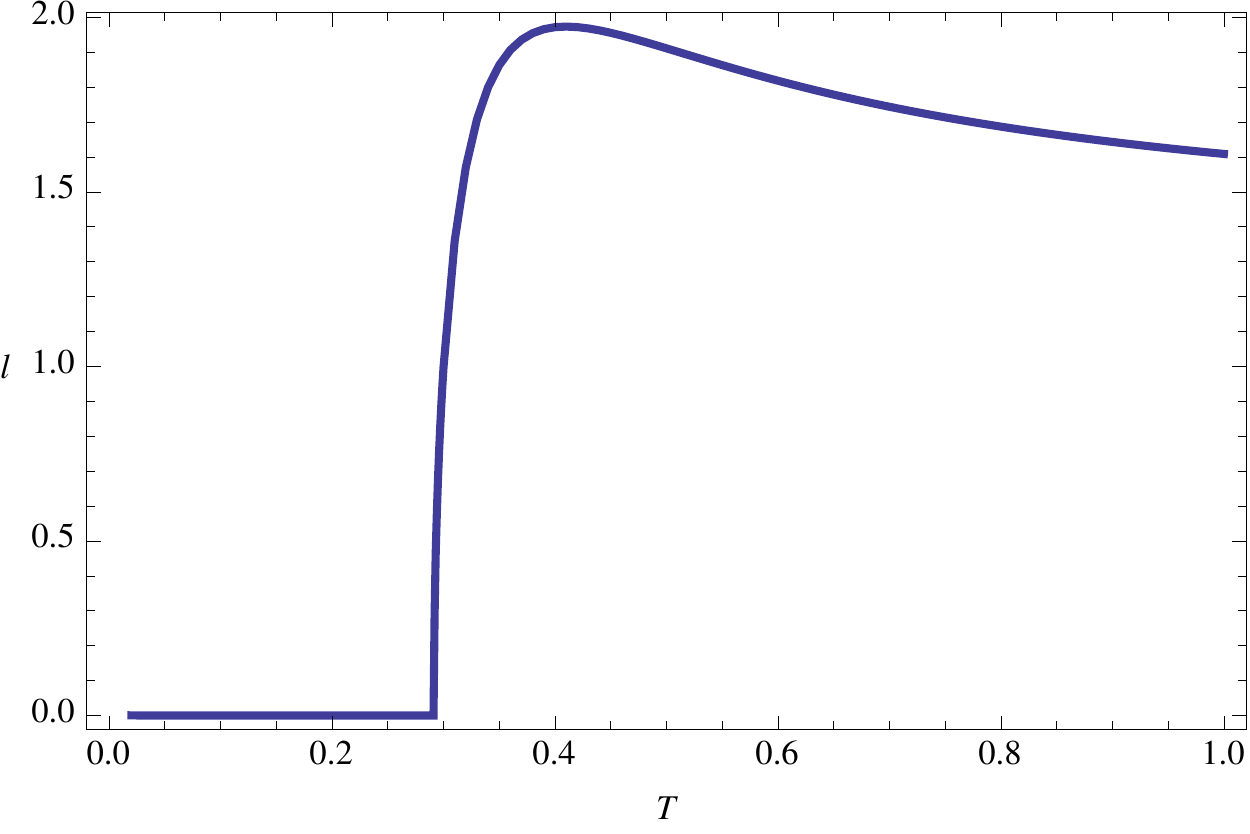}
   \caption{The gauge-invariant Polyakov loop \eqn{eq:decadix} at next-to-leading order as a function of the temperature with the parameters of \Fig{fig_rT}.}
  \label{fig_ploop}
\end{figure}

The background field $r=r_{\rm min}(T)$ that minimizes the two-loop potential of Ref.~\cite{Reinosa:2014zta} is shown in \Fig{fig_rT} as a function of the temperature. It presents the characteristic cusp at the critical temperature of the second order transition of the $SU(2)$ theory. We also show the behavior of the dimensionful background $rT$, that is, the effective frequency shift for charged color modes, which contributes a term $(rT)^2$ to the tree-level square mass of the corresponding propagators at zero frequency; see Eqs.~\eqn{eq:lefantome} and \eqn{eq_decompo_propag}. { For completeness, we also show the behavior of the gauge-invariant order parameter, the Polyakov loop \eqn{eq:decadix}, computed at next-to-leading order, in \Fig{fig_ploop}; see Ref.~\cite{Reinosa:2014zta} for details.}

\subsection{Gluon susceptibilities}

{ Before discussing the complete momentum dependence of the various propagators, we consider the electric and magnetic gluon susceptibilities of the neutral gluon mode, which have simple expressions and which already exhibit the most salient features of the influence of the background field on the correlation functions. Such susceptibilities describe the response of the system to a static homogeneous source linearly coupled to the neutral gluon mode and are respectively given by the longitudinal and transverse gluon propagator at vanishing frequency and momentum:
\begin{align}
  \chi^0_{L/T}=G^{0}_{L/T}(\omega=0,k\to0),
\end{align}
where $k=|{\bf k}|$. As discussed below, these quantities can be given relatively simple expressions at one-loop order because they involve loop integrals with vanishing external frequency and momentum. Moreover, their temperature dependence entirely comes from thermal loop contributions and they thus provide a direct measure of the nontrivial structure of the propagator beyond tree level. 

By analogy, we also consider the charged gluon propagators at vanishing shifted frequency
\begin{align}
  \chi^\lambda_{L/T}=G^{\lambda}_{L/T}(\omega=-\lambda rT,k\to0),
\end{align}
where the right-hand side is to be understood as the renormalized propagators \eqn{eq:rengluon} properly continued to arbitrary (i.e., non Matsubara) Euclidean frequencies.\footnote{The analytic continuation has to be understood after the Matsubara sums have been performed and the external Matsubara frequency has been removed from the thermal factors by means of the identity $n_{\varepsilon+i\omega_n}=n_\varepsilon$.}

Although these cannot be directly interpreted as susceptibilities because they involve a nonzero Euclidean frequency, they are interesting because, by definition, their temperature dependence entirely comes from thermal loop effects. They thus provide a new source for comparison between various continuum approaches. Besides, just as for the neutral mode, their one-loop expressions involve diagrams with vanishing external frequency and momentum and can thus be given simple expressions; see below. Finally, because such quantities involve an Euclidean frequency, they may be computed in lattice simulations \cite{Pawlowski:2016eck}.\footnote{The above continuation is also natural in view of the interpretation of the background as an imaginary chemical potential for the color charge(s) measured by the generators of the Cartan subalgebra, see above. In fact, in the presence of a real chemical potential associated to a charge $Q$, similar continuations allow to access the response functions for Heisenberg fields evolving according to the Hamiltonian $H$ of the system, which are the physical response functions, from the response functions associated to Heisenberg fields evolving according to the shifted Hamiltonian $H-\mu Q$, which arise more naturally within the functional integral formulation.} 

In what follows we shall use a more standard terminology and refer to the Debye and magnetic  screening square masses, respectively defined as $M^2_{{\rm D},\lambda}\equiv 1/\chi^\lambda_L$ and $M^2_{{\rm mag},\lambda}\equiv 1/\chi^\lambda_T$. We thus have 
\begin{align}
\label{eq:electric_masses}
M_{{\rm D},\lambda}^2(T)&= m^2+g^2 \Pi^{\lambda,{\rm th}}_{L}(\omega=-\lambda rT, k\to0)\,,\\
\label{eq:electric_masses2}
M_{{\rm mag},\lambda}^2(T)&= m^2+g^2 \Pi^{\lambda,{\rm th}}_{T}(\omega=-\lambda rT, k\to0)\,.
\end{align}
}

\subsubsection{Neutral mode}\label{sec:sn}

As emphasized above, the square masses \eqn{eq:electric_masses} and \eqn{eq:electric_masses2} involve sum-integrals with vanishing external momentum and frequency, which, at one-loop order, can be written in terms of simple tadpolelike sum-integrals. The relevant calculations are detailed in Appendix~\ref{app:gluon_su}. We get
\beq\label{eq:longzero}
\Pi_L^{0,{\rm th}}(0,k\to0)\,\,\hat{=}\,\,12m^2 S_m^++12 J_m^+-J_0^++2\frac{N_m^+-N_0^+}{m^2}
\eeq
and
\beq\label{eq:transzero}
\Pi_T^{0,{\rm th}}(0,k\to0)\,\,\hat{=}\,\,J_0^+-2J_m^++\frac{2}{3}\frac{N_0^+-N_m^+}{m^2}\,,
\eeq
where we introduced the integrals
\beq
\label{eq:Jm}
J_{m}^\lambda=\int_Q G_{m}(Q^\lambda)\,\hat{=}\,\frac{1}{2\pi^2}\int_0^\infty\!\! dq \frac{q^2}{\varepsilon_{m,q}} {\rm Re}\,n_{\varepsilon_{m,q}-i\lambda rT},
\eeq
\beq
N_{m}^\lambda=\int_Q q^2 G_{m}(Q^\lambda)\,\hat{=}\,\frac{1}{2\pi^2}\int_0^\infty \!\!dq \frac{q^4}{\varepsilon_{m,q}} {\rm Re}\,n_{\varepsilon_{m,q}-i\lambda rT},
\eeq
and
\beq
S_{m}^\lambda=\int_Q G^2_{m}(Q^\lambda)\,\hat{=}\,\frac{1}{4\pi^2}\int_0^\infty \!\!dq \frac{1}{\varepsilon_{m,q}} {\rm Re}\,n_{\varepsilon_{m,q}-i \lambda rT}\,.
\eeq
Here, the symbol $\hat{=}$ means that we discard the vacuum ($\smash{T=0}$) contributions since they are not needed in (\ref{eq:electric_masses}) and (\ref{eq:electric_masses2}). We have also defined $\varepsilon_{m,q}=\sqrt{q^2+m^2}$ and $n_z=(e^{\beta z}-1)^{-1}$ is the Bose-Einstein function. We have, explicitly, for $\varepsilon,r\in\mathds{R}$,
\beq
  {\rm Re}\,n_{\varepsilon-irT}=\frac{e^{\beta\varepsilon}\cos r-1}{e^{2\beta\varepsilon}-2e^{\beta\varepsilon}\cos r+1}.
\eeq
\vglue2mm
The massless integrals $J_0^+$ and $N_0^+$ can be determined analytically. We have $J^\lambda_0/T^2=P_2(\lambda r)/(2\pi^2)$ and $N^\lambda_0/T^4=P_4(\lambda r)/(2\pi^2)$, where $P_{2n+2}(z)={\rm Re}\,\int_0^\infty dx\,x^{2n+1}/[\exp (x-iz)-1]$. In the interval $z\in[0,2\pi]$, these integrals can be expressed as simple polynoms; see, e.g., Ref.~\cite{Reinosa:2014zta}. For $r\in[0,2\pi]$, we get
\begin{align}
\frac{J_0^\lambda}{T^2} & = \frac{1}{8}\left[\left(1-\frac{\lambda r}{\pi}\right)^2-\frac{1}{3}\right]\!,\\
\frac{N_0^\lambda}{T^4} & = -\frac{\pi^2}{16}\left[\left(1-\frac{\lambda r}{\pi}\right)^4-2\left(1-\frac{\lambda r}{\pi}\right)^2+\frac{7}{15}\right]\!.
\end{align}
The electric and magnetic screening square masses in the neutral sector thus read, explicitly,
\begin{align}
 \label{eq:AAA}
 M_{{\rm D},0}^2(T)&=m^2-\frac{g^2T^2}{8}\bigg\{\left(1-\frac{r}{\pi}\right)^2-\frac{1}{3}\nn
 &-\frac{\pi^2T^2}{m^2}\left[\left(1-\frac{r}{\pi}\right)^4-2\left(1-\frac{r}{\pi}\right)^2+\frac{7}{15}\right]\bigg\}\nn
 &+{g^2m^2\over\pi^2}\int_0^\infty\!\!dq\,{{\rm Re}\,n_{\varepsilon_{m,q}-irT}\over\varepsilon_{m,q}}\left(3+6{q^2\over m^2}+{q^4\over m^4}\right)
\end{align}
and
\begin{align}
  \label{eq_mag_mass}
  M_{\text {mag},0}^2(T)&=m^2+\frac{g^2T^2}{8}\bigg\{\left(1-\frac{r}{\pi}\right)^2-\frac{1}{3}\nn
  &-\frac{\pi^2T^2}{3m^2}\left[\left(1-\frac{r}{\pi}\right)^4-2\left(1-\frac{r}{\pi}\right)^2+\frac{7}{15}\right]\bigg\}\nn
 &-{g^2m^2\over\pi^2}\int_0^\infty\!\!dq\,{{\rm Re}\,n_{\varepsilon_{m,q}-irT}\over\varepsilon_{m,q}}\left({q^2\over m^2}+{q^4\over 3m^4}\right).
\end{align}
Here, it is understood that $r$ is to be taken at the minimum $r_{\rm min}(T)$ of the background potential. It is interesting to compare these expressions with those obtained in the Landau gauge; see Eqs.~(31) and (35) of Ref.~\cite{Reinosa:2013twa} with $N=2$, which we recover by evaluating Eqs.~(\ref{eq:AAA}) and (\ref{eq_mag_mass}) for $r=0$.
 \begin{figure}[t]
  \centering
  \includegraphics[width=0.9\linewidth]{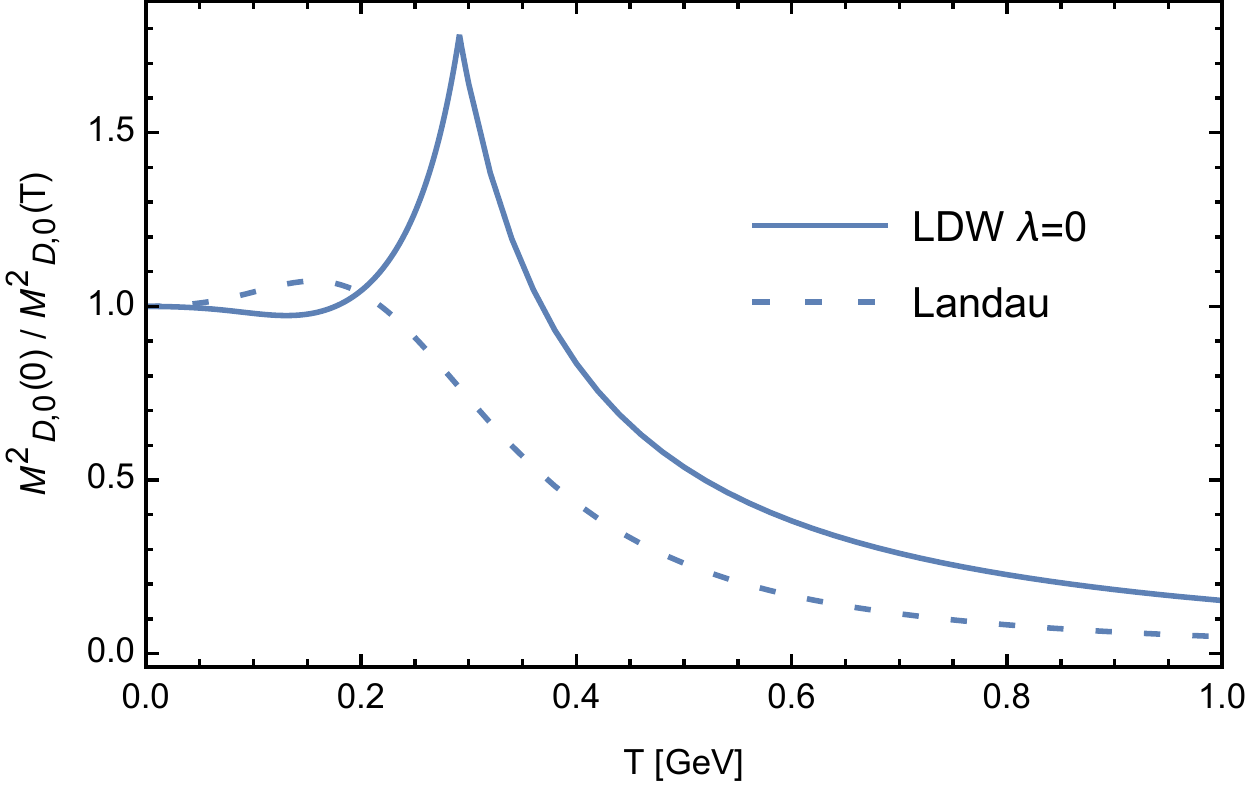}\\
  \vglue4mm
  \includegraphics[width=0.9\linewidth]{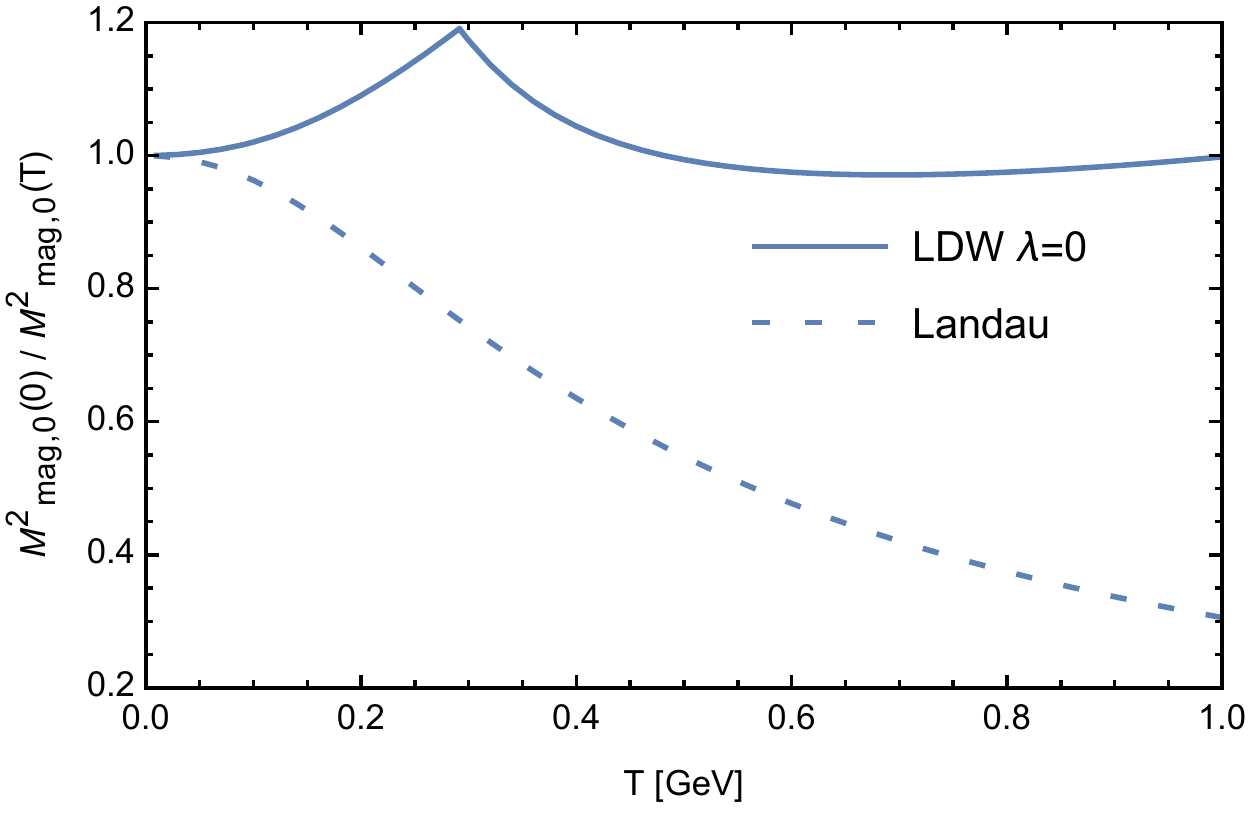}
   \caption{Temperature dependence of the electric and magnetic inverse square masses in the neutral mode normalized to their common zero temperature value. To emphasize the effect of the Polyakov loop, we compare with the corresponding results at vanishing background field ($r=0$), which corresponds to the Landau gauge.}
  \label{fig_k0masses}
\end{figure}

The temperature dependence of the corresponding inverse screening square masses (susceptibilities) across the phase transition is shown in \Fig{fig_k0masses}. We also show the corresponding results in the Landau gauge ($r=0$), to quantify the effect of the nontrivial background. We observe that the magnetic susceptibility is monotonously { increasing} with the temperature below $T_c$, whereas the electric one first slightly decreases at low temperatures and then increases to its maximum value at $T=T_c$. Both present a clear cusp at the transition. The electric square mass rapidly approaches a quadratic behavior $M_{{\rm D},0}^2\propto T^2$, whereas the magnetic mass remains essentially bounded in the range of temperature considered here.

 The cusp reflects the nonanalytic behavior of the order parameter $r_{\rm min}(T)$ across the transition and is in sharp contrast with the corresponding perturbative results in the Landau gauge. The electric susceptibility in the Landau gauge showed a slight nonmonotonous behavior below $T_c$, but no cusp.  As for the magnetic susceptibility, both the present perturbative approach and gauge-fixed lattice simulations show a smooth monotonous behavior in the Landau gauge, with a rapid decrease above $T_c$. This contrasts with the present results where  the magnetic susceptibility is essentially constant above the transition; see below. 

The main characteristics of the results described here can be understood in more detail as follows. First, in the low temperature, symmetric phase, the minimum of the potential is at the center symmetric value $\smash{r_{\rm min}(T\le T_c)=\pi}$. The distribution function of charged modes becomes a negative Fermi-Dirac distribution 
\beq
  {\rm Re}\,n_{\varepsilon-i\pi T}=-\frac{1}{e^{\beta \varepsilon}+1}\equiv -f_{\varepsilon}\,.
\eeq
For $T\le T_c\approx 0.38 m$, one can write $f_\varepsilon= \exp(-\beta\varepsilon)+{\cal O}(e^{-2\beta \varepsilon})$. The momentum integrals can then be expressed in terms of Bessel functions:
\begin{align}
 \label{eq:AAAlowT}
 &M_{{\rm D},0}^2(T\le T_c)\approx m^2+\frac{g^2T^2}{24}\left(1+{7\pi^2\over5}{T^2\over m^2}\right)\nn
 &-{3g^2m^2\over\pi^2}\left[K_0\left(\frac{m}{T}\right)+\frac{2T}{m}K_1\left(\frac{m}{T}\right)+\frac{T^2}{m^2}K_2\left(\frac{m}{T}\right)\right],
\end{align}
and
\begin{align}
  \label{eq_mag_masslowT}
  M_{\text {mag},0}^2(T\le T_c)&\approx m^2-\frac{g^2T^2}{24}\left(1+\frac{7\pi^2} {15}{T^2\over m^2}\right)\nn
   &+\frac {g^2mT}{\pi^2}\left[K_1\left(\frac{m}{T}\right)+\frac{T}{m}K_2\left(\frac{m}{T}\right)\right],
\end{align}
up to corrections $\propto \exp(-2m/T)$.

The high temperature regime ($T/m\gg1$) of the masses \eqn{eq:AAA} and \eqn{eq_mag_mass} is studied in detail in Appendix~\ref{appsec:highT}. Here, we summarize the leading behaviors:
\begin{align}
\label{eq:asympt_mass_el}
M_{{\rm D},0}^2(T) &=\frac{2g^2 T^2}{3} \left(1-\frac{3r}{\pi}+\frac{3r^2}{2\pi^2}\right)+{\cal O}(m^2)\,, \\ 
\label{eq:asympt_mass_mag}
M_{{\rm mag},0}^2 (T)&= m^2 \left[1+\frac{3 g^2}{16 \pi r} + {\cal O}\left(g^2\ln\frac{T}{m} \right)\right] ,
\end{align}
where the second line assumes $m\ll rT$ and is, thus, only valid for nonzero background. The electric mass is dominated by thermal effects which give rise to the standard behavior $M_{{\rm D},0}\propto gT$, where the numerical prefactor depends on the value of the background. At sufficiently large temperatures the background approches a nonzero value $r_\infty$ whose next-to-leading order (two-loop) expression is recalled in Appendix~\ref{appsec:highT}; see \Eqn{appeq:rinf}. In the weak coupling limit, it reads $r_\infty= 3 g^2/(8 \pi)+{\cal O}(g^4)$ and one recovers the standard expression of the SU($N$) Debye mass $M_{{\rm D}}^2= Ng^2T^2/3$, as in the Landau gauge \cite{Reinosa:2013twa}. 

In contrast, the magnetic sector is dramatically different as compared to the Landau gauge result. In presence of a nonzero background, the thermal corrections remain bounded, as discussed in Appendix~\ref{appsec:highT}. The high-temperature expression \eqn{eq:asympt_mass_mag} is valid in the regime $r/\pi \ll 1/\ln(T/m)$, which is satisfied, e.g. in the weak coupling regime; see the discussion in Appendix \ref{appsec:highT}. Using, as above, the two-loop asymptotic value of $r_\infty$, we get $M_{{\rm mag},0}^2=3m^2/2$. We show in Appendix~\ref{appsec:highT} that the magnetic mass remains bounded from above for a wide range of parameters. This is to be compared with the result in the absence of background  $M_{{\rm mag},0}^2|_{r=0}\propto mT$, which grows unbounded at large temperature \cite{Reinosa:2013twa}. As discussed later in this paper, this has important consequences in the ghost sector.

\subsubsection{Charged modes}

\begin{figure}[t]
  \centering
  \includegraphics[width=0.9\linewidth]{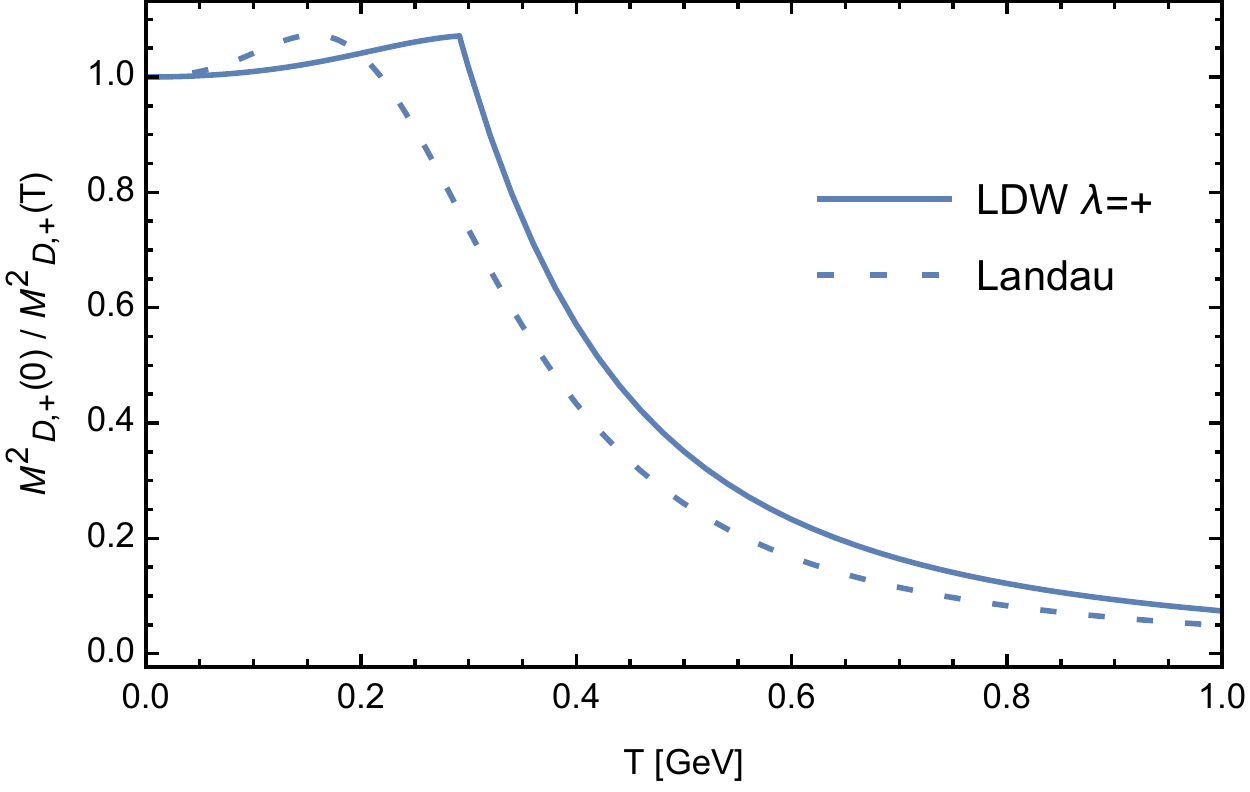}\\
  \vglue4mm
  \includegraphics[width=0.9\linewidth]{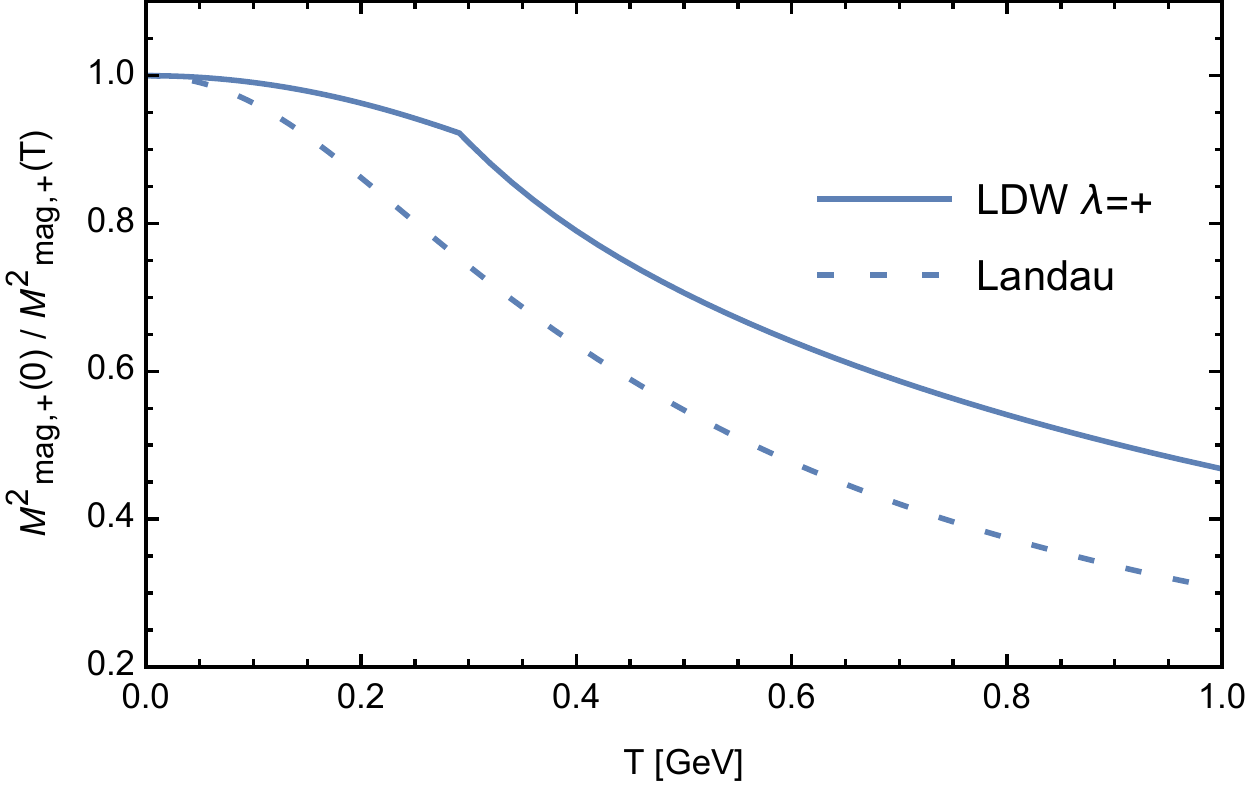}
   \caption{Same as \Fig{fig_k0masses} for the charged gluon modes, where the corresponding square masses are defined in Eqs.~\eqn{eq:electric_masses} and \eqn{eq:electric_masses2}}
  \label{fig_kpmasses}
\end{figure}

As emphasized before, the one-loop self-energies for charged modes at vanishing shifted frequency $\omega^\lambda=\omega+\lambda rT=0$ also have simple expressions; see Appendix \ref{app:gluon_su}. We obtain, after similar calculations as in the neutral case,
\begin{widetext}
\begin{align}
 \label{eq:AAAch}
 M_{{\rm D},\pm}^2(T)&=m^2-\frac{g^2T^2}{16}\bigg\{\left(1-\frac{r}{\pi}\right)^2+\frac{1}{3}-\frac{\pi^2T^2}{m^2}\left[\left(1-\frac{r}{\pi}\right)^4-2\left(1-\frac{r}{\pi}\right)^2-\frac{1}{15}\right]\bigg\}\nn
 &+{g^2m^2\over2\pi^2}\int_0^\infty\!\!dq\,\frac{n_{\varepsilon_{m,q}}+{\rm Re}\,n_{\varepsilon_{m,q}-irT}}{\varepsilon_{m,q}}\left(3+6{q^2\over m^2}+{q^4\over m^4}\right)
\end{align}
and
\begin{align}
  \label{eq_mag_massch}
  M_{{\rm mag},\pm}^2(T)&=m^2+\frac{g^2T^2}{16}\bigg\{\left(1-\frac{r}{\pi}\right)^2+\frac{1}{3}-\frac{\pi^2T^2}{3m^2}\left[\left(1-\frac{r}{\pi}\right)^4-2\left(1-\frac{r}{\pi}\right)^2-\frac{1}{15}\right]\bigg\}\nn
 &-{g^2m^2\over2\pi^2}\int_0^\infty\!\!dq\,\frac{n_{\varepsilon_{m,q}}+{\rm Re}\,n_{\varepsilon_{m,q}-irT}}{\varepsilon_{m,q}}\left({q^2\over m^2}+{q^4\over 3m^4}\right).
\end{align}
\end{widetext}
Equation (\ref{eq:LT}) guarantees that $M_{{\rm D},+}^2=M_{{\rm D},-}^2$ and $M_{{\rm mag},+}^2=M_{{\rm mag},-}^2$. 
For $r=0$, we check that $M_{{\rm D},\pm}^2=M^2_{{\rm D},0}$ and $M_{{\rm mag},\pm}^2=M^2_{{\rm mag},0}$. The square masses \eqn{eq:AAAch} and \eqn{eq_mag_massch} are plotted as functions of the temperature and compared to their value in the absence of background in \Fig{fig_kpmasses}.

The various features of the curves shown in \Fig{fig_kpmasses} can be readily understood from the results of the previous section by noting that, at one-loop order,
\beq
\label{eq:half-sum}
 M_{{\rm D},\pm}^2=\frac{M_{{\rm D},0}^2+\left.M_{{\rm D},0}^2\right|_{r=0}}{2}
\eeq
and similarly for the magnetic mass.
In the symmetric phase, $r=\pi$ and, using $n_\varepsilon-f_\varepsilon=2n_{2\varepsilon}\approx 2e^{-2\varepsilon}$, we have
\begin{align}
  \label{eq_ch_mass_el}
  M_{{\rm D},\pm}^2(T\le T_c)&\approx m^2-\frac{g^2T^2}{48}\left(1+\frac{\pi^2} {5}{T^2\over m^2}\right)\!,\\
  \label{eq_ch_mass_mag}
  M_{{\rm mag},\pm}^2(T\le T_c)&\approx m^2+\frac{g^2T^2}{48}\left(1+\frac{\pi^2} {15}{T^2\over m^2}\right)\!,
\end{align}
up to corrections $\propto \exp(-2m/T)$. At large temperature we get
\begin{align}
 M_{{\rm D},\pm}^2(T)&=\frac{2g^2 T^2}{3} \left(1-\frac{3r}{2\pi}+\frac{3r^2}{4\pi^2}\right)+{\cal O}(mT),\\
 M_{{\rm mag},\pm}^2(T)&=\frac{g^2mT}{6\pi}+{\cal O}(m^2).
\end{align}

\subsection{Gluon propagators}

We now study the momentum-dependent gluon and ghost propagators at zero Matsubara frequency. We first consider the gluon sector.

\subsubsection{Neutral mode}

\begin{figure}[ht]
  \centering
  \includegraphics[width=1\linewidth]{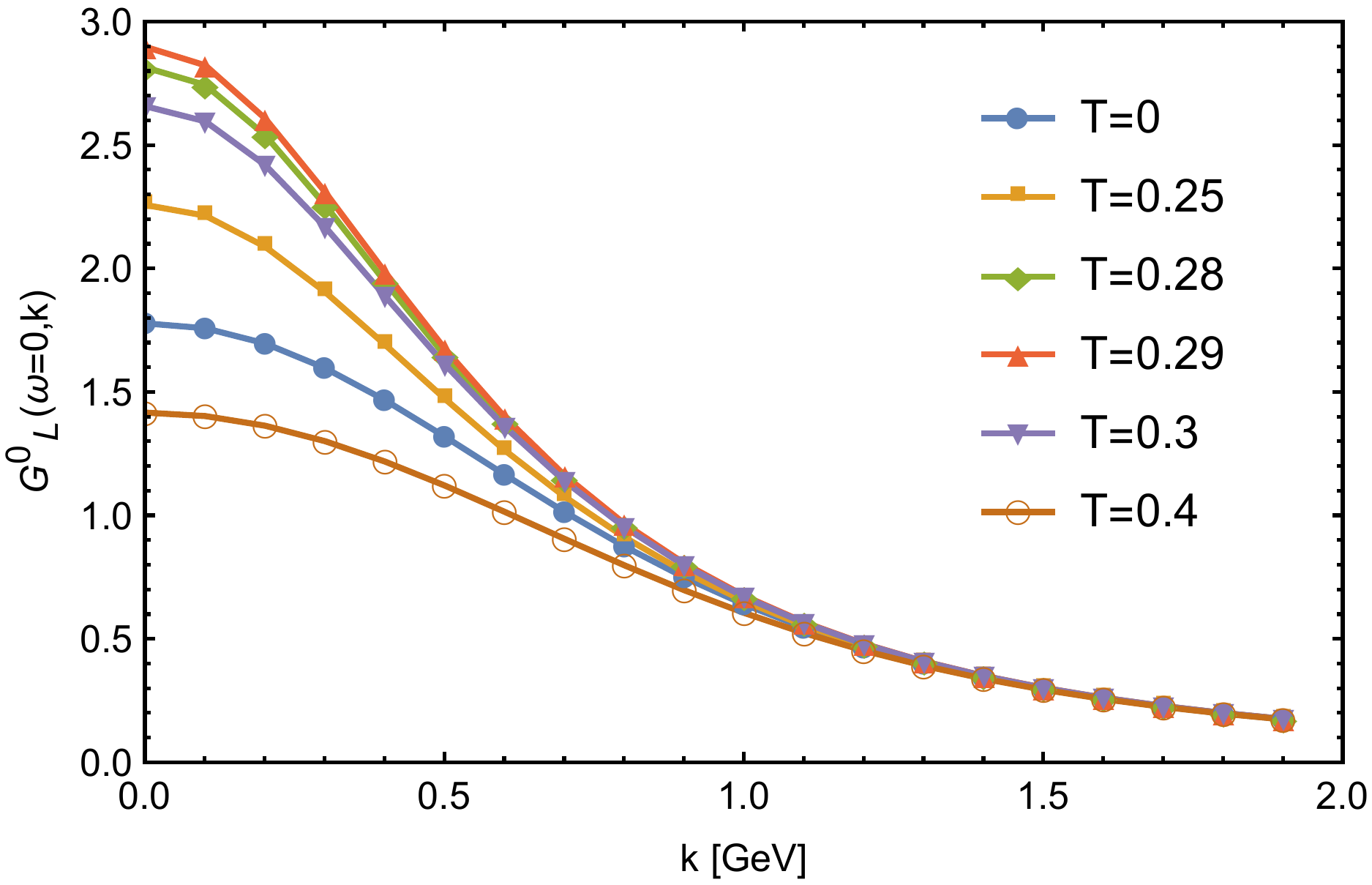}\quad
  \includegraphics[width=1\linewidth]{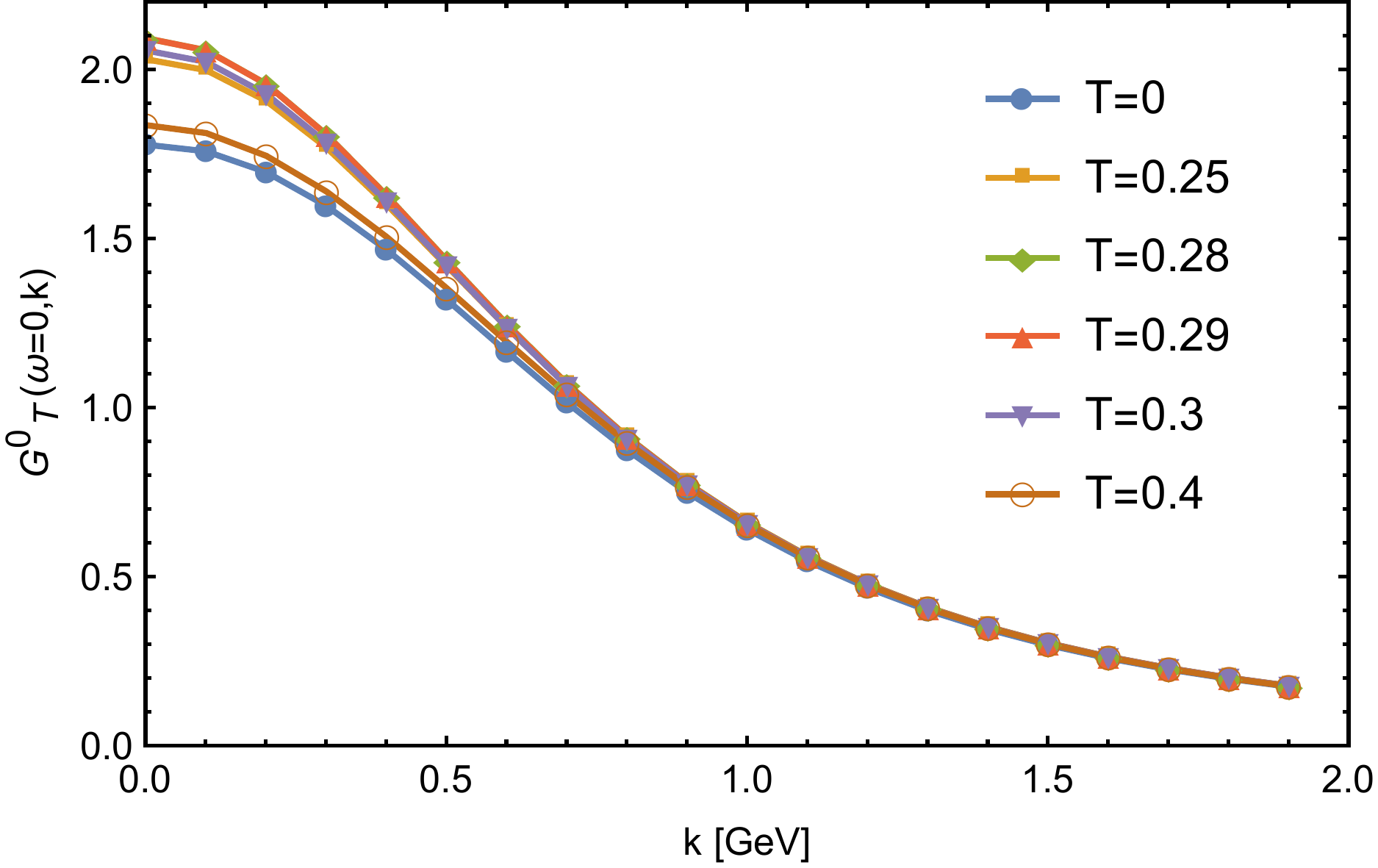}\quad
   \caption{The one-loop electric and magnetic propagators in the neutral sector at vanishing frequency as functions of the spatial momentum $k$ for various temperatures.}
  \label{fig_gluon_prop_k0}
\end{figure}

We plot the electric and magnetic propagators of the neutral color mode ${\cal G}_{L/T}^0(0,k)$ as functions of $k$ for various temperatures on \Fig{fig_gluon_prop_k0}. Both are smooth, monotonously decreasing functions of $k$. This is to be contrasted with the corresponding results in the Landau gauge, where both propagators exhibited a non monotonous behavior, more pronounced for higher temperatures, eventually resulting in { an effective} $3d$ behavior in the magnetic sector. In the present case, the main effect of the temperature can be read off the value of the propagators at vanishing momentum, respectively given by the susceptibilities $M_{{\rm D},0}^{-2}$ and $M_{{\rm mag},0}^{-2}$ and discussed in detail in the previous section. For instance, this explains the relative weak dependence of the magnetic propagator with temperature.

\subsubsection{Charged}

\begin{figure}[t]
  \centering
  \includegraphics[width=1\linewidth]{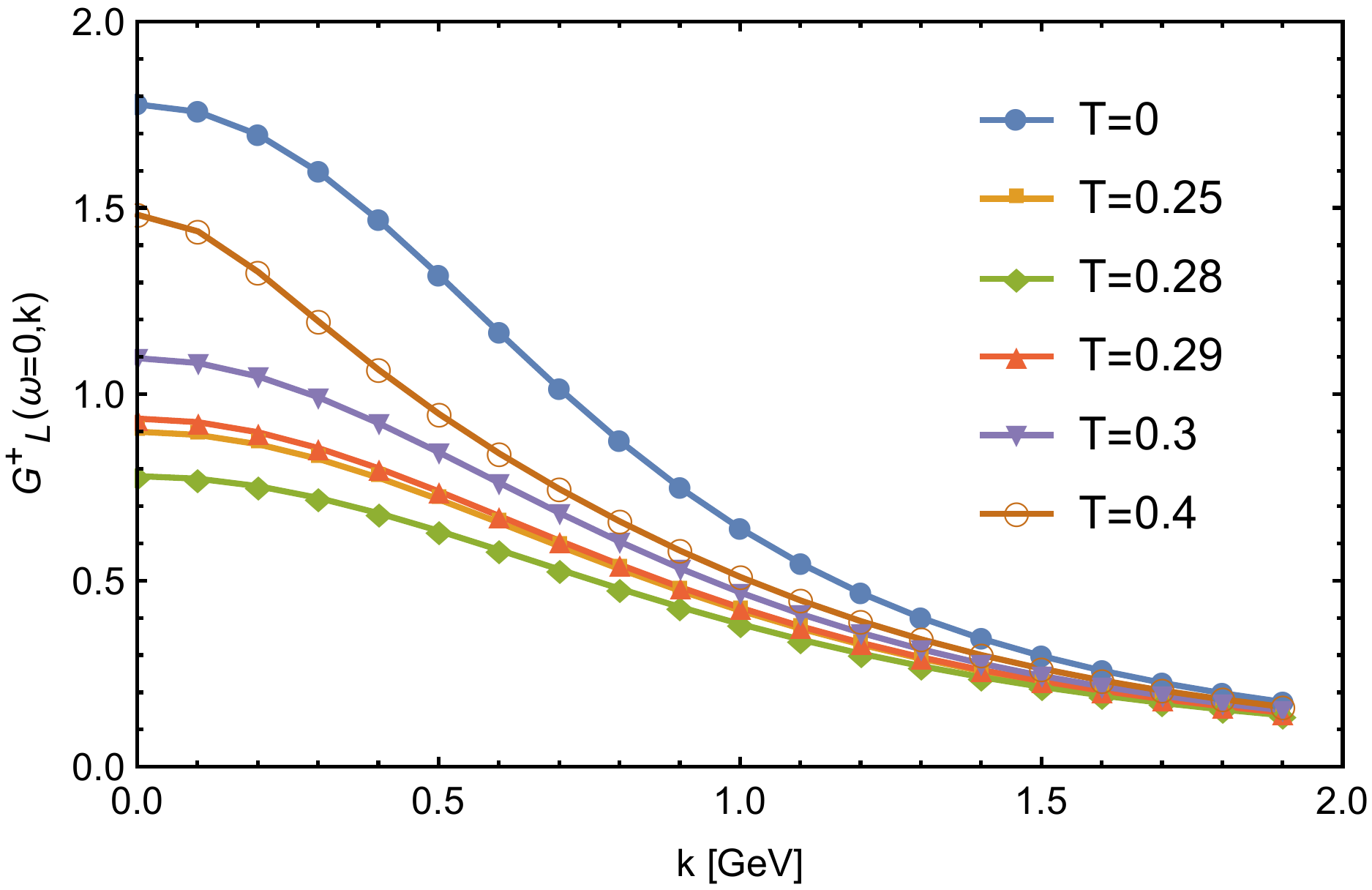}\quad
  \includegraphics[width=1\linewidth]{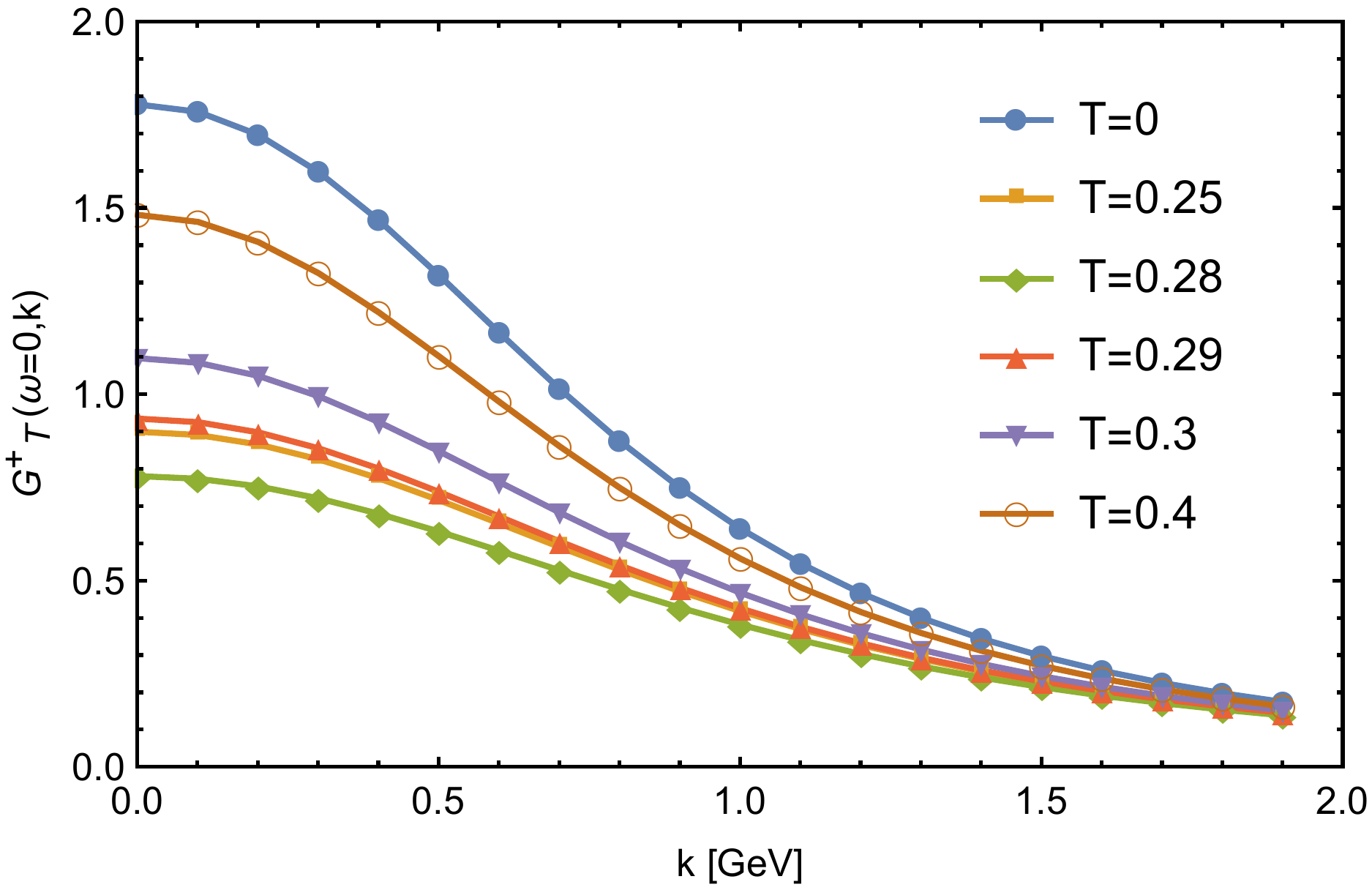}\quad
   \caption{The one-loop electric and magnetic propagators of the charged gluon at vanishing frequency as functions of the spatial momentum $k$ for various temperatures.}
  \label{fig_gluon_prop_kp}
\end{figure}

\begin{figure}[t]
  \centering
  \includegraphics[width=1\linewidth]{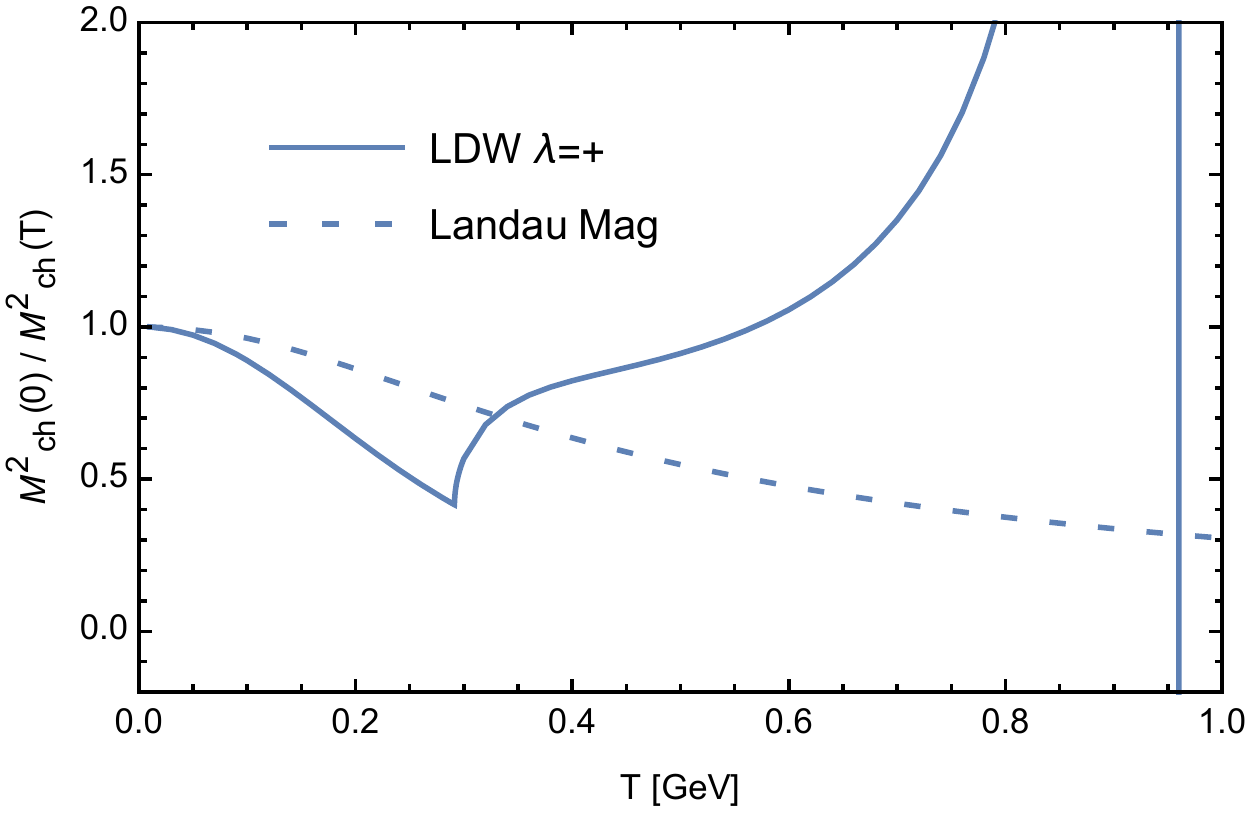}\quad\\
   \caption{Normalized inverse of the zero-momentum gluon square mass in any of the charged modes (this differs from the charged susceptibilities defined above).}
  \label{fig_kmasses}
\end{figure}

We now come to the charged sector. From the property \eqn{eq:LT}, and using spatial isotropy, we conclude that the charged propagators at vanishing frequency are degenerate:
\beq
 {\cal G}_{T/L}^+(0, k)={\cal G}_{T/L}^-(0,k).
\eeq
The electric and magnetic propagators of charged color modes  are plotted for various temperatures in \Fig{fig_gluon_prop_kp}. 
As before, it is interesting to plot the value of the propagators at vanishing momentum as a function of temperature. It is a simple exercise to show that the transverse and longitudinal components of the charged propagators are equal at vanishing momentum:\footnote{Any tensor $T_{\mu \nu}(K)$ can be decomposed in terms of four independent scalar functions as 
\begin{align*}
\label{eq_decompo}
T_{\mu \nu}(K)&= \delta_{\mu \nu} T^\lambda_\delta(K) +\left(n_\mu K^\lambda_\nu+ n_\nu K^\lambda_\mu\right)T^\lambda_{nK}(K)\nn
&+ K^\lambda_\mu K^\lambda_\nu T^\lambda_{KK}(K)+n_\mu n_\nu T^\lambda_{nn}(K),
\end{align*}
where $n_\mu = \delta_{\mu0}$ characterizes the thermal bath rest frame. Defining the transverse and longitudinal projections as in Eqs.~\eqn{eq_projT} and \eqn{eq_projL}, we have
\begin{align*}
T^\lambda_{T}(K) & =  T^\lambda_\delta(K),\\
T^\lambda_{L}(K) & =  T^\lambda_\delta(K)+ \frac{{k}^2}{(\omega+\lambda rT)^2+{k}^2}  T_{nn}^\lambda (K).
\end{align*}
For charged modes, $\lambda\neq0$, this implies
\beq
T^\pm_{L}(0, k\to0 )= T^\pm_{T}(0, k\to0 ).\nonumber
\eeq
}
\beq
\label{eq:chargedmassequality1}
{\cal G}^\pm_{L}(0, k\to0 )= {\cal G}^\pm_{T}(0, k\to0 ).
\eeq
Accordingly, we define ${\cal G}_{T/L}^\pm(0,k\to0)=1/M^2_{\rm ch}$, that is,
\beq
\label{eq:chargedmassedef}
 M_{\rm ch}^2(T)={\cal G}^{-1}_{\rm vac}\left((rT)^2\right)+g^2\Pi_{T/L}^{\pm,{\rm th}}(0,k\to0).
\eeq
One easily checks from \Eqn{eq:PLT} that $M_{\rm ch}^2$ reduces to the magnetic square mass at vanishing background (this is because $I_L^+(0,k\to0)$ converges to $I_T(0,k\to0)$ in the Landau gauge).
The temperature dependence of both square masses is shown in \Fig{fig_kmasses}. 

Again here, the presence of a cusp at the critical temperature is inherited from that of the background. Note that, in contrast with what happens with the susceptibilities, the cusp in the inverse zero-momentum square charged mass is oriented downwards. This can be understood from the fact that, $M_{\rm ch}^2$ possesses a background-dependent tree-level part, namely $m^2+(rT)^2$, which first increases up to $T_c$ and then decreases up to $T\approx 0.4~\mbox{GeV}$ as displayed in \Fig{fig_rT}.

We observe a rapid increase of $1/M_{\rm ch}^2$ above $2T_c$ and a pole at about $T\approx0.9$~GeV. This results from the competition between the (positive) vacuum contribution and the (negative) thermal contribution in \Eqn{eq:chargedmassedef}: $M_{\rm ch}^2$ turns negative in a finite range of temperature, before the positive vacuum contribution (which depends on the temperature through the nonzero frequency) dominates again at asymptotically large temperature. For the parameters used here this range is $m\lesssim T\lesssim 10m$. This unphysical behavior may simply be an artefact of the present perturbative calculation which could be resolved at higher orders  (which become relevant anyway at high temperatures) and/or by taking into account a possible temperature dependence of the parameters, as already mentioned. However, we cannot exclude the possibility that this behavior is a sign of a deeper problem. For instance, one can imagine that for sufficiently high temperatures, one explores field configurations beyond the first Gribov region, for which the functional measure is not positive anymore due to the sign of the Faddeev-Popov determinant. A study of these questions is certainly needed but it is beyond the scope of the present work. Here, we have simply checked that the temperature where the square mass turns negative is pushed to higher values when the coupling is decreased.

\subsection{Ghost propagators}
We now turn to the ghost sector. As before, we discuss the neutral and charged color modes separately.

\subsubsection{Neutral modes}

Despite the presence of a nonvanishing background, there remains an antighost shift symmetry for the neutral mode $\bar c^0\to\bar c^0 +{\rm const}$ which, together with spatial isotropy, implies
\beq
 \Sigma^0(0,k)=k^2\sigma(k),
\eeq
where $\sigma(0)<\infty$. We define the ghost dressing function at vanishing frequency as $F(k)=k^2 {\cal G}^0(0,k)$, that is,
\beq\label{eq:def_dressing}
F^{-1}(k)=1+g^2\sigma(k)=F^{-1}_{\rm vac}(k^2)+g^2\sigma^{\rm th}(k).
\eeq

\begin{figure}[t]
  \centering
  \includegraphics[width=1\linewidth]{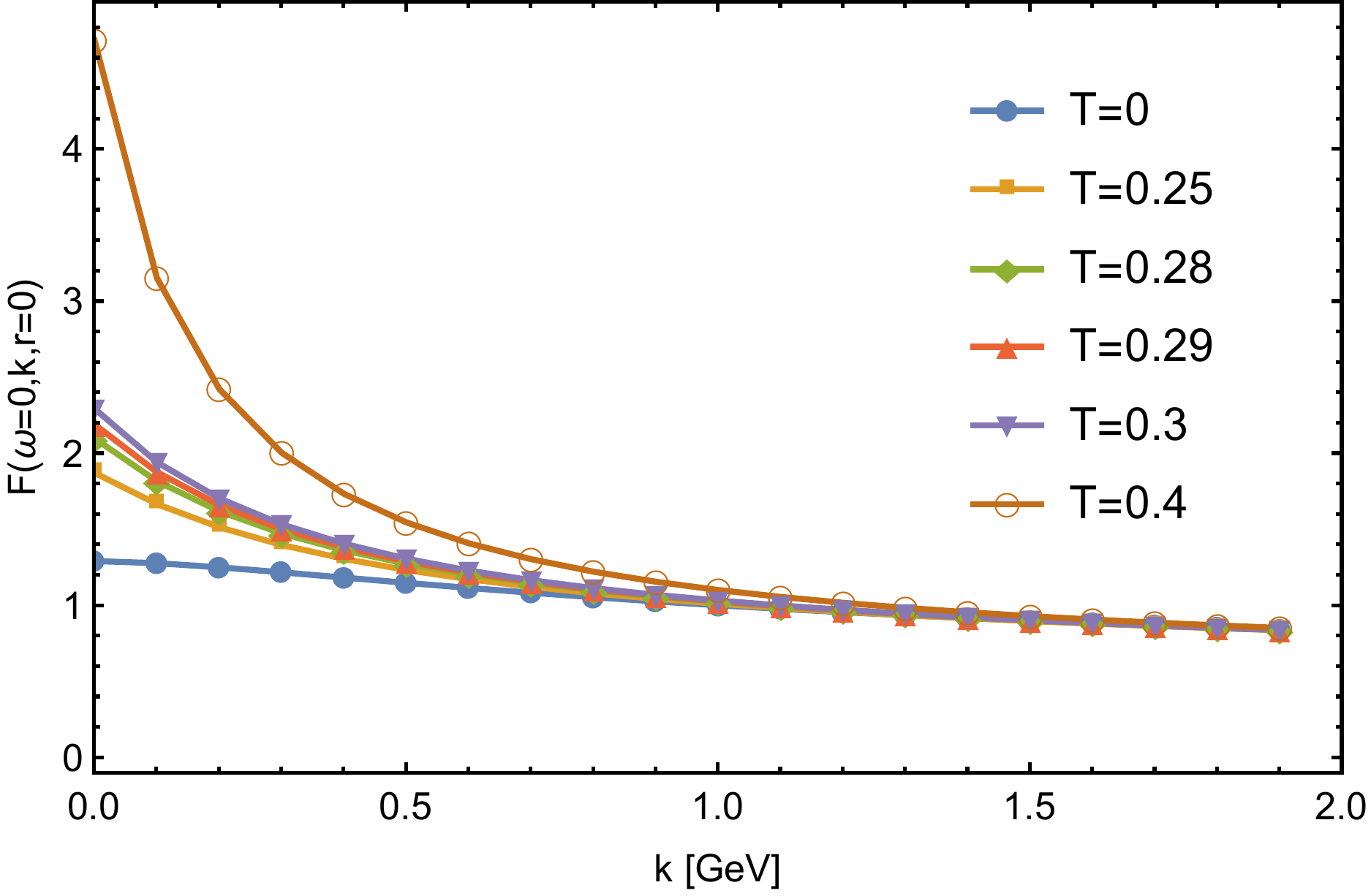}\quad
   \caption{The ghost dressing function \eqn{eq:def_dressing} at vanishing background field as a function of the momentum $k$ for various temperatures. As discussed in \cite{Reinosa:2013twa}, for temperature-independent parameters, at sufficiently high temperature, the ghost dressing function develops a pole.}
  \label{fig_dressLandau}
\end{figure}

\begin{figure}[t]
  \centering
  \includegraphics[width=1\linewidth]{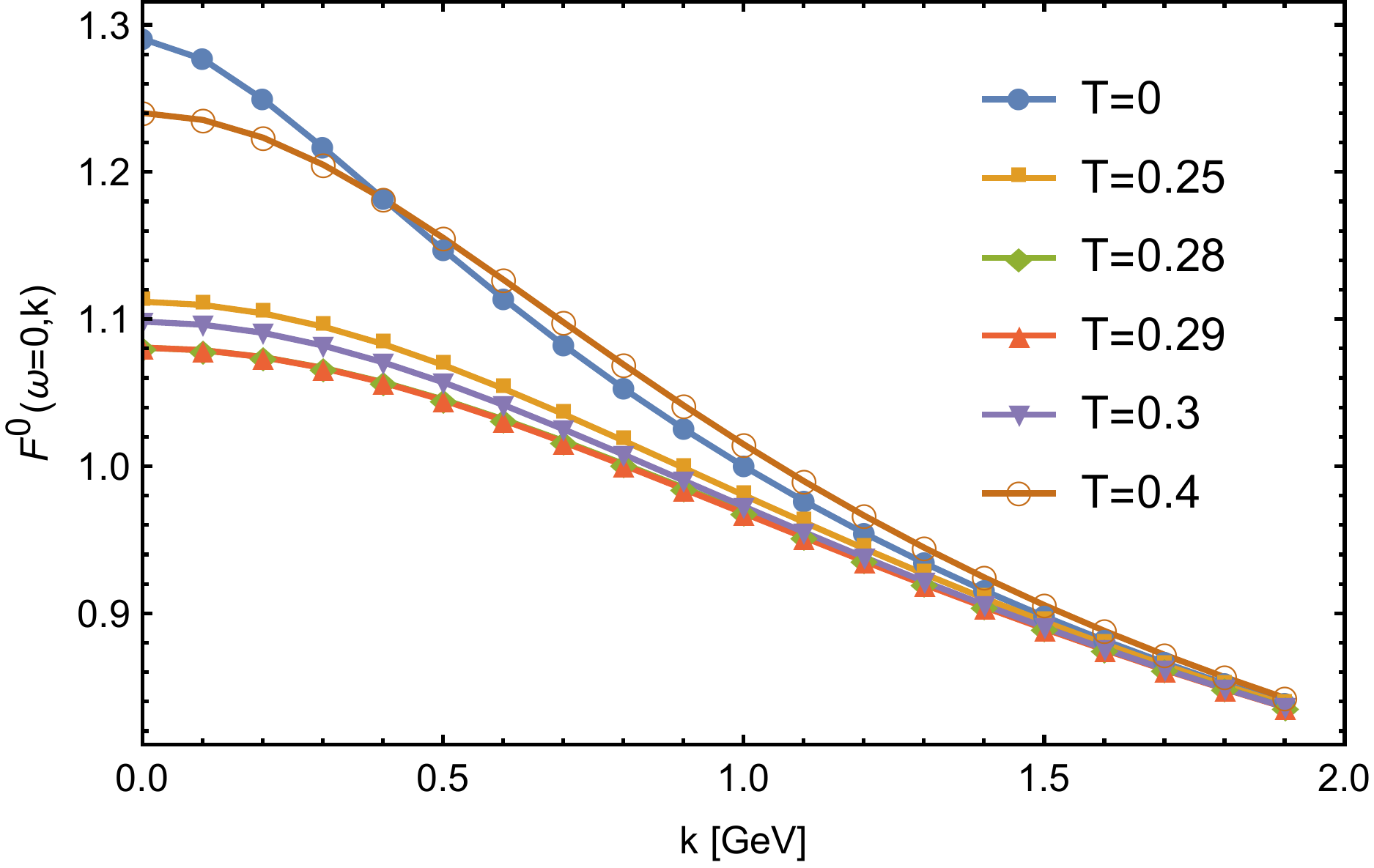}\quad
   \caption{The dressing function \eqn{eq:def_dressing} of the neutral ghost color mode as a function of the momentum $k$ for various temperatures. The pole of the vanishing-background case is absent.}
  \label{fig_dressk0}
\end{figure}

{ One striking result in the case of vanishing background is the fact that the ghost dressing function develops a pole for sufficiently high temperatures; see \Fig{fig_dressLandau}. As discussed in Ref.~\cite{Reinosa:2013twa}, this is a direct consequence of the Slavnov-Taylor identities of the present model in the Landau gauge and of the fact that the magnetic mass grows unbounded with temperature. Moreover, the fact that the ghost propagator explores negative values can be interpreted as a sign that field configurations beyond the first Gribov region (where the FP operator is strictly positive definite) are being explored \cite{Vandersickel:2012tz}. In the case of the Landau gauge, this pole is at odds with the results from lattice calculations, which, by construction, are restricted to the first Gribov region. In Ref.~\cite{Reinosa:2013twa}, this issue could be resolved by allowing temperature-dependent parameters. 

In the LDW gauge with a nontrivial background field, the situation is very different and we do not observe any pole, neither in the dressing function of the neutral ghost mode, shown in \Fig{fig_dressk0}, nor in the charged sector, discussed in the next subsection. For the neutral mode, this can again be understood from the Slavnov-Taylor identities and the behavior of the magnetic susceptibility of the  color-neutral gluon mode discussed above.}
Indeed, the discussion of Ref.~\cite{Reinosa:2013twa} at vanishing background can be easily generalized to the present case in the neutral color sector. The LDW generalization of Eq.~($40$) of Ref.~\cite{Reinosa:2013twa} yields
\beq \label{eq:ward}
 {\cal G}^0_{T,B}(K)F^0_{B}(K)\left|_{\omega=0, k\to0}= 1/m^2_0 , \right.
\eeq
where the index $B$ denotes bare correlators. In the renormalization scheme considered here, this identity becomes, at one-loop order,
\beq
m^2\sigma^{{\rm th}}(0) = -\Pi^{0,{\rm th}}_{T}(0,k \to 0).
\eeq
Finally, we have
\beq\label{eq:neutral_dressing_0}
F^{-1}(0)={ F_{R,\rm vac}^{-1}(0)}+1-\frac{M^2_{{\rm mag},0}}{m^2},
\eeq
where $F_{R,\rm vac}^{-1}(0)\approx .75$ for the present set of parameters.

As mentioned above, at vanishing background, the magnetic mass grows linearly with the temperature, which eventually leads to a pole in the ghost dressing function, with $F^{-1}(0)<0$. {As discussed in Sec.~\ref{sec:sn}, the situation is different in the presence of a nontrivial background, where $M^2_{{\rm mag},0}$ remains bounded from above, thus preventing the appearance of a pole in the neutral ghost dressing function. Incidentally, this suggests that the present perturbative expansion around the nontrivial background does not explore field configurations beyond the first Gribov region as evoked in the previous subsection, although one should keep in mind that the absence of pole in the ghost propagator is not completely conclusive for this question.

Finally, we see from \Eqn{eq:neutral_dressing_0} that the value of the neutral ghost dressing function at vanishing momentum is controlled by that of  the neutral gluon magnetic mass. In particular, the nonmonotonic behavior of the latter at $T_c$ (see \Fig{fig_k0masses}) is directly visible in  \Fig{fig_dressk0}. }

\subsubsection{Charged modes}

 \begin{figure}[t]
  \centering
  \includegraphics[width=1\linewidth]{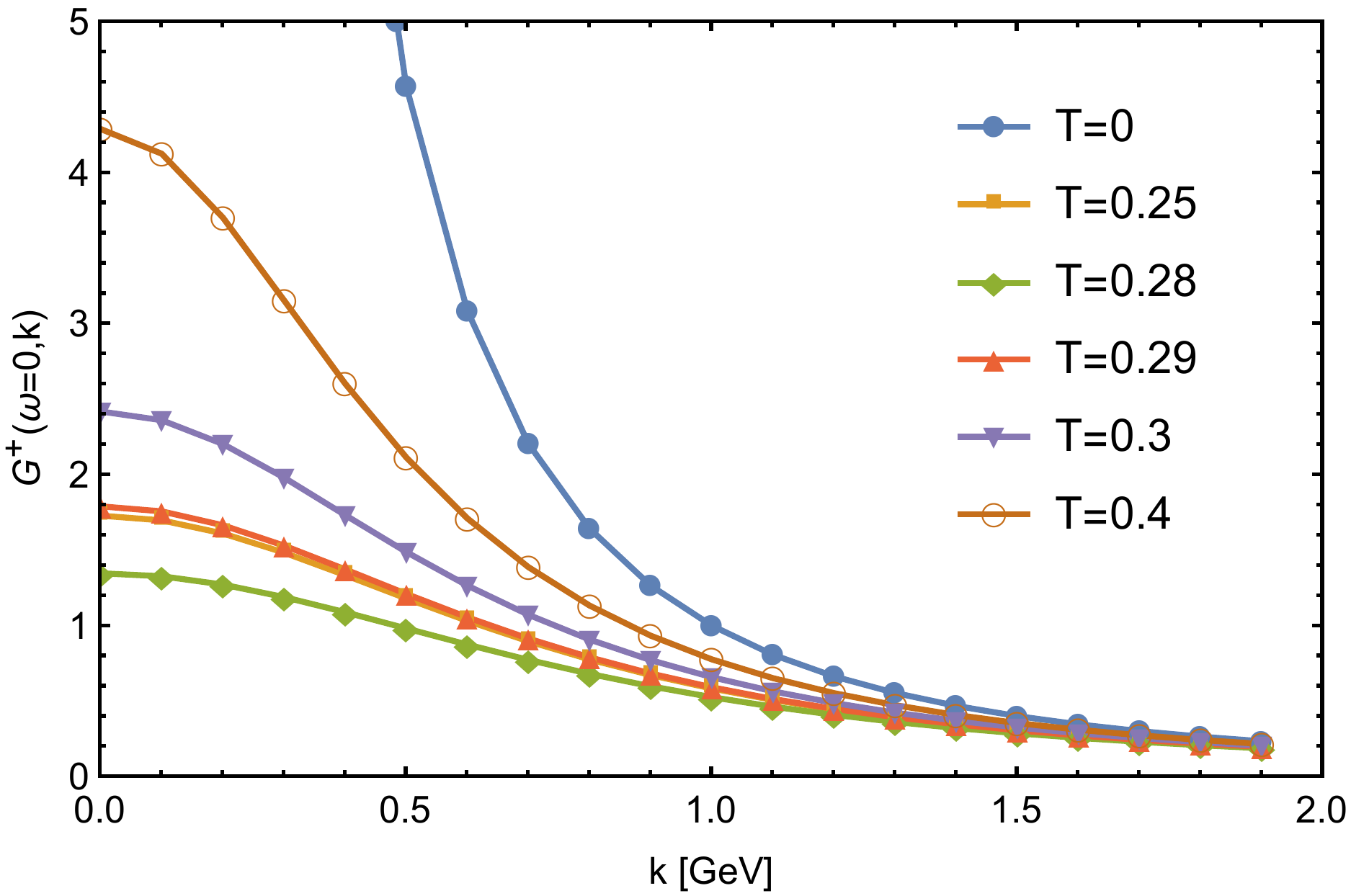}\quad
   \caption{The ghost propagator at vanishing frequency in the charged color sector as a function of momentum $k$ for various temperatures.}
  \label{fig_dressk+}
\end{figure}
 \begin{figure}[t]
  \centering
  \includegraphics[width=1\linewidth]{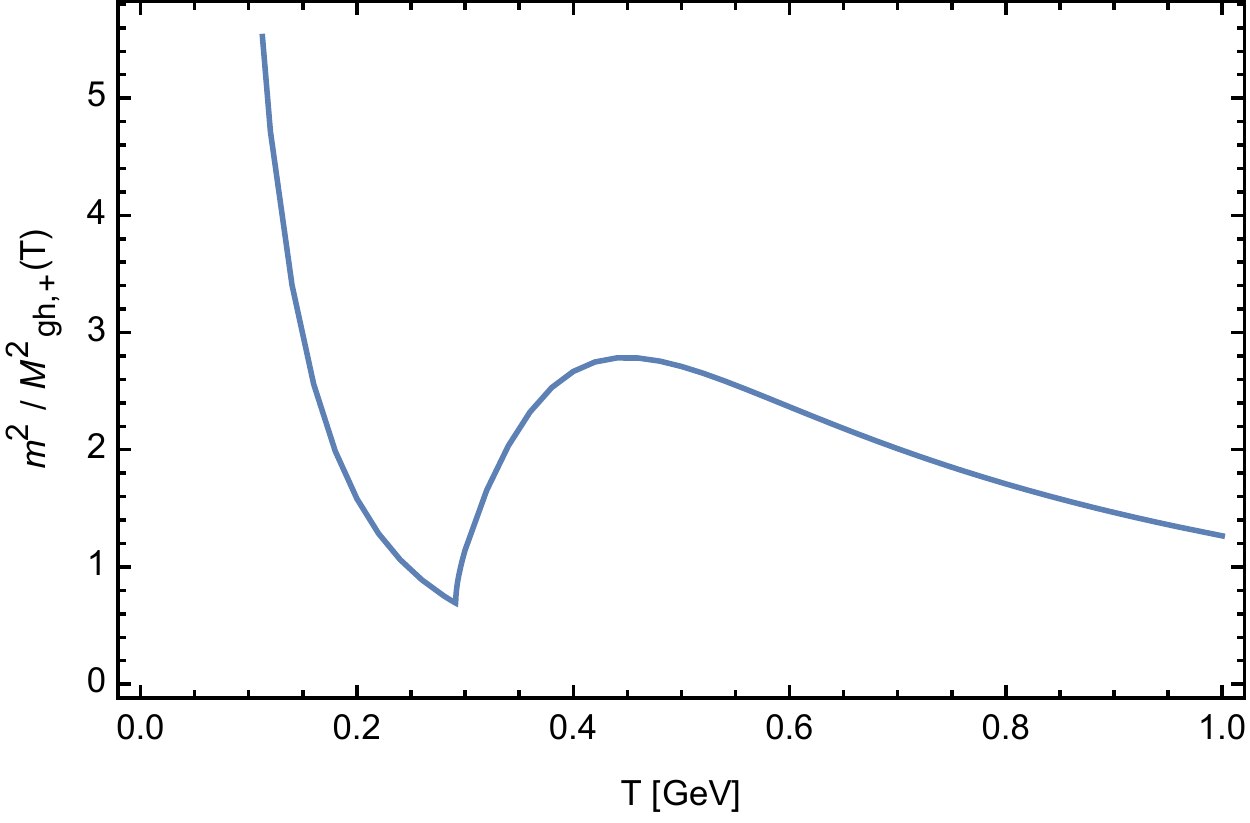}\quad
   \caption{The (normalized) ghost propagator for charged color modes at vanishing frequency and zero momentum as a function of temperature.}
  \label{fig_dresskMASS}
\end{figure}
As for the gluon case, the charged ghost modes at zero Matsubara frequency are degenerate for a charge-conjugation invariant system, 
\beq
 {\cal G}^+(0,k)={\cal G}^-(0,k),
\eeq
as follows from \Eqn{eq:prop1} and spatial isotropy. In the case of a nonvanishing background field, there is no antighost shift symmetry in the charged sector and we shall thus directly study the propagators. The charged ghost propagator at zero frequency is shown as a function of $k$ for various temperatures in \Fig{fig_dressk+}. As for the neutral mode, it presents a nonmonotonic behavior in temperature with two changes of monotony at $T_c$ and around $T=0.4$ GeV; see also \Fig{fig_dresskMASS}. This corresponds to the change of monotony of its effective tree level mass $rT$. In the limit $T\to 0$, the background field $\bar A\propto r T\to0$ and we recover the original antighost shift symmetry at vanishing background, which implies that the ghost propagator diverges at zero momentum.

{
\section{Discussion}

The most salient feature of the present results is the clear signature of the phase transition in the various two-point correlators of the theory, { and particularly in the neutral electric component,} due to the influence of the nontrivial order parameter. In the SU($2$) theory, where the transition is second order, the susceptibility of the order parameter, i.e., the Polyakov loop, should diverge at the transition and one may wonder whether such a divergence should be visible in the basic correlators of the theory, as computed here. In fact, this issue is even more pronounced if one remembers that the background gluon field is itself an order parameter for the transition and that the correlator of the neutral gluon color mode precisely probes the fluctuations of the latter. In other words, how can one reconcile the fact that the second derivative of the (temporal) background field potential vanishes at the transition with the fact that the (electric) gluon square mass in the neutral sector remains finite? This can be understood from the relations between the vertices of the background effective action \eqn{eq:bckfunc} with the vertices of the theory at fixed background, \Eqn{eq:vertexfunc}.

In Ref.~\cite{Reinosa:2015gxn}, it was shown that the effective action at fixed background field satisfies the following identity
\beq
\label{eq:bfindep}
\left. \frac{\delta\Gamma[\bar A,a]}{\delta\bar A_\mu^a} \right|_{a_{\rm min}}=-m_0^2\,a_{{\rm min},\mu}^a
\eeq
where $a_{\rm min}\equiv a_{\rm min}[\bar A]$ obeys
\beq
 \left.\frac{\delta\Gamma[\bar A,a]}{\delta a_\mu^a} \right|_{a_{\rm min}}=0.
\eeq
This shows how the bare mass term in the present model spoils the exact background-field independence of the partition function\footnote{In principle, the background field only arises through the gauge fixing condition and the partition function ($\ln Z\propto \Gamma[\bar A,a_{\rm min}]$) should thus be independent of $\bar A$. { The way to restore this important property in the present approach is not known. This requires a dedicated study, beyond the scope of the present work.}}. For $m_0\neq0$, the background-field independence is only verified locally, for self-consistent backgrounds, for which $a_{\rm min}=0$. Exploiting the above equations, one can derive relations between the functional derivatives of $\tilde \Gamma[\bar A]$ and those of $\Gamma[\bar A,a]$ with respect to $a$ at fixed $\bar A$. 

Let us focus on the case of interest here, where the background field is taken of the form \eqn{eq:form}. Clearly, the minimum $a_{\rm min}$ has the same symmetries as the background field and we thus have $a_{{\rm min},\mu}^a(\bar A,x)=a_{\rm min}(r)\delta_{\mu0}$, where $a_{\rm min}(r)$ belongs to the Cartan subalgebra of the group. In the SU($2$) case, $a_{\rm min}(r)$ is a single function of a single variable. After simple alegbra, we obtain the following identity at the minimum $r_{\rm min}$ of the potential~\eqn{eq:potential}:
\beq
 V''(r_{\rm min})=M^2_{{\rm D},0}\left[a'_{\rm min}(r_{\rm min})\right]^2-m_0^2\,a'_{\rm min}(r_{\rm min})
\eeq
At $T=T_c$, the left-hand side vanishes but we see that this does not imply that the electric square mass $M_{{\rm D},0}^2$ vanishes as well. Instead, we have explicitly checked that, at one-loop order, $a'_{\rm min}(r_{\rm min})=0$ at $T=T_c$ \cite{Vrs}.

We stress, however, that the present considerations follow from \Eqn{eq:bfindep} which expresses the fact that the partition function of the present massive model is not exactly independent of the background field. We do not exclude the possibility that in, say a lattice implementation, where the background-field independence of the partition function should hold, the neutral Debye mass actually vanishes at the transition and that the cusp observed here in \Fig{fig_k0masses} turns into an actual divergence. We postpone a detailed discussion of these aspects to a later work.}

\section{Conclusions}\label{sec_app_A}

We have studied the influence of the nontrivial order parameter of the deconfinement transition on the two-point correlators of the basic (gluon and ghost) degrees of freedom of Yang-Mills theories. We have given the expressions of the correlators for a broad class of gauge groups at one-loop order in a perturbative expansion in the context of the massive extension of background field techniques proposed in Ref.~\cite{Reinosa:2014ooa}. We have considered explicitly the SU($2$) case and we have shown that the presence of the nontrivial background gluon field dramatically affects the correlators in various ways. We stress that most of these effects have not been taken into account in some previous works \cite{Braun:2007bx,Braun:2010cy,Quandt:2016ykm}, where the LDW propagators are simply modelled by using the zero-temperature Landau gauge propagators with shifted momentum; see however \cite{Fister:2013bh,Herbst:2015ona}. It remains to be studied how much the  effects obtained in the present work affect the results of those studies.

The most stringent feature is that the nonanalytic behavior of the order parameter at the phase transition is directly imprinted in the temperature dependence of the correlators. In the SU($2$) case, where the transition is continuous, this results in a very distinct cusp in, say the correlators at vanishing momentum, { in particular in the neutral electric component}. It is to be expected that in the case of a first order transition, e.g. in the SU($3$) theory, the correlators will exhibit a discontinuity at the transition. We plan to study the SU($3$) case in a further work. It is also interesting to extend the present study to real-time response functions, such as spectral functions. Another possible line of investigation would be to perform a similar study in QCD, e.g., in the case of heavy quarks, where one could also study the effect of the Polyakov loop on the quark propagator. 

Finally, we mention that the present work, together with our previous studies, suggests that the calculation of the finite temperature correlators may be better controlled in the LDW gauge with the class of backgrounds considered here, than in the Landau gauge. In a sense, the nonzero background efficiently selects the relevant field configurations corresponding to a given value of the Polyakov loop, around which the fluctuations are not too large. It would be of definite interest to try to implement this class of background field gauge in lattice simulations.\\

\section*{Acknowledgments}
We are grateful to M.~Pel\'aez and N.~Wschebor for many useful discussions.\\

\appendix

{ \section{Sum-integrals}\label{app:sum_int}
In what follows, we define $\kappa^jr^j T\equiv \kappa^j\hat r^j\equiv \kappa\cdot\hat r $. The Matsubara sums of all the elementary integrals defined in the main text can be performed using standard integration contour techniques, see for instance \cite{Reinosa:2013twa}. For the tadpolelike sum-integrals, this yields
\beq
J^\kappa_m\,\hat{=} \int_{\bf q}\, {\rm Re}\,\frac{n_{\varepsilon_{m,q}-i\kappa\cdot\hat r}}{\varepsilon_{m,q}}=\frac{1}{2\pi^2}\int_0^\infty dq\,q^2{\rm Re}\,\frac{n_{\varepsilon_{m,q}-i\kappa\cdot\hat r}}{\varepsilon_{m,q}}
\eeq
and
\begin{align}
\label{appeq:A2}
\tilde J^\kappa_m & = \int_{\bf q}\, {\rm Im}\,n_{\varepsilon_{m,q}-i \kappa\cdot\hat r}=\frac{1}{2\pi^2}\int_0^\infty dq\,q^2{\rm Im}\,n_{\varepsilon_{m,q}-i\kappa\cdot\hat r}\,,
\end{align}
where $\int_{\bf q}=\int\frac{d^3q}{(2\pi)^3}$, $\varepsilon_{m,q}=\sqrt{q^2+m^2}$, and $n_z=(\exp \beta z -1)^{-1}$ is the Bose-Einstein distribution function, which satisfies $n_{-z}=-1-n_z$. The symbol $\hat{=}$ means that we disregard vacuum contributions defined as the limit of the above expressions as $T\to 0$ for fixed $\hat r$. The reason why we can do so is that, as explained in the main text, the vacuum contributions to the self-energies can be very easily obtained from the results of \cite{Tissier:2010ts}.\\

For the bubblelike sum-integrals, we obtain,  similarly,
\begin{widetext}
\bea
I^{\kappa\tau}_{m_1m_2}(K) & \hat{=} & \int_{\bf q}{\rm Re}\!\left[\frac{n_{\varepsilon_{m_1,q}+i\kappa\cdot \hat{r}}}{\varepsilon_{m_1,q}}G_{m_2}(\omega^\lambda+i\varepsilon_{m_1,q},l)+\,(m_1,\kappa\leftrightarrow m_2,\tau)\right]\,,\\
2\{I^\lambda_T\}^{\kappa\tau}_{m_1m_2}(K) & \hat{=} & \int_{\bf q}\left(q^2-\frac{({\bf k}\cdot{\bf q})^2}{k^2}\right){\rm Re}\left[\frac{n_{\varepsilon_{m_1,q}+i\kappa\cdot \hat{r}}}{\varepsilon_{m_1,q}}G_{m_2}(\omega^\lambda+i\varepsilon_{m_1,q},l)+\,(m_1,\kappa\leftrightarrow m_2,\tau)\right]\,,\\
k^2K^2_\lambda \{I^\lambda_L\}^{\kappa\tau}_{m_1m_2}(K) & \hat{=} & \int_{\bf q}\,\,{\rm Re}\!\left[\frac{n_{\varepsilon_{m_1,q}+i\kappa\cdot \hat{r}}}{\varepsilon_{m_1,q}}\left(-ik^2\varepsilon_{m_1,q}+\omega^\lambda({\bf k}\cdot{\bf q})\right)^2\!G_{m_2}(\omega^\lambda+i\varepsilon_{m_1,q},l)+\,(m_1,\kappa\leftrightarrow m_2,\tau)\right]
\eea
where we have introduced $\lambda=-\tau-\kappa$ and $\omega^\lambda=\omega+\lambda\cdot\hat r$. An angular integration leads then to
\bea
I^{\kappa\tau}_{m_1m_2}(K)&\hat{=}&-\frac{1}{8\pi^2k}\int_0^\infty \!\!\!dq\,q\,{\rm Re}\left[\frac{n_{\varepsilon_{m_1,q}+i\kappa\cdot \hat{r}}}{\varepsilon_{m_1,q}}g_{m_2}(\omega^\lambda+i\varepsilon_{m_1,q};q,K_\lambda)+(m_1,\kappa\leftrightarrow m_2,\tau)\right]\,,\\
\{I_T^\lambda\}^{\kappa\tau}_{m_1m_2}(K) &\hat{=} &\frac{1}{64\pi^2k^3}\!\int_0^\infty \!\!dq\,q\,{\rm Re}\Big[\frac{n_{\varepsilon_{m_1,q}+i\kappa\cdot \hat{r}}}{\varepsilon_{m_1,q}}\!\Big\{4kq\left(K^2_\lambda+2i\omega^\lambda\varepsilon_{m_1,q}+m^2_2-m^2_1\right)\nonumber\\
& & \hspace{4.8cm}+\,\ell_T^{m_2}(\omega^\lambda+i\varepsilon_{m_1,q};q,K_\lambda)\Big\}+(m_1,\kappa\leftrightarrow m_2,\tau)\Big],\\
\{I_L^\lambda\}^{\kappa\tau}_{m_1m_2}(K)&\hat{=}& -\frac{\omega^2_\lambda}{32\pi^2K^2_\lambda k^3}\!\int_0^\infty\!\!dq\,q\,{\rm Re}\Big[\frac{n_{\varepsilon_{m_1,q}+i\kappa\cdot \hat{r}}}{\varepsilon_{m_1,q}}\Big\{4kq\left(K^2_\lambda+m^2_2-m^2_1+2i\omega^\lambda\varepsilon_{m_1,q}\left(1-\frac{2kq}{\omega^2_\lambda}\right)\right)\nonumber\\
& & \hspace{5.7cm}+\,\ell_L^{m_2}(\omega^\lambda+i\varepsilon_{m_1,q};q,K_\lambda)\Big\}+(m_1,\kappa\leftrightarrow m_2,\tau)\Big],
\eea
\end{widetext}
where we introduced the functions
\beq
 g_\beta(z;q,K)=\ln\frac{z^2+\varepsilon^2_{\beta,k-q}}{z^2+\varepsilon^2_{\beta,k+q}},
\eeq
\beq
 \ell_T^\beta(z;q,K)=\left(\varepsilon_{\beta,k+q}^2+z^2\right)\left(\varepsilon_{\beta,k-q}^2+z^2\right)\,g_\beta(z;q,K),
\eeq
and
\beq
 \ell_L^\beta(z;q,K)=\left[\varepsilon_{\beta,q}^2+z^2+k^2\left({2z\over\omega}-1\right)\right]^{2}\!g_\beta(z;q,K).
\eeq}

{ \section{Details on the evaluation of the gluon self-energy}\label{app:gl_self}
The gluon self-energy has three one-loop contributions, $\Pi_{\mu\nu}^{{\rm tad},\lambda}$, $\Pi_{\mu\nu}^{{\rm gh},\lambda}$ and $\Pi_{\mu\nu}^{{\rm gl},\lambda}$, which stand respectively for the tadpole diagram, the ghost bubble diagram and the gluon bubble diagram. A direct application of the Feynman rules given in the main text leads to
\begin{widetext}
\bea
\Pi_{\mu\nu}^{\rm gl,\lambda}(K) & = & \sum_{\kappa,\tau}{\cal C}_{\kappa\lambda\tau}\frac{1}{2}\left\{\left[-\frac{1}{2}\int_Q (Q^\kappa-L^\tau)_\mu(Q^\kappa-L^\tau)_\nu {\rm Tr}\,\big[P_\perp(Q^\kappa)\cdot P_\perp(L^\tau)\big]G_m(Q^\kappa)G_m(L^\tau)\right.\right.\nonumber\\
& & \hspace{1.5cm}-\,4\int_Q \Big[Q^\kappa\cdot P_\perp(L^\tau)\cdot Q^\kappa\Big]P_{\mu\nu}^\perp(Q^\kappa)\,G_m(Q^\kappa)G_m(L^\tau)\nonumber\\
& & \hspace{1.5cm}+\,2\int_Q \left\{(Q^\kappa-L^\tau)_\mu \Big[Q^\kappa\cdot P_\perp(L^\tau)\cdot P_\perp(Q^\kappa)\Big]_\nu+(\mu\leftrightarrow \nu)\right\} G_m(Q^\kappa)G_m(L^\tau)\nonumber\\
& & \hspace{1.5cm}\left.\left.+\,4\int_Q \Big[L^\tau\cdot P_\perp(Q^\kappa)\Big]_\mu \Big[Q^\kappa\cdot P_\perp(L^\tau)\Big]_\nu G_m(Q^\kappa)G_m(L^\tau)\right]+(\kappa\leftrightarrow\tau)\right\},
\eea
where, for convenience, we have symmetrized the summand in $\kappa\leftrightarrow\tau$ by using that ${\cal C}_{\kappa\lambda\tau}$ is totally symmetric. Evaluating the trace and using $(Q^\kappa-L^\tau)_\mu(Q^\kappa-L^\tau)_\nu=2L^\tau_\mu L^\tau_\nu+2Q^\kappa_\mu Q^\kappa_\nu-K^\lambda_\mu K^\lambda_\nu$, the factor multiplying $G_m(Q^\kappa)G_m(L^\tau)$ in the first integral becomes (for this integral, symmetrization in $\kappa\leftrightarrow\tau$ does not change anything)
\beq
-\frac{1}{2}(2Q^\kappa_\mu Q^\kappa_\nu+2L^\tau_\mu L^\tau_\nu-K^\lambda_\mu K^\lambda_\nu)\left(d-2+\frac{(Q^\kappa\cdot L^\tau)^2}{Q^2_\kappa L^2_\tau}\right).
\eeq
The similar factor in the second integral becomes, after symmetrization,
\beq
-2\delta_{\mu\nu}\left(Q^2_\kappa+L^2_\tau-\frac{(Q^\kappa\cdot L^\tau)^2}{L^2_\tau}-\frac{(Q^\kappa\cdot L^\tau)^2}{Q^2_\kappa}\right)+2\left(Q_\mu^\kappa Q_\nu^\kappa+L_\mu^\tau L_\nu^\tau\right)\left(1-\frac{(Q^\kappa\cdot L^\tau)^2}{Q^2_\kappa L^2_\tau}\right).
\eeq
For the third integral, we obtain
\beq
\left(Q_\mu^\kappa Q_\nu^\kappa+3L_\mu^\tau L_\nu^\tau-K_\mu^\lambda K_\nu^\lambda\right)\frac{Q^\kappa\cdot L^\tau}{L^2_\tau}+\left(3Q_\mu^\kappa Q_\nu^\kappa+L_\mu^\tau L_\nu^\tau-K_\mu^\lambda K_\nu^\lambda\right)\frac{Q^\kappa\cdot L^\tau}{Q^2_\kappa}+\left(4Q_\mu^\kappa Q_\nu^\kappa+4L_\mu^\tau L_\nu^\tau-2K_\mu^\lambda K_\nu^\lambda\right)\frac{(Q^\kappa\cdot L^\tau)^2}{Q^2_\kappa L^2_\tau}
\eeq
and for the fourth
\beq
2\left(K_\mu^\lambda K_\nu^\lambda-Q^\kappa_\mu Q_\nu^\kappa-L_\mu^\tau L^\tau_\nu\right)\left(1+\frac{(Q^\kappa\cdot L^\tau)^2}{Q^2_\kappa L^2_\tau}\right)-4 Q_\mu^\kappa Q_\nu^\kappa\frac{Q^\kappa\cdot L^\tau}{Q^2_\kappa}-4 L_\mu^\tau L_\nu^\tau\frac{Q^\kappa\cdot L^\tau}{L^2_\tau}\,.
\eeq
Putting all these pieces together, we arrive at
\begin{align}
\Pi_{\mu\nu}^{\rm gl,\lambda}(K)  = \sum_{\kappa,\tau}{\cal C}_{\kappa\lambda\tau}\Bigg\{&-4\delta_{\mu\nu}\int_Q\left(L^2_\tau-\frac{(Q^\kappa\cdot L^\tau)^2}{Q^2_\kappa}\right)G_m(Q^\kappa)G_m(L^\tau)\nonumber\\
& +2\int_Q(-Q_\mu^\kappa Q_\nu^\kappa+L_\mu^\tau L_\nu^\tau-K_\mu^\lambda K_\nu^\lambda)\frac{Q^\kappa\cdot L^\tau}{Q^2_\kappa}\,G_m(Q^\kappa)G_m(L^\tau)\nonumber\\\label{eqLLLL}
& +\int_Q\left(\frac{1}{2}K_\mu^\lambda K_\nu^\lambda-Q^\kappa_\mu Q_\nu^\kappa-L_\mu^\tau L^\tau_\nu\right)\frac{(Q^\kappa\cdot L^\tau)^2}{Q^2_\kappa L^2_\tau}\,G_m(Q^\kappa)G_m(L^\tau)\nn& +\int_Q\left[\left(\frac{d-2}{2}+2\right)K_\mu^\lambda K_\nu^\lambda-(d-2)(Q_\mu^\kappa Q_\nu^\kappa +L_\mu^\tau L_\nu^\tau)\right]G_m(Q^\kappa)G_m(L^\tau)\Bigg\}.
\end{align}
The next step uses the identity
\beq
\frac{Q^\kappa\cdot L^\tau}{Q^2_\kappa}G_m(Q^\kappa)G_m(L^\tau) = \frac{K^2_\lambda+m^2}{2m^2}G_0(Q^\kappa)G_m(L^\tau)-\frac{K^2_\lambda+2m^2}{2m^2}G_m(Q^\kappa)G_m(L^\tau)-\frac{1}{2m^2}\big[G_0(Q^\kappa)-G_m(Q^\kappa)\big]\,,
\eeq
as well as
\begin{align}
\frac{(Q^\kappa\cdot L^\tau)^2}{Q^2_\kappa L^2_\tau}G_m(Q^\kappa)G_m(L^\tau) & = \frac{(K^2_\lambda+2m^2)^2}{4m^4} G_m(Q^\kappa)G_m(L^\tau)+\frac{K^4_\lambda}{4m^4} G_0(Q^\kappa)G_0(L^\tau)\nonumber\\
& - \frac{(K^2_\lambda+m^2)^2}{4m^4}\Big[G_m(Q^\kappa)G_0(L^\tau)+G_0(Q^\kappa)G_m(L^\tau)\Big]\nonumber\\
& + \frac{1}{4m^2}\Big[G_0(Q^\kappa)+G_0(L^\tau)-G_m(Q^\kappa)-G_m(L^\tau)\Big],
\end{align}
and
\begin{align}
\left(L^2_\tau-\frac{(Q^\kappa\cdot L^\tau)^2}{Q^2_\kappa}\right)G_m(Q^\tau)G_m(L^\tau) & = \frac{K^2_\lambda(K^2_\lambda+4m^2)}{4m^2}G_m(Q^\kappa)G_m(L^\tau)-\frac{(K^2_\lambda+m^2)^2}{4m^2}G_0(Q^\kappa)G_m(L^\tau)-\frac{1}{4}G_m(L^\tau)\nonumber\\
& - \frac{K^2_\lambda}{4m^2}G_m(Q^\kappa)+\frac{K^2_\lambda+m^2}{4m^2}G_0(Q^\kappa)- \frac{(K^\lambda\cdot Q^\kappa)}{2m^2}\big[G_0(Q^\kappa)-G_m(Q^\kappa)\big].
\end{align}
These identities allow us to express \Eqn{eqLLLL} in terms of the sum-integrals \eqn{eq_integrals}, \eqn{eq_integralsss}, and \eqn{eq:hehehe}. Using the symmetry properties of the latter and of the tensor ${\cal C}_{\kappa\lambda\tau}$, we obtain
\begin{align}
&\Pi_{\mu\nu}^{\rm gl,\lambda}(K)  = \sum_{\kappa,\tau}{\cal C}_{\kappa\lambda\tau}\Bigg\{\delta_{\mu\nu}\left[\frac{(K^2_\lambda+m^2)^2}{m^2}I_{m0}^{\kappa\tau}(K)-\frac{K^2_\lambda(K^2_\lambda+4m^2)}{m^2}I_{mm}^{\kappa\tau}(K)+\,\frac{K^2_\lambda+m^2}{m^2}(J^\kappa_m-J^\kappa_0)+2\frac{\omega^\lambda}{m^2}(\tilde J^\kappa_0-\tilde J^\kappa_m)\right]\nonumber\\
&+K_\mu^\lambda K_\nu^\lambda\left[\left(\frac{d-2}{2}+\frac{(K^2_\lambda+6m^2)^2}{8m^4}\right)I_{mm}^{\kappa\tau}(K)+\frac{K^4_\lambda}{8m^4}I_{00}^{\kappa\tau}(K)-\frac{(K^2_\lambda+m^2)(K^2_\lambda+5m^2)}{4m^4}I_{m0}^{\kappa\tau}(K)+\frac{5}{4m^2}(J^\kappa_0-J^\kappa_m)\right]\nonumber\\
&+\left(4-2d-\frac{(K^2_\lambda+2m^2)^2}{2m^4}\right)\left\{I_{\mu\nu}\right\}_{mm}^{\kappa\tau}(K)-\frac{K^4_\lambda}{2m^4}\left\{I_{\mu\nu}\right\}_{00}^{\kappa\tau}(K)+\frac{(K^2_\lambda+m^2)(K^2_\lambda+3m^2)}{2m^4}\left\{I_{\mu\nu}\right\}_{m0}^{\kappa\tau}(K)\nonumber\\
& +\frac{(K^4_\lambda-m^4)}{2m^4}\left\{I_{\mu\nu}\right\}_{0m}^{\kappa\tau}(K)+\frac{1}{2m^2}\int_Q (Q_\mu^\kappa Q_\nu^\kappa-3L_\mu^\tau L_\nu^\tau) [G_0(Q^\kappa)-G_m(Q^\kappa)]\Bigg\}.
\end{align}
To treat the last integral, we use $Q_\mu^\kappa Q_\nu^\kappa-3L_\mu^\tau L_\nu^\tau=-2Q_\mu^\kappa Q_\nu^\kappa-3K_\mu^\lambda K_\nu^\lambda-3(K_\mu^\lambda Q_\nu^\kappa+K_\nu^\lambda Q_\mu^\kappa)$ to get
\begin{align}
\Pi_{\mu\nu}^{\rm gl,\lambda}(K)  &= \sum_{\kappa,\tau}{\cal C}_{\kappa\lambda\tau}\Bigg\{\delta_{\mu\nu}\left[\frac{(K^2_\lambda+m^2)^2}{m^2}I_{m0}^{\kappa\tau}(K)-\frac{K^2_\lambda(K^2_\lambda+4m^2)}{m^2}I_{mm}^{\kappa\tau}(K)+\,\frac{K^2_\lambda+m^2}{m^2}(J^\kappa_m-J^\kappa_0)+2\frac{\omega^\lambda}{m^2}(\tilde J^\kappa_0-\tilde J^\kappa_m)\right]\nonumber\\
&+K_\mu^\lambda K_\nu^\lambda\left[\left(\frac{d-2}{2}+\frac{(K^2_\lambda+6m^2)^2}{8m^4}\right)I_{mm}^{\kappa\tau}(K)+\frac{K^4_\lambda}{8m^4}I_{00}^{\kappa\tau}(K)-\frac{(K^2_\lambda+m^2)(K^2_\lambda+5m^2)}{4m^4}I_{m0}^{\kappa\tau}(K)-\frac{J^\kappa_0-J^\kappa_m}{4m^2}\right]\nonumber\\
&+\left(4-2d-\frac{(K^2_\lambda+2m^2)^2}{2m^4}\right)\left\{I_{\mu\nu}\right\}_{mm}^{\kappa\tau}(K)-\frac{K^4_\lambda}{2m^4}\left\{I_{\mu\nu}\right\}_{00}^{\kappa\tau}(K)+\frac{(K^2_\lambda+m^2)(K^2_\lambda+3m^2)}{2m^4}\left\{I_{\mu\nu}\right\}_{m0}^{\kappa\tau}(K)\nonumber\\
& +\frac{(K^4_\lambda-m^4)}{2m^4}\left\{I_{\mu\nu}\right\}_{0m}^{\kappa\tau}(K)-\frac{3}{2}\!\left(K_\mu^\lambda n_\nu+K_\nu^\lambda n_\mu\right)\!\frac{\tilde J_0^\kappa-\tilde J_m^\kappa}{m^2}-\frac{\left\{J_{\mu\nu}\right\}_{0}^{\kappa}-\left\{J_{\mu\nu}\right\}_{m}^{\kappa}}{m^2}\Bigg\},
\end{align}
where we defined
\beq
 \left\{J_{\mu\nu}\right\}_m^\kappa = \int_Q Q^\kappa_\mu Q^\kappa_\nu G_m(Q^\kappa).
\eeq
Similarly, we obtain
\beq
\Pi_{\mu\nu}^{{\rm tad},\lambda}=\sum_{\kappa,\tau}{\cal C}_{\kappa\lambda\tau}\left[(d-2)\delta^{\mu\nu}J^\kappa_m+\frac{\left\{J_{\mu\nu}\right\}_{0}^{\kappa}-\left\{J_{\mu\nu}\right\}_{m}^{\kappa}}{m^2}\right],
\eeq
and
\beq
\Pi_{\mu\nu}^{{\rm gh},\lambda}(K)=\sum_{\kappa,\tau}{\cal C}_{\kappa\lambda\tau}\left[\left\{I_{\mu\nu}\right\}_{00}^{\kappa\tau}(K)-\frac{K_\mu^\lambda K_\nu^\lambda}{2} I_{00}^{\kappa\tau}(K)\right].
\eeq
The total gluon self-energy is then
\begin{align}
\label{appeq:lastone}
\Pi_{\mu\nu}^\lambda(K) =\sum_{\kappa,\tau}{\cal C}_{\kappa\lambda\tau}\Bigg\{&\delta_{\mu\nu}\left[\frac{(K^2_\lambda+m^2)^2}{m^2}I_{m0}^{\kappa\tau}(K)-\frac{K^2_\lambda(K^2_\lambda+4m^2)}{m^2}I_{mm}^{\kappa\tau}(K)+\frac{K^2_\lambda+m^2}{m^2}(J^\kappa_m-J^\kappa_0)+2\frac{\omega^\lambda}{m^2}(\tilde J^\kappa_0-\tilde J^\kappa_m)\right]\nn
+&\,\delta_{\mu\nu}(d-2)J_m^\kappa+\left(4-2d-\frac{(K^2_\lambda+2m^2)^2}{2m^4}\right)\left\{I_{\mu\nu}\right\}_{mm}^{\kappa\tau}(K)-\left(\frac{K^4_\lambda}{2m^4}-1\right)\left\{I_{\mu\nu}\right\}_{00}^{\kappa\tau}(K)\nonumber\\
+&\,\frac{(K^2_\lambda+m^2)}{2m^4}\Big[(K^2_\lambda+3m^2)\left\{I_{\mu\nu}\right\}_{m0}^{\kappa\tau}(K)+(K^2_\lambda-m^2)\left\{I_{\mu\nu}\right\}_{0m}^{\kappa\tau}(K)\Big]\nonumber\\
 +&\,K_\mu^\lambda K_\nu^\lambda\left[\left(\frac{d-2}{2}+\frac{(K^2_\lambda+6m^2)^2}{8m^4}\right)I_{mm}^{\kappa\tau}(K)+\left(\frac{K^4_\lambda}{8m^4}-\frac{1}{2}\right)I_{00}^{\kappa\tau}(K)-\frac{J^\kappa_0-J^\kappa_m}{4m^2}\right]\nonumber\\
-&\,K_\mu^\lambda K_\nu^\lambda\frac{(K^2_\lambda+m^2)(K^2_\lambda+5m^2)}{4m^4}I_{m0}^{\kappa\tau}(K)-\frac{3}{2}\!\left(K_\mu^\lambda n_\nu+K_\nu^\lambda n_\mu\right)\!\frac{\tilde J_0^\kappa-\tilde J_m^\kappa}{m^2}\Bigg\}.
\end{align}
 \end{widetext}
The transverse and longitudinal projections of this formula lead to \Eqn{eq:PLT} after using $\{I^\lambda_{L,T}\}_{m0}^{\kappa\tau}(K)=\{I^\lambda_{L,T}\}_{0m}^{\tau\kappa}(K)$ and the fact that ${\cal C}_{\kappa\lambda\tau}$ is totally symmetric. Note, in particular, that the last two lines in \Eqn{appeq:lastone} are not transverse (with respect to the generalized momentum $K^\lambda$) and, hence, do not contribute to $\Pi_{T/L}^\lambda(K)$.
}

\section{Gluon susceptibilities}
\label{app:gluon_su}

In this section, we derive the expressions of the gluon susceptibilities \eqn{eq:electric_masses} and \eqn{eq:electric_masses2} in terms of simple one-dimensional integrals. We consider the neutral and charged color sectors separately. 

\subsection{Neutral sector}
From Eq.~(\ref{eq:PLT}), we obtain
\begin{align}
\Pi^0_{T/L}(0,0) & = 2(d-1)J_m^+-2J_0^+ +2m^2I_{m0}^{+-}(0,0)\nn
&+2\{I_{T/L}^0\}_{00}^{+-}(0,0)+2\{I_{T/L}^0\}_{m0}^{+-}(0,0)\nn
&-4(d-1)\{I_{T/L}^0\}_{mm}^{+-}(0,0),\label{eq:uu}
\end{align}
where we used the symmetry properties of the integrals \eqn{eq_integrals}--\eqn{eq_integralsss} and \eqn{eq:integrals2}--\eqn{eq:hehehe}. Here and in the following, we write $0$ for $k\to 0$ for simplicity but we warn the reader that it is sometimes important to take this limit after setting $\omega =0$; see below.

This can be simplified as follows. First, using the identity \eqn{eq:idmom} as well as $G_m(Q)=G_m(-Q)$, we have
\begin{align}
I_{m_1 m_2}^{+-}(0,0) &= \int_Q G_{m_1}(Q^+)G_{m_2}(Q^+),\\
\left\{I_T^0\right\}_{m_1 m_2}^{+-}(0,0) & = \frac{1}{d-1}\int_Q q^2 G_{m_1}(Q^+)G_{m_2}(Q^+),
\end{align}
and
\begin{align}
&\left\{I_L^0\right\}_{m_1 m_2}^{+-}(0,0)  =\int_Q \left(Q_+^2-q^2\right)\! G_{m_1}(Q^+)G_{m_2}(Q^+)\nn
&= J_{m_1}^+-m_2^2I_{m_1 m_2}^{+-}(0,0)-(d-1)\left\{I_T^0\right\}_{m_1 m_2}^{+-}(0,0).
\end{align}
Note that 
\beq
 \left\{I_{T/L}^0\right\}_{m_1 m_2}^{+-}(0,0)=\left\{I_{T/L}^0\right\}_{m_2 m_1}^{+-}(0,0).
\eeq
 For $m_1\neq m_2$, we use the identity 
\beq
 G_{m_1}(Q)G_{m_2}(Q)=-\frac{G_{m_1}(Q)-G_{m_2}(Q)}{m_1^2-m_2^2}
\eeq
which leads to
\begin{align}
 I_{m0}^{+-}(0,0)&=\frac{J_0^+-J_m^+}{m^2}\,,\label{eq:u1}\\
 \left\{I_T^0\right\}_{m0}^{+-}(0,0)&=\frac{1}{d-1}\frac{N_0^+-N_m^+}{m^2}\,,\label{eq:u2}\\
 \left\{I_L^0\right\}_{m0}^{+-}(0,0)&=J_m^++\frac{N_m^+-N_0^+}{m^2}\,,\label{eq:u3}
\end{align}
and allows us to rewrite the above integrals in terms of
\beq
J_{m}^+\equiv\int_Q G_{m}(Q^+)\,\hat{=}\,\frac{1}{2\pi^2}\int_0^\infty dq \frac{q^2}{\varepsilon_{m,q}} {\rm Re}\,n_{\varepsilon_{m,q}-irT}
\eeq
and
\beq
N_{m}^+\equiv\int_Q q^2 G_{m}(Q^+)\,\hat{=}\,\frac{1}{2\pi^2}\int_0^\infty dq \frac{q^4}{\varepsilon_{m,q}} {\rm Re}\,n_{\varepsilon_{m,q}-irT}\,,
\eeq
where the symbol $\hat{=}$ means that we only keep the thermal contributions, since our renormalization scheme is anyway such that the vacuum corrections to the gluon masses of the neutral mode, are zero. 

When $m_1=m_2=m$, we can use
\begin{align}
\int_Q q^{2n} G_{m}^2(Q^+) & =-\int_Q q^{2n}\frac{dG_{m}(Q_+)}{dq^2}\nonumber\\
& \,\,\hat{=}\,\, \frac{2n+1}{2}\int_Q q^{2n-2} G_{m}(Q^+)\nonumber\\
\label{eq:ipp}
&\,\, \hat{=}\, \,\frac{2n+1}{4\pi^2}\int_0^\infty dq \frac{q^{2n}}{\varepsilon_{m,q}}{\rm Re}\,n_{\varepsilon_{m,q}-irT}\,.
\end{align}
In particular, in addition to $J_{m}^+$ and $N_{m}^+$, we are lead to consider
\beq
S_{m}^+\equiv\int_Q G^2_{m}(Q^+)\,\hat{=}\,\frac{1}{4\pi^2}\int_0^\infty dq \frac{1}{\varepsilon_{m,q}} {\rm Re}\,n_{\varepsilon_{m,q}-irT}\,.
\eeq
We have then
\begin{align}
 \left\{I_T^0\right\}_{mm}^{+-}(0,0)&\,\hat =\,\frac{1}{2}J_m^+\,,\\
 \left\{I_L^0\right\}_{mm}^{+-}(0,0)&\,\hat=\,-\frac{1}{2}J_m^+-m^2S_m^+\,,
\end{align}
which, together with Eqs.~(\ref{eq:u1})--(\ref{eq:u3}) allow us to rewrite Eq.~(\ref{eq:uu}) as in Eq.~(\ref{eq:longzero}). We mention that the same results can be obtained from the formulae derived in Appendix~\ref{app:sum_int} after performing the Matsubara sums but it is then important to take the limit $k\to0$ only after setting $\omega=0$.

\subsection{Charged sector}

Similarly, we obtain
\begin{widetext}
\begin{align}
\Pi^+_{T/L}(-rT,0) & = 2\{I_{T/L}^+\}_{00}^{0-}(-rT,0)+\{I_{T/L}^+\}_{m0}^{0-}(-rT,0)+\{I_{T/L}^+\}_{0m}^{0-}(-rT,0)\nn
& -4(d-1)[\{I_{T/L}^+\}_{mm}^{0-}(-rT,0)+m^2\left[I_{m0}^{0-}(-rT,0)+I_{0m}^{0-}(-rT,0)\right]\nonumber\\
& + (d-1)\left(J_m^++J_m^0\right)-J_0^+-J_0^0\,,
\end{align}
\end{widetext}
where all quantities are appropriate analytic continuations (after the Matsubara sums have been performed and the external Matsubara frequency has been removed from the thermal factors using $n_{\varepsilon+i\omega_n}=n_\varepsilon$) evaluated at $\omega=-rT$ and $k\to0$. As before, we write $0$ for $k\to0$ for simplicity but it is important to perform the continuation and set $\omega=-rT$ before taking the limit $k\to0$. For $m_1\neq m_2$, we obtain
\begin{align}
I^{0-}_{m_1m_2}(-rT,0) &\,\hat =\, \frac{J^+_{m_2}-J^0_{m_1}}{m^2_1-m^2_2}\,,\\
\left\{I^+_T\right\}^{0-}_{m_1m_2}\!\!(-rT,0) &\,\hat =\, \frac{1}{d-1}\frac{N^+_{m_2}-N^0_{m_1}}{m^2_1-m^2_2}\,,\\
\left\{I^+_L\right\}^{0-}_{m_1m_2}\!\!(-rT,0) &\,\hat =\, \frac{N^+_{m_2}\!\!-\!N^0_{m_1}+m^2_2J^+_{m_2}\!\!-\!m^2_1J^0_{m_1}}{m^2_2-m^2_1},
\end{align}
where we have used\footnote{We exploit the fact that the replacement $\omega\to-rT$ can be done before the Matsubara sum for any occurrence of $\omega$, except that in the denominators.}
\begin{align}
&\left\{I_L^+\right\}_{\!m_1 m_2}^{0-}\!\!(-rT,0)=\int_Q \omega_n^2 \,G_{m_1}(Q)G_{m_2}(Q+K^+),
\end{align}
where $K^+=(-rT,k\to0)$ on the right-hand side.
For $m_1=m_2=m$, we obtain (note that these are not the limits $m_1\to m_2$ of the formulae given above)
\begin{align}
 \left\{I_T^+\right\}_{mm}^{0-}(-rT,0)&\,\hat=\,\frac{J_m^++J_m^0}{4}\,,\\
 \left\{I_L^+\right\}_{mm}^{0-}(-rT,0)&\,\hat =\,-\frac{J_m^++J_m^0+2m^2\!\left(S_m^++S_m^0\right)}{4}\,,
 \end{align}
We finally obtain
\begin{align}
\label{eq:longch}
 &\Pi_{L}^+(- rT,0)\,\,\hat{=}\,\,6m^2 \left(S_m^0+S_m^+\right)\nn
 &+6(J_m^0+J_m^+)-\frac{1}{2}(J_0^0+J_0^+)+\frac{N_m^0+N_m^+-N_0^0-N_0^+}{m^2}
\end{align}
and
\begin{align}
\label{appeq:tralali}
 \Pi_{T}^+(- rT,0)&\,\,\hat{=}\,\, \frac{1}{2}\left(J_0^0+J_0^+\right) - \left(J_m^0+J_m^+\right) \nn
 &+ \frac{1}{3}\left(N_0^0 +N_0^+- N_m^0 - N_m^+\right).
\end{align}
which are nothing but the average between the respective neutral components and the corresponding expressions at $r=0$. We thus obtain \Eqn{eq:half-sum}.\\

\section{Neutral ghost dressing function}
The neutral ghost self-energy at zero frequency reads
\begin{align}
\Sigma^0(0,k) & = \frac{k^2-m^2}{2m^2} \left(J_m^+-J_0^+ \right)+\frac{k^4}{2m^2}I_{00}^{+-}(0,k)\nonumber\\
& - \frac{\left(k^2+m^2\right)^2}{4m^2} \left[I_{m0}^{+-}(0,k)+I_{m0}^{-+}(0,k)\right]\,.
\end{align}
Using $I_{m0}^{+-}(0,0)=I_{m0}^{-+}(0,0)=(J_0^+-J_m^+)/m^2$, it is easily checked that $\Sigma^0(0,0)=0$ as follows from the anti-ghost shift symmetry. The next term in the expansion of $\Sigma(0,k)$ around $k=0$ is $k^2\sigma(0)$ where $\sigma(0)$ is given by
\bea
 \sigma(0) & = & \frac{J_m^+-J_0^+}{2m^2}-\int_QG_0(Q_+)G_m(Q_+)\nn
 & - & \frac{2m^2}{d-1}\int_Q q^2 G_0^3(Q_+)G_m(Q_+)\nonumber\\
 & + & \frac{m^2}{2}\int_Q G_0^2(Q_+)G_m(Q_+),
\eea
where we have used
\bea
& & I_{m0}^{+-}(0,k)=I_{m0}^{-+}(0,k)\nonumber\\
& & \hspace{0.2cm}=\,\int_Q \frac{1}{Q_+^2+m^2}\frac{1}{Q_+^2}\left(1-\frac{k^2}{Q^2_+}+\frac{4}{d-1}\frac{q^2k^2}{Q_+^4}+\dots\right).\nonumber\\
\eea
Using $G_mG_0=(G_0-G_m)/m^2$ and
\beq
 \int_Q\frac{q^2}{Q_+^6}=-\int_Q\frac{q^2}{Q_+^2}\frac{\partial}{\partial q^2}\frac{1}{Q_+^2}\,\,\hat{=}\,\,\frac{3}{4}\int_Q\frac{1}{Q_+^4},
\eeq
we finally arrive at
\beq
\sigma(0)=\frac{2J_m^+-J_0^+}{m^2}+\frac{2}{3}\frac{N_m^+-N_0^+}{m^4}.
\eeq
We thus explicitly check the identity \eqn{eq:neutral_dressing_0} at one-loop order.


\section{High-temperature expansions}
\label{appsec:highT}

Here, we study in detail the high-temperature expansion of the various gluon susceptibilities \eqn{eq:electric_masses} and \eqn{eq:electric_masses2}. In particular, we wish to study the leading asymptotic behavior of the magnetic susceptibility in the neutral sector, which, for $r\neq0$, has neither $T^2$, nor $mT$ contributions and thus requires a detailed analysis of the various sum-integrals at play. This allows us to get the subleading asymptotic behaviors of all the generalized susceptibilities \eqn{eq:electric_masses}--\eqn{eq:electric_masses2} for free, which we show here for completeness.

We begin our discussion with the neutral sector. This requires some care because the various sum-integrals entering Eqs. \eqn{eq:electric_masses} and \eqn{eq:electric_masses2} are not analytic in $m^2/T^2$. Following the strategy of Ref.~\cite{Laine:2016hma}, we treat separately the contributions from the zero and the nonzero frequency modes in Matsubara sums. The former is infrared sensitive and one must keep track of the full mass dependence. The (regularized) sum over nonzero Matsubara modes can be safely expanded in $m^2/T^2$. Considering a sum-integral $\int_Qf(Q,m)$ whose primitive degree of divergence is zero in $d=4$, this is summarized as
\begin{align}
 &\int_Qf(Q,m)=T\mu^{2\epsilon}\sum_{n\in\mathds{Z}}\int_{\bf q}f(\omega_n,{\bf q},m)\nn
 &=T\!\int_{\bf q}f(0,{\bf q},m)+T\mu^{2\epsilon}\!\sum_{n\neq0}\int_{\bf q}f(\omega_n,{\bf q},m)\nn
 \label{eq:strategy}
 &=T\!\int_{\bf q}f(0,{\bf q},m)+T\mu^{2\epsilon}\!\sum_{n\neq0}\int_{\bf q}f(\omega_n,{\bf q},0)+{\cal O}\left(\frac{m^2}{T^2}\right)\!,
\end{align}
where the three-dimensional spatial integral $\int_{\bf q} f(Q,m)$ is finite by assumption so that we can safely send the ultraviolet regulator $\epsilon\to0$ in the zero-frequency contribution. All the tadpolelike sum-integrals discussed below can be reduced to such a case by first subtracting certain (infrared finite) massless sum-integrals.
Finally, to obtain the relevant thermal contributions, we  subtract the corresponding vacuum parts, 
\beq
 \left.\int_Qf(Q,m)\right|_{\rm vac}=\mu^{2\epsilon}\int\frac{d^dq}{(2\pi)^d}f(q,m)
\eeq
which bring additional logarithmic factors $\ln (T/m)$.

Let us first consider
\beq
 J_m^\lambda=\int_Q\frac{1}{Q_\lambda^2+m^2}.
\eeq 
Subtracting the leading $T^2$ contribution $J_0^\lambda$, we get
\beq
  \frac{J_m^\lambda-J_0^\lambda}{m^2}=-\int_Q\frac{1}{Q_\lambda^2(Q_\lambda^2+m^2)},
\eeq
which is of the form considered in \eqn{eq:strategy}.
The spatial momentum integrals appearing in \Eqn{eq:strategy} are easily computed using ($a,b\in\mathds{R}$)
\beq
 \int_{\bf q}\frac{1}{(q^2+a^2)(q^2+b^2)}=\frac{1}{4\pi\left(|a|+|b|\right)}+{\cal O}(\epsilon)
\eeq
and
\beq
 \int_{\bf q}\frac{1}{(q^2+a^2)^n}=\frac{1}{(4\pi)^{d-1\over2}}\frac{\Gamma\left(n-\frac{d-1}{2}\right)}{\Gamma(n)}\frac{1}{(a^2)^{n-\frac{d-1}{2}}}.
\eeq
Denoting the shifted Matsubara frequency as $\omega_n^\lambda=(2\pi n+\lambda r)T$, we have
\begin{widetext}
\begin{align}
  \frac{J_m^\lambda-J_0^\lambda}{m^2}&= -\frac{T}{4\pi\left(M_\lambda+|\lambda rT|\right)}-\frac{T\mu^{2\epsilon}\Gamma({1\over2}+\epsilon)}{(4\pi)^{3/2-\epsilon}}\sum_{n\neq0}\frac{1}{|\omega_n^\lambda|^{1+2\epsilon}}+{\cal O}\left(\frac{m^2}{T^2}\right)\nonumber\\
  \label{eq:HHH}
  &= -\frac{T}{4\pi\left(M_\lambda+|\lambda rT|\right)}-\left(\frac{\mu^2}{\pi T^2}\right)^{\!\epsilon}\frac{\Gamma\left({1\over2}+\epsilon\right)}{16\pi^{5/2}}{\cal H}\left(1+2\epsilon,\frac{\lambda r}{2\pi}\right)+{\cal O}\left(\frac{m^2}{T^2}\right),
\end{align}
with $M_\lambda\equiv \sqrt{m^2+(\lambda rT)^2}$ and where we defined\footnote{Due to the symmetries of the background field potential (see Sec.~\ref{sec:sym}), it is sufficient to consider $r\in[0,\pi]$. The second argument of the function ${\cal H}(s,z)$ in \Eqn{eq:HHH} is, thus, such that $|z|<1$.}
\beq
 {\cal H}(s,z)=\sum_{n\neq0}\frac{1}{|n+z|^{s}}=\zeta(s,1+z)+\zeta(s,1-z),
\eeq
where
\beq
 \zeta(s,z)=\sum_{n\ge0}\frac{1}{(n+z)^s}
\eeq
is the Hurwitz generalized zeta function. Using
\begin{align}
\Gamma\left({1\over2}+\epsilon\right) & = \sqrt{\pi}\Big[1-\epsilon(\gamma+2\ln 2)+{\cal O}(\epsilon^2)\Big],\\
  \zeta(1+2\epsilon,z)&=\frac{1}{2\epsilon}-\psi(z)+{\cal O}(\epsilon),
\end{align}
where $\psi(z)=\Gamma'(z)/\Gamma(z)$, we obtain
\beq
\frac{J_m^\lambda-J_0^\lambda}{m^2} =  -\frac{T}{4\pi\left(M_\lambda+|\lambda rT|\right)}+\frac{1}{16\pi^2}\left[-\frac{1}{\epsilon}+\ln\frac{4\pi T^2}{\mu^2}+\Psi\left(\frac{\lambda r}{2\pi}\right)+\gamma\right]+{\cal O}\left(\frac{m^2}{T^2},\epsilon\right).
\eeq
where we have introduced 
\beq
 \Psi(z)=\psi(1+z)+\psi(1-z)
\eeq
Subtracting the zero-temperature contribution, which reads, in dimensional regularization,
\beq
\left.\frac{J_m^\lambda-J_0^\lambda}{m^2}\right|_{\rm vac}=\frac{\mu^{2\epsilon}}{m^2}\int\frac{d^dq}{(2\pi)^d}\frac{1}{q^2+m^2}= -\frac{1}{16\pi^2}\left(\frac{1}{\epsilon}+\ln\frac{4\pi\mu^2}{m^2}-\gamma+1\right)+{\cal O}(\epsilon),
\eeq
we obtain, for the thermal part,
\beq
\label{eq:highT-J}
\left.\frac{J_m^\lambda-J_0^\lambda}{m^2}\right|_{\rm th}=-\frac{T}{4\pi\left(M_\lambda+|\lambda rT|\right)}+\frac{1}{16\pi^2}\left[\ln\left(\frac{4\pi T}{m}\right)^{\!\!2}+\Psi\left(\frac{\lambda r}{2\pi}\right)+1\right]+{\cal O}\left(\frac{m^2}{T^2}\right).
\eeq

Next, we consider
\beq
 N_m^\lambda=\int_Q\frac{q^2}{\left(Q_\lambda^2+m^2\right)}
\eeq
After subtracting the leading $T^4$ and $m^2T^2$ contributions, we are led to consider the following sum-integral
\beq
\label{eq:rhs}
  \frac{N_m^\lambda-N_0^\lambda}{m^4}+\frac{1}{m^2}\int_Q\frac{q^2}{Q_\lambda^4}=\int_Q\frac{q^2}{Q_\lambda^4(Q_\lambda^2+m^2)}.
\eeq
In dimensional regularization, $J_0^\lambda|_{\rm vac}=0$ so that, using \eqn{eq:ipp}, the second term on the left-hand side rewrites
\beq
 \int_Q\frac{q^2}{Q_\lambda^4}=\frac{3}{2}J_0^\lambda.
\eeq
It is also easy to check that
\beq
\left.\frac{N_m^\lambda-N_0^\lambda}{m^4}\right|_{\rm vac}=-\frac{d-1}{d}\left.\frac{J_m^\lambda-J_0^\lambda}{m^2}\right|_{\rm vac}
\eeq
Finally, the right-hand side of \Eqn{eq:rhs} can be treated as in \Eqn{eq:strategy} using  
\beq
 \int_{\bf q}\frac{q^2}{(q^2+a^2)^2(q^2+b^2)}=\frac{|a|+2|b|}{8\pi\left(|a|+|b|\right)^2}
\eeq
for the zero-mode contribution and
\beq
 \int_{\bf q}\frac{q^2}{(q^2+a^2)^{n+1}}=\frac{1}{(4\pi)^{d-1\over2}}\frac{d-1}{2}\frac{\Gamma\left(n-\frac{d-1}{2}\right)}{\Gamma(n+1)(a^2)^{n-\frac{d-1}{2}}}
\eeq
for the nonzero modes. After similar calculations as in the previous case, we get
\beq
\label{eq:highT-N}
\left.\left(\frac{N_m^\lambda-N_0^\lambda}{m^4}+\frac{3}{2m^2}J_0^\lambda\right)\right|_{\rm th}=\frac{T\left(2M_\lambda+|\lambda rT|\right)}{8\pi\left(M_\lambda+|\lambda rT|\right)^2}-\frac{3}{64\pi^2}\left[\ln\left(\frac{4\pi T}{m}\right)^{\!\!2}+\Psi\left(\frac{\lambda r}{2\pi}\right)+\frac{3}{2}\right]+{\cal O}\left(\frac{m^2}{T^2}\right)\,.
\eeq
One easily checks that the high-temperature expansions \eqn{eq:highT-J} and \eqn{eq:highT-N} satisfy the identity
\beq
  \left.\frac{\partial N_m^\lambda}{\partial m^2}\right|_{\rm th}=\left.-\frac{3}{2}J_m^\lambda\right|_{\rm th},
\eeq
which follows from \Eqn{eq:ipp}.
The last case we need to consider is
\beq
 S_m^\lambda=\int_Q\frac{1}{\left(Q_\lambda^2+m^2\right)^2}=-\frac{\partial J_m^\lambda}{\partial m^2}.
\eeq
We get 
\beq
\left.S_m^\lambda\right|_{\rm th}=\frac{T}{8\pi M_\lambda}-\frac{1}{16\pi^2}\left[\ln\left(\frac{4\pi T}{m}\right)^{\!\!2}+\Psi\left(\frac{\lambda r}{2\pi}\right)\right]+{\cal O}\left(\frac{m^2}{T^2}\right)\,.
\eeq

Using the above results in Eqs.~\eqn{eq:longzero} and \eqn{eq:transzero}, we finally obtain, for the high-temperature expansions of the neutral gluon masses, 
\begin{align}
\label{appeq:eef}
 &M^2_{{\rm D},0}(T)=g^2T^2\left[\left(\frac{r}{\pi}-1\right)^2-\frac{1}{3}\right]\nn
 &+m^2\left\{1+\frac{g^2T\left[M_+(2M_++rT)-6m^2\right]}{4\pi M_+(M_++rT)^2}-\frac{3g^2}{32\pi^2}\left[\ln\left(\frac{4\pi T}{m}\right)^{\!\!2}+\Psi\left(\frac{ r}{2\pi}\right)-\frac{13}{2}\right]+{\cal O}\left(\frac{m^2}{T^2}\right)\right\}
\end{align}
and
\beq\label{eq:micheline}
 M^2_{{\rm mag},0}(T)=m^2\left\{1+\frac{g^2T(4M_++5rT)}{12\pi (M_++rT)^2}-\frac{3g^2}{32\pi^2}\left[\ln\left(\frac{4\pi T}{m}\right)^{\!\!2}+\Psi\left(\frac{ r}{2\pi}\right)+\frac{5}{6}\right]+{\cal O}\left(\frac{m^2}{T^2}\right)\right\}\,.
\eeq
A similar analysis can be performed for the charged gluon masses defined in Eqs.~\eqn{eq:electric_masses} and \eqn{eq:electric_masses2}. Using the relation \eqn{eq:half-sum}, valid at one-loop order, we have
\begin{align}
 &M^2_{{\rm D},\pm}(T)=\frac{g^2T^2}{2}\left[\left(\frac{r}{\pi}-1\right)^2+\frac{1}{3}\right]-\frac{g^2mT}{2\pi}\nn
 &+m^2\left\{1+\frac{g^2T\left[M_+(2M_++rT)-6m^2\right]}{8\pi M_+(M_++rT)^2}-\frac{3g^2}{32\pi^2}\left[\ln\left(\frac{4\pi T}{m}\right)^{\!\!2}+\frac{1}{2}\Psi\left(\frac{ r}{2\pi}\right)-\gamma-\frac{13}{2}\right]+{\cal O}\left(\frac{m^2}{T^2}\right)\right\}
\end{align}
\beq
\label{appeq:casimir}
 M^2_{{\rm mag},\pm}(T)=\frac{g^2mT}{6\pi}+m^2\left\{1+\frac{g^2T(4M_++5rT)}{24\pi (M_++rT)^2}-\frac{3g^2}{32\pi^2}\left[\ln\left(\frac{4\pi T}{m}\right)^{\!\!2}+\frac{1}{2}\Psi\left(\frac{ r}{2\pi}\right)-\gamma+\frac{5}{6}\right]+{\cal O}\left(\frac{m^2}{T^2}\right)\right\}\,.
\eeq
\end{widetext}

In the expressions \eqn{appeq:eef}--\eqn{appeq:casimir}, it is understood that $r=r_{\rm min}(T)$. We see, in particular, that for $r\neq0$, the leading-order behavior of the magnetic square mass (\ref{eq:micheline}) is $\sim m^2$, to be contrasted with $\sim g^2mT$ at $r=0$, which corresponds to the result in the Landau gauge \cite{Reinosa:2013twa}. As pointed out in Ref.~\cite{Reinosa:2014zta}, the background never vanishes for finite temperatures and, as can be seen in \Fig{fig_rT}, the two-loop solution quickly reaches its asymptotic value 
\beq
\label{appeq:rinf}
 \frac{r_\infty}{\pi}=1-\sqrt{\frac{8\pi^2+g^2}{8\pi^2+7g^2}}.
\eeq
Although this strict two-loop result neglects renormalization group and hard thermal loop effects, we expect that the background only reaches zero at asymptotically high temperatures, with the logarithmic running of the coupling. It thus make sense to analyze the above high temperature expressions using the value \eqn{appeq:rinf} for not too high temperatures. For instance, the leading behavior of the neutral electric square mass,
\beq
\label{eq:leading}
 M^{2,\infty}_{{\rm D},0}=\frac{2}{3}g^2T^2\frac{1-g^2/4\pi^2}{1+7g^2/8\pi^2}+{\cal O}(m^2),
\eeq
only makes sense if the (running) coupling at the relevant scale remains such $g<2\pi$. 

\begin{figure}[t!]  
\begin{center}
\epsfig{file=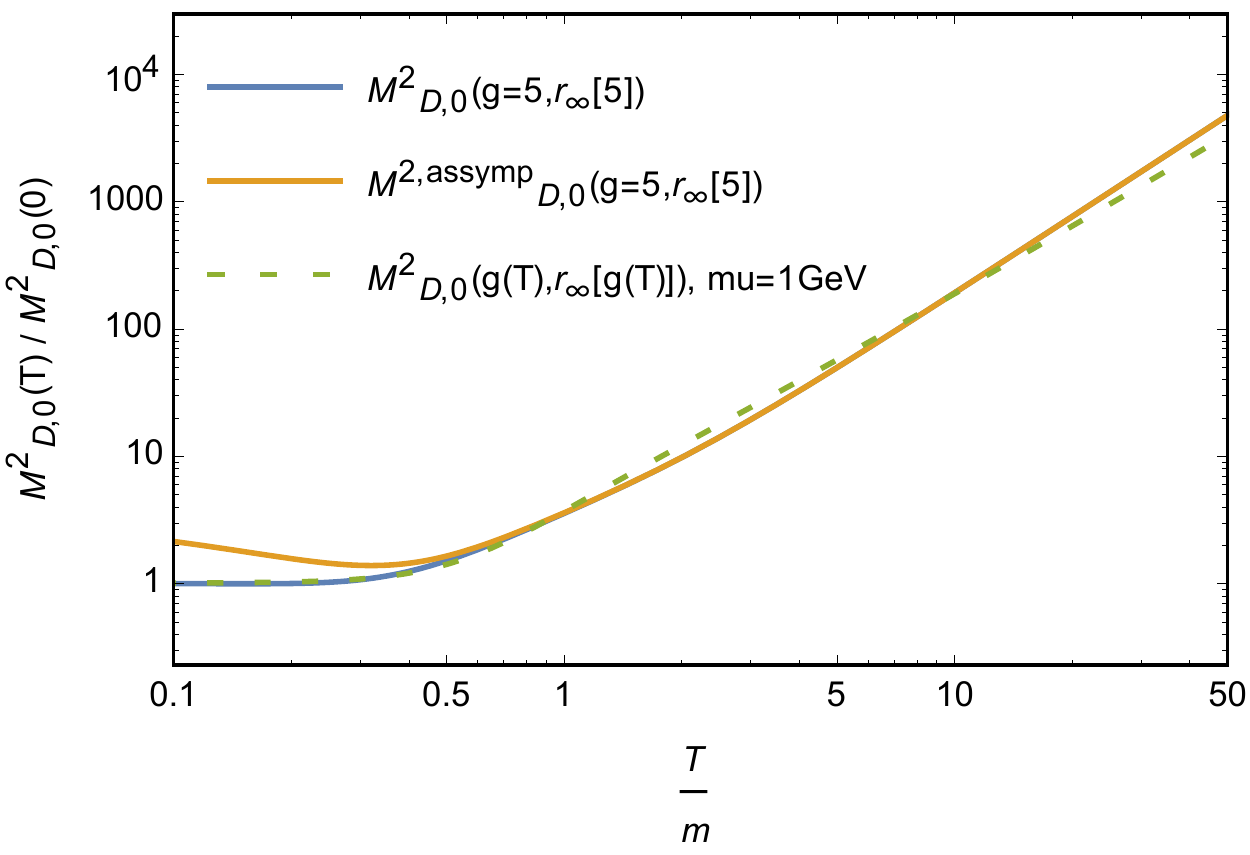,width=8cm}\\
\epsfig{file=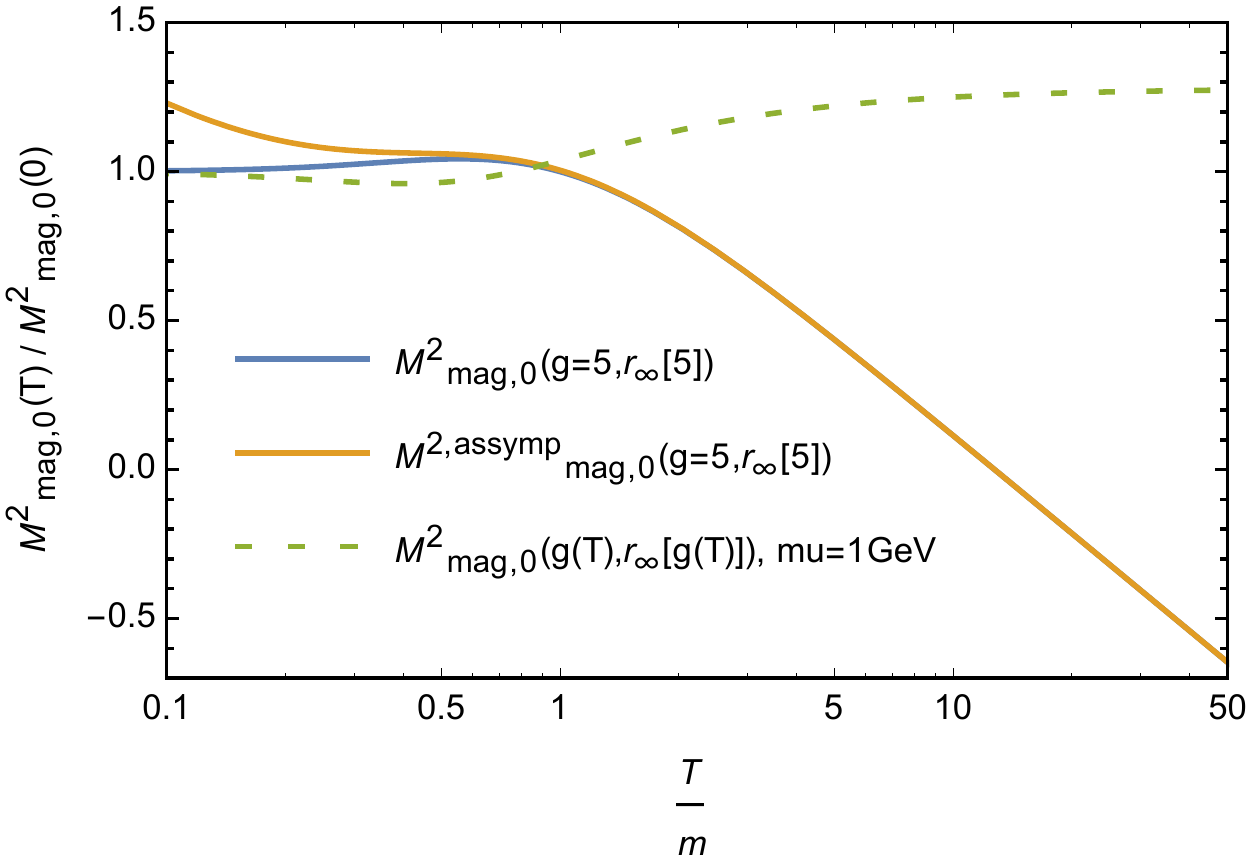,width=8cm}
 \caption{The electric (top) and magnetic (bottom) square masses in the neutral sector as functions of $T/m$ compared to the corresponding asymptotic expressions \eqn{appeq:eef} and  \eqn{eq:micheline} with $r\to r_\infty$ and $g=5$. One clearly observes the negative logarithmic behavior which drives the magnetic square mass to negative values  at large $T/m$. The dashed curve illustrates how this artifact is cured by the running of the coupling: We plot the magnetic square mass computed with a running coupling put by hand as $1/g^{2}(T)=1/g^{2}+{11\over 24\pi^2}\ln\left(\frac{T^2+m^2}{\mu^2}\right)$, with $m=0.75$~GeV, $g=5$, and $\mu=1$~GeV.}\label{fig:highT}
\end{center}
\end{figure}

We also note that the strict perturbative expression \eqn{eq:micheline} eventually becomes negative for any value of the coupling at large temperatures, driven by the negative logarithmic contribution. Again, neglecting renormalization group and hard thermal loop effects is not justified in this regime and one easily checks that the ultraviolet running of the coupling cures this unphysical behavior. The large-temperature behaviors of the neutral square masses, including a simple ansatz for the running coupling, are illustrated in \Fig{fig:highT}. 

\end{document}